\documentclass[aps,pre,amsmath,twocolumn,amssymb,nofootinbib, superscriptaddress]{revtex4}
\usepackage[colorlinks,citecolor=blue,bookmarks]{hyperref}
\usepackage{graphicx,amssymb,rotate}
\usepackage{mathrsfs}
\usepackage{color}
\usepackage{multirow}
\usepackage{float,lscape}
\usepackage{pstricks}
\usepackage[toc,page]{appendix}
\usepackage[mathscr]{euscript}
\usepackage{textcomp}
\usepackage{hhline}
\usepackage[ddmmyyyy,hhmmss]{datetime}
\def\ADNDT{{At. Data Nucl. Data Tables }}
\def\RPP{Rep. Prog. Phys. }
\def\NP{Nucl. Phys. A }
\def\PR{Phys. Rev. }
\def\PRL{Phys. Rev. Lett. }
\def\PL{Phys. Lett. }
\def\jpg{J. Phys. G: Nucl. Part. Phys. }
\def\PRC{Phys. Rev. C }
\def\NDT{{Nucl. Data Tables }}
\usepackage{rotating}
\usepackage{comment}
\usepackage{caption}
\usepackage{footnote}
\usepackage[labelformat=parens,labelsep=quad, skip=3pt]{caption}
\usepackage{subfig}
\usepackage{mathtools}
\usepackage{array}
\def\eg{ {\em e.g.\ }}
\def\etc{ {\em etc.\ }}
\def\ie{ {\em i.e.,\ }}
\def\viz{ {\em viz.\ }}
\def\etal{\!{\em et al.\ }}
\def\mpl{M_{\rm Pl}}
\def\hw{\qquad \boxed{\mathbf {HW}}}
\def\mET{E_T \hspace{-1.0em}/\;\:}
\def\Eqn#1{Eq.\ (\ref{#1})}
\def\Eqs#1#2{Eqs.\ (\ref{#1}) and (\ref{#2})}
\def\arrow{$\rightarrow$}
\def\logft{log$~ft$ }
\def\bta{$\beta^-$}
\newcolumntype{L}{>{$}c<{$}}
\newcolumntype{M}{>{\centering\arraybackslash}m{2.2cm}}
\newcolumntype{N}{>{\centering\arraybackslash}m{4cm}}

\begin{document}
\title{Allowed $\beta^-$ Decay of Bare Atoms (A$\approx$ 60\textendash  80) in Stellar Environments}

\author {Arkabrata Gupta}
\affiliation {Department of Physics, Indian Institute of Engineering Science and Technology, Shibpur, Howrah-711103, India}

\author {Chirashree Lahiri}
\affiliation  {Department of Physics, Surendranath Evening College, 24/2 M. G. Road, Kolkata- 700009, India}

\author {S. Sarkar}
 \thanks{ Corresponding author: ss@physics.iiests.ac.in}
\affiliation {Department of Physics, Indian Institute of Engineering Science and Technology, Shibpur, Howrah-711103, India}

\today 
\hspace*{.33cm}\currenttime

\vspace*{0.9 cm}

\begin{abstract}

We have calculated \bta decay rates to the continuum and bound states of some fully ionized atoms in the stellar s-process environment having free electron density and temperature in the range  $n_e = 10^{26} $ cm$^{-3}\textendash 10^{27} $ cm$^{-3}$ and $T = 10^8$  K\textendash $5 \times 10^{8}$ K, respectively. The presence of bare atoms in these particular situations has been confirmed by solving Saha ionization equation taking into account the ionization potential depression (IPD). At these temperatures, low lying excited energy levels of parent nuclei may have thermal equilibrium population and those excited levels may also decay via \bta emission. The Nuclear Matrix Element (NME) 
of all the transitions of the set of 15 nuclei is calculated using nuclear shell-model. These NME are then used to calculate the comparative half-life ($ft_{1/2}$) of the transitions. Calculated terrestrial half-lives of the \bta decays are in good agreement with the experimental results in most of the cases. Decay to bound and continuum states of bare atoms from ground/isomeric levels and excited nuclear levels have been calculated separately. The ratio of bound state to continuum state decay rates as a function of IPD modified $Q$-value reveals that bound state \bta decay rate may compete and even dominate for $Q$-value $<$ 100 keV. 
The importance of the bound state \bta decay in stellar situations has been shown explicitly. We have calculated total \bta decay rates (bound state plus continuum state) taking into account IPD corrected neutral atom $Q$-value as a function of density and temperature. We have also presented results for the stellar \bta half-lives and compared the ratio of neutral atom to bare atom half-lives for different density and temperature combinations. These results may be useful for s-process nucleosynthesis calculations.

\end{abstract}

\maketitle


\section{Introduction}

The \bta decay is a weak interaction process that allows the conversion of a neutron into a proton with the creation of an electron and an antineutrino in the continuum state. Terrestrial \bta decay has been studied through the decades both experimentally and theoretically since its discovery that enriched the knowledge of nuclear interaction process and nuclear structure. 
The terrestrial \bta decay of atomic nucleus occurs from the nuclear ground state and isomeric states. However, the scenario differs when parent and daughter atoms are in a stellar environment. It is possible even for the high Z ($\simeq$ 35) elements to be partially or fully ionized, due to high temperature ($\approx 10^8$ K), wherein free electron density also plays a role \cite{Donald_1983}.  This creates vacancy in atomic orbits. 
Also, the environmental condition leads to the depression of ionization potential which in turn not only changes the $Q$-value of \bta transitions but also affects the charge state distribution of the atoms. Availability of vacancy, i.e., free phase space in the atomic orbits may lead to another type of \bta decay, known as bound state \bta decay. In 1947, Daudel \etal \cite{daudel_JPR_1947} first theoretically predicted this new branch of \bta decay as the phenomenon of the creation of an electron in the empty bound atomic orbit. This is just the time reversed process of atomic orbital electron capture. Later, in the early 1960s, John Bahcall used renormalized V-A theory to calculate the bound state \bta decay rate \cite{bahcall_PR_1961}. After many years, in the '80s Takahashi \etal \cite{Takahashi_ADNDT_1987, Takahashi_NP_1983} made more elaborate studies of the bound state \bta decay in the context of nuclear astrophysics. The study of bound state decay has also become relevant in other context such as, in the study of atomic effects on \bta decay \cite{Budick_PRL_1983, Pyper_PRS_1988}. Takahashi and Yokoi \cite{Takahashi_NP_1983} calculated \bta decay rates including bound state \bta decay of a number of nuclei relevant for the s-processes. 
Takahashi and co-workers \cite{Takahashi_PRC_1987} also predicted a way to observe this phenomenon in a terrestrial laboratory. In the following decade, Jung \etal \cite{Jung_PRL_1992} first succeeded in experimentally observing this phenomenon in the case of the $^{163}$Dy atom  by storing the fully ionized parent atom in a heavy ion storage ring. After that, Bosch \etal \cite{Bosch_PRL_1996} studied the bound state \bta decay for bare $^{187}$Re which was helpful for the calibration of $^{187}$Re-$^{187}$Os galactic chronometer \cite{Yokoi_AA_1983}. Further experiment with bare $^{207}$Tl \cite{Ohtsubo_PRL_2005} showed the simultaneous measurement of bound and continuum state \bta decay rate. Experimental study of bound state \bta decay of $^{205}$Tl$^{81+}$ ions also has been done recently \cite{Litvinov_RPP_2011, Sidhu_Thesis_2021}.  However, to study the role of this phenomenon in the context of stellar processes, such as nucleosynthesis, one has to rely only on the theoretical predictions of the \bta decay rates that include both bound and continuum state decays.

We could not trace in literature any further theoretical study on bound state \bta decay in the context of nuclear astrophysics after the works of Takahashi and Yokoi \cite{Takahashi_ADNDT_1987}. 
With the availability of more accurate modern day experimental \bta decay half-lives, branching, energetics, we studied both the bound state and the continuum state \bta decay from the ground state and isomeric states of the parent nuclei in 2019 \cite{Gupta_PRC_2019}. In that study, we have shown the maximum possible \bta decay rate of bare atom in the mass range A = 60\textendash 240. Apart from this, we examined the case studies as mentioned in the work of Takahashi and Yokoi \cite{Takahashi_ADNDT_1987}. We had also shown for the first time, that for some nuclei, it is possible that the \bta branching may flip \cite{Gupta_PRC_2019} in comparison to the terrestrially measured branching, if the contribution from bound state \bta decay is taken into account. It was shown \cite{Takahashi_NP_1983, Gupta_PRC_2019} that bound state \bta decay is possible for the transitions which have low and negative $Q$-value if the binding energy of the atomic shell is large enough to make the $Q$-value positive. Following this, recently Liu \etal \cite{Liu_PRC_2021,Liu_CPC_2021} also studied bound state \bta half-lives for bare atoms.

Terrestrially, as mentioned above, only the ground state and a few of the isomeric levels decay via \bta emission. However, in proper stellar environments, there is a definite probability of thermal equilibrium population of higher excited nuclear levels. In that case \bta decay from those levels may come into play, if allowed by the energetics and \bta decay selection rules.  

In an earlier attempt, we reported \cite{NIC_2021} the calculated total \bta decay rates of an atom in its bare form to the bound and continuum states for the s-process situation using only experimentally available $ft_{1/2}$ (comparative half-life, commonly termed as $ft$) values. 

In the present attempt, we have calculated both bound and continuum state \bta decay for some fully ionized atoms in the mass range A $=$ 59{\textendash}81 in a stellar environment assumed to exist during the s-process. For example, one may consider \cite{Takahashi_NP_1983} that the environment mainly consists of 75\% bare H, $\approx$25\% bare He, and traces of heavy ions floating in the ionized sea of H and He.
 Temperature ($T$) and free electron number density ($n_e$) of the environment have been chosen, following Takahashi and Yokoi \cite{Takahashi_NP_1983}, in the range $T = 10^8$K\textendash $5{\times}10^{8}$K and  $n_e = 10^{26} $cm$^{-3} \textendash 10^{27} $cm$^{-3}$, respectively. Experimental \bta decay half-life and branching for transitions from the ground state and isomeric states in these nuclei are available \cite{NNDC} presently. Consequently, the experimental values of the comparative half-life for these transitions are available. However, the comparative half-lives corresponding to the \bta transitions from the nuclear excited levels are not available, since these decays do not occur terrestrially. 

In their work, Takahashi and Yokoi \cite{Takahashi_NP_1983, Takahashi_ADNDT_1987} had taken the contribution of the \bta decay rate from the nuclear excited states to calculate the total \bta decay rate of the parent nuclei. However, to calculate the decay rate from these nuclear excited states, they estimated the comparative half-lives in different ways. For example, in some cases, they had adopted average $ft$ values from older systematic, and even in some cases they had used a single $ft$ value for all transitions from a parent level. In this work, we have evaluated the $ft$ values for relevant allowed \bta transitions for all nuclei in the range by realistic nuclear shell-model calculations. 
 
 In this paper, we present bound and continuum state \bta decay rates separately to reveal the importance of the former rate for stellar evolution processes.
 
To confirm the presence of bare atom, Saha Ionization equation has been solved. The required ionization potential \cite{NIST} has been modified using the Ionization Potential Depression (IPD). IPD has been estimated using the fitted formula of Takahashi and Yokoi \cite{Takahashi_NP_1983} which was based on Stewart-Pyatt model \cite{Stewart_AJ_1966} to account for the environmental conditions. 

The paper is organized as follows: Sec. \ref{sec.2} contains the methodology of our entire calculation for \logft and bare atom \bta decay rates. In Sec. \ref{sec.3A}, we have discussed shell-model calculation of $ft$ values. The Sec. \ref{sec.3B} contains the discussion about the availability of bare atoms in different stellar environments, variation of decay rates with temperature, and density, variation of bound to continuum decay rate ratio, and total \bta decay rates in details. Also, in this section, we have presented the change in \bta half-life in stellar environment in comparison with terrestrial half-life. In Sec. \ref{sec.4} the conclusion of this work has been discussed. Later, in Appendix  \ref{app_shell} we have discussed briefly the procedure to choose the model space and Hamiltonian for calculation of \logft using nuclear shell-model. The method to find the GT quenching factor has been discussed in Appendix \ref{app_quenching}, and the Saha ionization equation in Appendix \ref{appB}.


\section{{Methodology} \label{sec.2}}

\subsection {{$\beta^-$ Decay Rate}\label{sec.2A}}

 In this work, we have dealt with the allowed \bta transitions of some nuclei involved in the s-process in the mass range $A = 59 \textendash 81$. The contributions of forbidden transitions, though not many in numbers, are negligible in the determination of the \bta decay rate for these nuclei and thus we have not taken those forbidden transitions into account. The experimental \logft values for the allowed transitions range from 4.26 to 8.72, whereas the experimental \logft values for the first forbidden \bta transitions are lying in the range 9.80\textendash 11.14. The nuclei in this range which have sizable contribution from non-unique or unique forbidden transitions are not taken in the present study. 

The transition rate (in s$^{-1}$) for an allowed (a) transition ($Z-1 \rightarrow Z $) is given by \cite{Takahashi_NP_1983}

\begin{eqnarray} \label{Eq1}
\lambda = [(\text {ln~} 2)/ft](f^{*}_{a}).  
\end{eqnarray}

Here $t$ is the partial half-life of the specific $\beta ^{-}$ transition and $f^{*}_{a}$ is the lepton phase volume part for allowed decays described below. The $ft$ values have been obtained via shell-model calculations. 

The lepton phase volume $f^{*}_{a}$ \cite{Takahashi_NP_1983} for the continuum state  $\beta^-$ decay can be expressed as,

\begin{eqnarray} \label{Eq2}
f^{*}_{a}(Continuum) = \int^{W_c}_1\sqrt{(W^2-1)}  \\ \nonumber
 \times  W (W_c-W)^2 F_0(Z,W) L_0 (1-f_{FD}(\eta,\beta)) dW
\end{eqnarray},
 
with,
 \begin{eqnarray}
f_{FD}(\eta,\beta) =\frac{1}{1+ \text{exp}(\beta(W-1)-\eta)}.
\end{eqnarray}

Here the factor $(1-f_{FD}(\eta,\beta))$ is taking care of Pauli's exclusion principle, $W_c = Q_c/m_ec^2 +1 $ is the maximum energy available to the emitted $\beta^-$ particle, $\beta = m_ec^2/k_BT$, $\eta$ is the electron degeneracy parameter (i.e., chemical potential without the rest mass divided by $k_BT$) and can be obtained from the electron number density $n_e$ where,

\begin{eqnarray}
n_e = \int^{\infty}_1 W \sqrt{W^2-1} [1+\text{exp}\{\beta(W-1)-\eta\}]^{-1} \\ \nonumber
\times  dW/(\pi^2 \lambdabar^3),
\end{eqnarray}.

Here $\lambdabar = \hbar/m_ec$. $Q_c$ is given by,

\begin{eqnarray} \label{Eq3}
Q_c = Q_m - \left[ B_n(Z) - B_n(Z-1)\right].
\end{eqnarray}

Where,

\begin{eqnarray} \label{Eq4}
Q_m = Q_n -( \sum_{j_D = 0}^{Z_{D}-1} \Delta_{j} - \sum_{j_P = 0}^{Z_{P}-1} \Delta_{j}).
\end{eqnarray}
     
$Q_n$ is the neutral atom $Q$-value of \bta transition and $Q_m$ is the IPD modified $Q$-value. $\Delta_{j}$ is the ionization potential depression \cite{Takahashi_NP_1983}. The term $\left[ B_n(Z) - B_n(Z-1)\right]$ denotes the difference of binding energies for bound electrons of the daughter and the parent atom. The experimental values for all the atomic data (binding energies/ionization potential) are availed from Ref. \cite{NIST}. 

Certain combinations of electron radial wave functions evaluated at nuclear radius R (in the unit of $\hslash/m_ec$) were first introduced by Konopinski and Uhlenbeck \cite{Konopinski_PR_1941} as $L_k$'s. Behrens and J\"anecke \cite{Behrens_Springer_1969} had made precise calculation of $L_0$ for nuclei close to the valley of stability. To cover all isotopes between the proton and neutron drip lines for $Z\leq 60$, Wilkinson \cite{Wilkinson_NIMP_1990} provided a momentum dependent fitted expression of $L_0$, 
 
\begin{eqnarray}
L_0 = 1 + \dfrac{13}{60} (\alpha Z)^2 - \dfrac{\alpha Z W R (41-26 \gamma)}{15(2 \gamma - 1)} - \\ \nonumber
\dfrac{\alpha Z R \gamma (17-2 \gamma)}{30 W (2 \gamma - 1)} + a_{-1} \dfrac{R}{W} + \sum_{n=0}^{5} a_n(WR)^n + \\ \nonumber
0.41 (R-0.0164)(\alpha Z)^{4.5},
\end{eqnarray}

with $\gamma = \sqrt {1-(\alpha Z)^2}$. Here, $\alpha$ is the fine structure constant. The parameter $a_n$ (for $n = -1$ to 5) is defined as,
\begin{eqnarray}
a_n = \sum_{x=1}^{6}b_{x,n}(\alpha z)^{x}
\end{eqnarray}
Details of $b_{x,n}$ has been discussed in \cite{Wilkinson_NIMP_1990, Hayen_RMP_2018}.

Quantities presented in this paper are also worked out with momentum independent $L_0$ \cite{Konopinski_PR_1941}
\begin{eqnarray}
L_0 = \dfrac{1+\sqrt{1-\alpha^2 Z^2}}{2}.
\end{eqnarray}
We have found that decay rates with this $L_0$ are within $1\% - 2\% $ of the results presented in this paper. The decay rates and related quantities corresponding to momentum independent $L_0$ has been presented in the supplemental material of this paper.

In Eq.\ref{Eq2}, $W$ is the total energy of the $\beta^-$ particle for a $Z-1 \rightarrow Z $ transition. Here the mass difference between initial (parent) and final (daughter) states of neutral atoms are expressed as the decay $Q$-value ($Q_n$ in keV). The term $F_0(Z,W)$ is the Fermi function for allowed transition given by \cite{Konopinski_PR_1941}

\begin{eqnarray}
F_0(Z,W)=\dfrac{4}{ \left|\Gamma \left( {1+2\gamma}\right)\right|^2}\\ \nonumber
\left(2R\sqrt{W^2-1}\right)^{2\left(\gamma -1\right)}\text{exp}\left[\dfrac{\pi \alpha Z W}{\sqrt{W^2-1}}\right] \\ \nonumber
\times \left|{\Gamma{\left(\gamma + i \dfrac{\alpha Z W}{\sqrt{W^{2}-1}} \right)}}\right |^2.
\end{eqnarray}

Further, for the bound state  $\beta^-$ decay of the bare atom  $f^{*}_{a}$ takes the form \cite{Takahashi_NP_1983}

\begin{eqnarray}
f^{*}_{a}(Bound) = \sum_x \sigma_x \left(\pi/2\right) g_x^2 b_x^2 ,\\ \nonumber
\left(\text {for  } x=ns_{1/2},np_{1/2}\right).
\end{eqnarray}

Here $g_x$ is the large component of electron radial wave function evaluated at the nuclear radius $R$ of the daughter for the orbit $x$. The $g_x$ is obtained by solving Dirac radial wave equations \cite{Bethe}. Here $x$ is taken as $1s_{1/2}, 2s_{1/2}, 3s_{1/2}$ and $4s_{1/2}$. Effect of $np_{1/2}$ wave functions is negligible. In this case, $\sigma_x$ is the relative vacancy of the orbit, that in the case of bare atoms is 1, and $b_x = Q_b/m_ec^2$ where,

\begin{eqnarray} \label{Eq8}
Q_b = Q_m - \left[ B_n(Z) - B_n(Z-1)\right]+B_{shell}(Z).
\end{eqnarray}
 
For example, in case of a bare atom, if the emitted $\beta^-$ particle gets absorbed in the atomic K  shell, then the last  term of Eq.\ref{Eq8} will be the ionization potential for the K electron denoted by $B_K(Z)$, and a positive value for it has been used. In Ref. \cite{Gupta_PRC_2019}, Eq. (15) is the same as Eq.\ref{Eq8} of this paper. In Ref. \cite{Gupta_PRC_2019} we used a negative value for the ionization potential. The effect of IPD on $f^{*}_{a}$ has been discussed in Sec. \ref{3B-II}.

\subsection {{Population of Excited Nuclear Energy Levels in the Thermodynamic Equilibrium}\label{sec.2B}}

In stellar environment due to high temperature, there is a definite probability of an equilibrium population of excited nuclear levels given by the Boltzmann distribution. These excited levels may also decay via $\beta^-$ emission. Thus to incorporate these decays, we take the equilibrium population derived from

\begin{eqnarray} \label{Boltzman}
\frac{n_{ik+1}}{n_{ik}} = \frac{b_{ik+1}}{b_{ik}} ~~\text{exp} (-\Delta E_{ik}/k_BT).
\end{eqnarray}

Where, the fractional population of the element $i$ in its $k$-th nuclear state is expressed as $n_{ik}$.  $b_{ik}$ is the multiplicity of the k-th state and $\Delta E_{ik}$ is the energy difference between $k$-th and ($k+1$)-th nuclear levels. Change in the ground state and excited state population in thermal equilibrium in parent due to the reverse $\beta$ decay of the daughter is not possible for any of the nuclei considered. 

The total stellar \bta decay rate ($\lambda_{bare(s)}$) of a bare atom is given by,
\begin{equation}
  \lambda_{bare(s)} = \sum_{k}(n_{ik} \sum_{m} \lambda_{km}).
\end{equation}

Where, $\lambda_{km} = \lambda_b + \lambda_c$ is \bta decay rate of bare atoms from k-th level of the parent nucleus to the m-th level of the daughter nucleus. $\lambda_b$ and $\lambda_c$ are the rates for the decays to the bound and continuum state, respectively. 

\subsection {{\logft Calculation}\label{sec.2C}}

In the case of $\beta^-$ decay the comparative half-life $ft$ corresponding to each transition can be calculated using \cite{Brown_ADNDT_1985}

\begin{eqnarray} \label{Eq10}
ft = \frac{6177}{((g_A/g_v)q)^2  B(GT) + B(F)} ~\text{s}.
\end{eqnarray}

The factor $B(GT)$ is the reduced Gamow-Teller strength, from which one can define a matrix element $M(GT)$ as \cite{Brown_ADNDT_1985}

\begin{eqnarray}
M(GT) = \sqrt{(2J_i + 1) B(GT)}.
\end{eqnarray}

$B(F)$ is the reduced Fermi strength. $J_i$ is the total angular momentum of the parent state. $g_A$ and $g_V$ are the weak interaction vector and axial-vector coupling constants for the decay of a neutron to proton, respectively. The ``free nucleon'' (i.e., for the decay of a neutron to proton) value of $|g_A/g_V|$ is \cite{Wilkinson_NP_1982}

\begin{eqnarray}
|g_A/g_V| = 1.2606 \pm 0.0075. 
\end{eqnarray} 
 The concept of quenching of GT strength arose out of the fact that the sum rule observed in experiment is in general less than that predicted by shell-model calculations. 
The ratio of observed strength and predicted strength is taken as the q-factor. Quenching factor q is used as the normalization of the GT operator which is understood as general inadequacies inherent in the truncated shell-model calculation.


Thus comparison between experimental and theoretical $M(GT)$ leads to the Gamow-Teller strength quenching factor $q$. This is further discussed in Appendix \ref{app_quenching}.


\section{{Results and Discussion} \label{sec.3}}

\subsection {{Shell-Model Calculations of $ft$ Values}\label{sec.3A}}


\begin{figure}[H]
{\includegraphics[width=85mm,height=65mm]{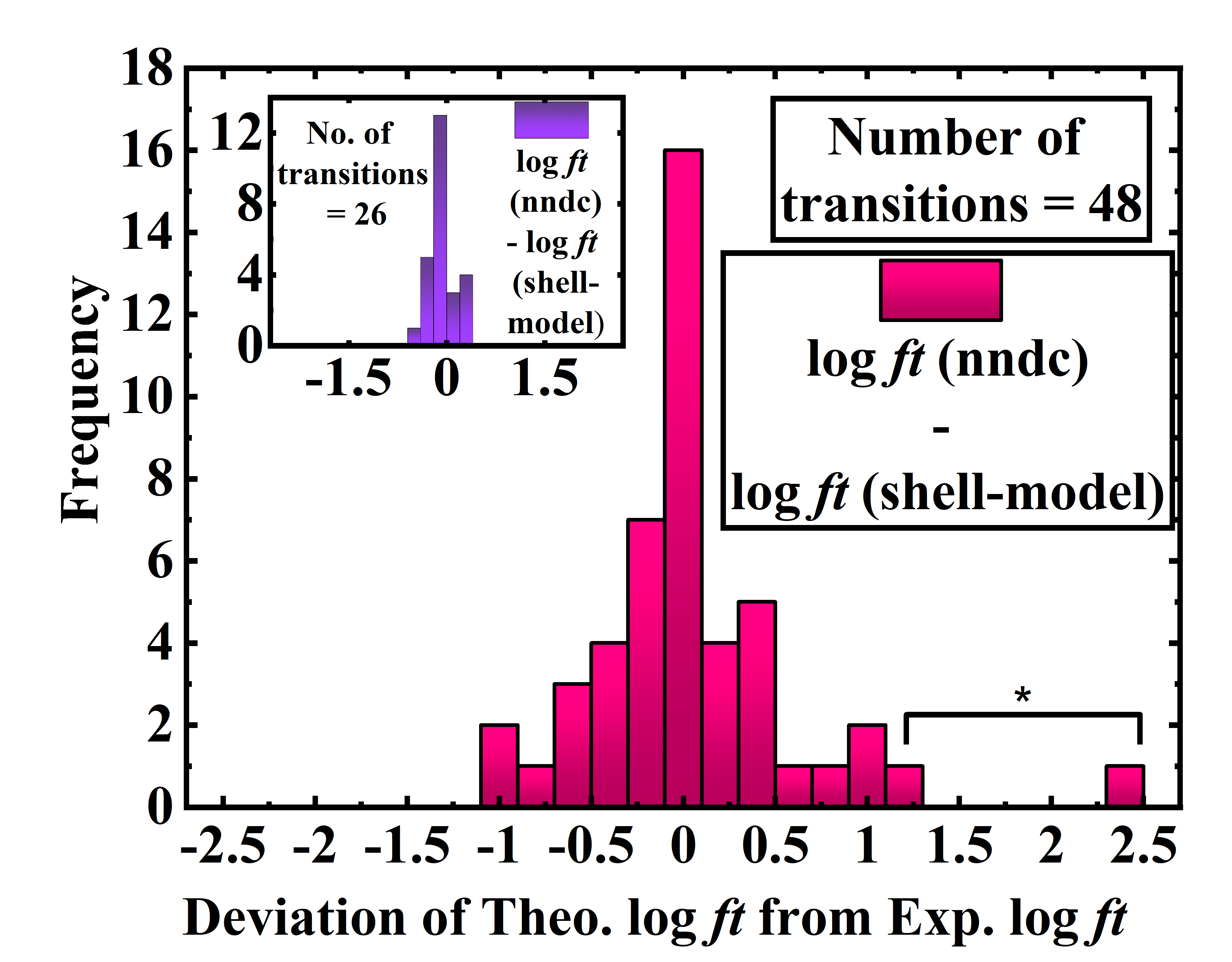}}
\vspace{0.2cm}
\caption{(Color online) The number of \logft values as frequency vs difference between experimental and theoretical \logft values of all the known transitions (48 in numbers) with a bin size of 0.3. See text for details. 
\label{figure-logft}}
\end{figure}  
\vspace{0.2cm}


\begin{figure}[H]
{\includegraphics[width=85mm,height=65mm]{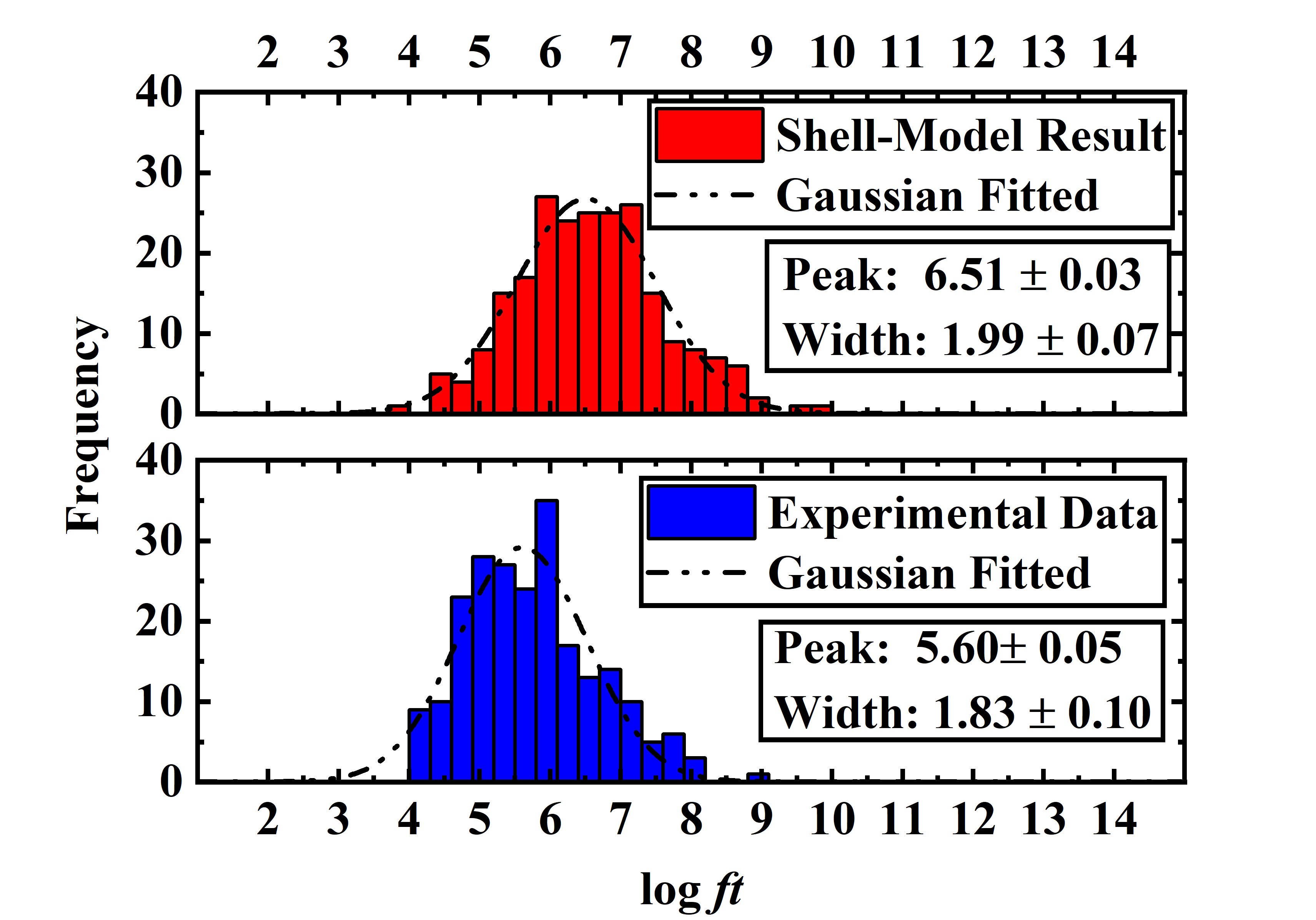}}
\vspace{0.2cm}
\caption{(Color online) The number of \logft values as frequency is plotted as a function of \logft values with a bin size of 0.3. Histograms are fitted with the Gaussian curves in both of the plots, as shown. An overall similarity and agreement of the two distributions can be noted. See text for details.
\label{figure-logft-nndc-shell}}
\end{figure}  
\vspace{0.2cm}

In the terrestrial scenario, the \logft values of $\beta^-$ transitions from parent ground state/ isomeric state to the daughter states are available for most of the cases considered in the mass region A = $59 \textendash 81$. However, in stellar s-process situation, as discussed before, thermally populated excited states may also undergo $\beta^-$ decay to various daughter levels. Evidently, no experimental data corresponding to these $\beta^-$ decays from the excited states are available. So, one has to rely on theoretical predictions. Therefore, we have performed shell-model calculations in the ($1f_{7/2} 2p_{3/2} 1f_{5/2} 2p_{1/2}$) and ($2p_{3/2} 1f_{5/2} 2p_{1/2} 1g_{9/2}$) valance spaces with $^{40}$Ca and $^{56}$Ni cores, respectively. These two model spaces cover all nuclei in the considered range. The shell-model calculations have been carried out with the OXBASH \cite{Oxbash} and the NuShellX \cite{NushellX} codes. The energy eigenfunctions obtained for the parent and the daughter nuclei have been used to calculate the reduced Fermi (B(F)) and the Gamow-Teller (B(GT)) strengths. These matrix elements have been then used to calculate the $ft$ values for each transition (Eq.\ref{Eq10}), allowed by the $Q$-value of the decay. 
In these model spaces, shell-model calculations are difficult to perform in some cases, because of prohibitively large dimensionalities. For those cases we have performed moderately Large Basis Shell Model (LBSM) calculations with reasonable and judicious truncation (See Appendix \ref{app_shell}). 

In this work our primary aim is to reproduce experimental \logft values theoretically within the shell-model framework. Thus for each nucleus these values have been calculated with various available interaction Hamiltonians. Later, for each nucleus, an appropriate interaction has been chosen which reproduced the data most successfully.

We have checked that the use of average quenching factors for fp and fpg spaces have limited predictability \cite{Kumar_JPG_2016} of \logft values for GT transitions in various nuclei. One may also expect a dependence of the quenching factor on A, (N-Z), and the shell closure \cite{Brown_ADNDT_1985}. So, we have found it advantageous to determine quenching factor for each nucleus while calculating \logft values.

The method to obtain the quenching factor q of the GT strength is as follows.

$\bullet$ In case of single \bta transition from ground state/ isomeric state quenching factor is chosen as one, and same q is used for \bta transition from excited states of that parent nucleus, if any. 

$\bullet$ In case of multiple \bta transitions from ground state/ isomeric state of parent, quenching factor has been obtained from the slope of the fitting of $M(GT)_{exp}$ with $M(GT)_{theo}$. Same q is used for \bta transitions from the excited levels of the same nucleus. See Appendix \ref{app_quenching} for details (Exception: $^{61}$Co, $^{78}$Ge).

For the cases of $^{61}$Co and $^{78}$Ge, we have taken q = 1, since for each of them, experimental \logft  for only two transitions are available. 
In the case of $^{61}$Co, the $7/2^-_1$ ground state can decay to five levels of $^{61}$Ni. However, experimental \logft for only two transitions are available. Similarly, for the case of $^{78}$Ge, the $0^+_1$ ground state can decay to three $1^+$ states of $^{78}$As. But experimental \logft values for the first and second $1^+$ states are only available.  Thus the experimental information is incomplete to obtain the q-values.

We have presented, in Table \ref{table-logft}, only the calculated \logft values which closely agree with the experimental ones. The \logft values of transitions in a nucleus is calculated with a single Hamiltonian appropriate for that nucleus as selected from the comparisons as discussed above. 
In this table, we have shown the results for energy eigenvalues of parent and daughter also, which are relevant here, along with the derived quenching factor q. 

In Figure \ref{figure-logft}, we have shown a summary of the results  based on Table \ref{table-logft}, in the form of a statistics of the deviation of the calculated \logft value from the corresponding experimental one. The figure shows that the predicted \logft values agree with the experimental results in most cases. Larger deviations are found to be associated with the very weak \bta branchings. For example, for the $^{72}$Zn $ \rightarrow ^{72}$Ga \bta decay, two transitions $ 0^+_1 \rightarrow (0^+_1)$ and $ 0^+_1 \rightarrow (1, 2)$, \logft values (Ref. \cite{NNDC}) are given as $> 8.6$ and 7.2, respectively. However, it is mentioned in Ref. \cite{NNDC} that the existence of these branches (0.01\% and 0.21(3)\%, respectively) are questionable. These two \logft values deviate most from the theoretical ones are marked with an asterisk in Figure \ref{figure-logft}. We have selected, out of the 48 transitions, 26 transitions those have \bta branching $>5\%$ and have plotted the frequency distribution in the inner panel of Figure \ref{figure-logft}. One can see that the calculated \logft values are much closer to the experimental values. The deviation range reduced from (-1.2, + 1.2) to (-0.6, +0.4).  

In Figure \ref{figure-logft-nndc-shell}, the frequency distribution of the calculated \logft of all 223 transitions have been compared with that of the experimentally available allowed 225 \logft values,  of the same mass region. The histograms clearly indicate that the calculated \logft values are in good agreement with the experimental values. The statistical properties, the peaks, centroids, and widths, calculated from the two histograms or from the Gaussian fits shown in the figure are similar. The upper panel of Figure \ref{figure-logft-nndc-shell} includes transitions from the parent excited levels also. Whereas, in the lower panel, transitions from the ground and isomeric levels of the nuclei are plotted. Close similarity and statistical behavior indicates the reliability of the calculated \logft values for the \bta decay from excited nuclear levels. 

In Table \ref{table-halflife} we have compared the terrestrial half-life obtained from theoretical \logft values with the experimentally measured half-lives, for the set of nuclei. The good agreement between these, once again indicates the acceptability of the calculated \logft values.

We have noted that in some cases energy eigenvalues predicted by the shell-model calculations are not in agreement with the experimental level energies. However, the eigenfunctions of parent and daughter reproduce \logft values which are reliable, as is evident from the agreement of the predicted half-lives with the experimental ones. 

It is to be noted that for the phase space calculations, we have used experimental \cite{NNDC} $Q$-values and level energies.

One finds from Table \ref{table-halflife} that the agreement of the calculated half-lives with the experimental values are good for most of the cases. However, for $^{60}$Co to $^{60}$Ni decay we have got the terrestrial half-life $T_{1/2(t)} $ = 3.285 years instead of 5.275 years \cite{NNDC}. In the case of \bta decay from the isomeric  $2^+_1$ state of the same parent nucleus to the daughter states, the calculated $T_{1/2(t)}$ is 5.41 days, whereas the experimental value is 2.91 days \cite{NNDC}. As Shown in Ref. \cite{NNDC} these measurements are quite old. Moreover, the \bta decay from the isomeric state constitutes only 0.25\% (0.1\% in a measurement of the year 2010 \cite{Wauters_PRC_2010})\cite{NNDC} is also a measurement of 1963 \cite{Schmidt_Physik_1963}. Because of these uncertainties in the measured values, it is difficult to comment on the results for these transitions.  Similar is the case for the decay from the isomeric level of $^{75}$Ge, which has \bta branching 0.03\% only (measured in 1976 \cite{Bhattacharyya_Nuovo_1976}), the disagreement of the calculated and measured $T_{1/2(t)} $ is quite large. For $^{63}$Ni to $^{63}$Cu decay, we have got $T_{1/2(t)} $ = 44.69 years, if GT quenching factor q is taken as 1. If the globally accepted quenching factor q = 0.77 was taken then the $T_{1/2(t)}$ would have been about 75 years. It is well known that there is a difficulty in the measurement of long half-life as was the case of $^{44}$Ti. Our calculated half-life for $^{63}$Ni is not close to the measured \cite{NNDC} $T_{1/2(t)}$ = 101.2(15) years. So, it is difficult to comment on this disagreement. We have also noted that there are slight disagreements of $T_{1/2(t)} $ for the cases of \bta decay from $^{67}$Cu and $^{72}$Zn. In both of the cases, the measurements are old (1953 \cite{Harry_NP_1953} and 1968 \cite{Kjelberg_NP_1968}, respectively), and the measured branchings are quite uncertain. 

\subsection {{$\beta^-$ Decay In Stellar Environment}\label{sec.3B}}
As mentioned before, our calculation of \bta decay rate is based on the s-process environment having the temperature and free electron density ranges between  $10^8$ K to $5\times 10^8$ K and $10^{26}$cm$^{-3} \textendash 10^{27}$cm$^{-3}$, respectively. In such  conditions the ionization of atoms and \bta decay from the excited nuclear levels, change the total \bta decay rate noticeably. In the following subsections, we have discussed these effects.

\subsubsection {{Ionization of Atoms: Presence of Bare Atoms}\label{3B-I}}

The probability of bound state \bta decay is directly related to the availability of phase space volume of the atomic shells of the parent atom. Relativistic solution of the electronic radial function indicates that bound state decay probability is highly dominated by the creation of electron in atomic K shell, followed by L, M, and N shells. As a consequence, to get the bound state $\beta^-$ decay rate it is essential to have the exact information about the charge state, more specifically, occupancies of the electronics shells of the parent atom.  

\begin{figure*}

{\includegraphics[width=160mm,height=85mm]{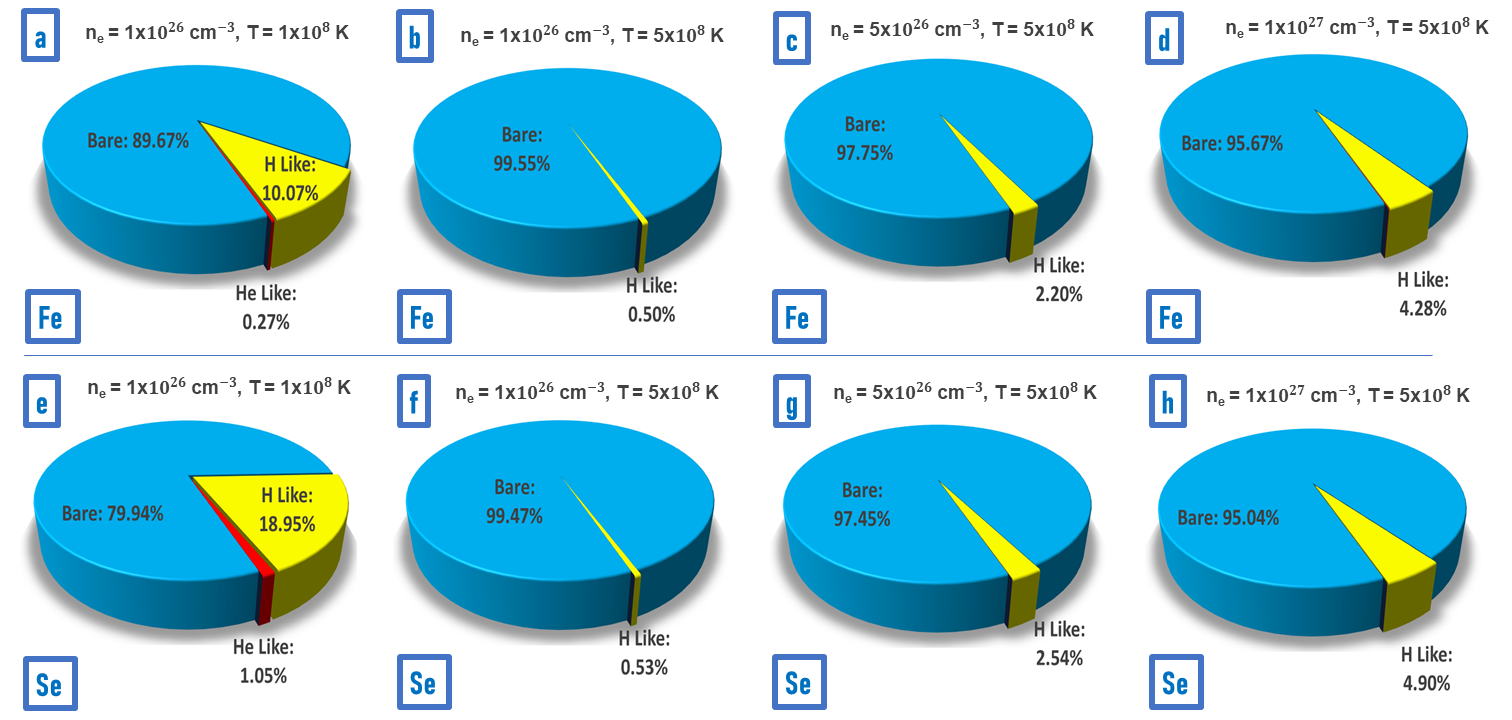}}
\caption{(Color online) Percentage of the different ionization states for Fe (upper panel) and Se (lower panel) at different free electron density ($n_e$) and temperature (T) situations. 
\label{Fe_Saha}}
\end{figure*}

In stellar conditions mentioned above, the atoms get ionized. The charge state of the ionized atoms depends on various factors like temperature of the environment, free electron density, and ionization potential \cite{Takahashi_NP_1983}. Saha Ionization equation \cite{Takahashi_ADNDT_1987} provides a clear view of the ionization scenario as a ratio of two different charge states of an atom in thermodynamic equilibrium (see Appendix \ref{appB} for details). It can be shown that, with the increase in temperature of the stellar site, atoms tend to be in higher and higher charge states. Whereas an increase of the free electron density inhibits the ionization of the atoms, while, on the other hand, in such circumstances, ionization potential of an atom embedded in matter in thermodynamic equilibrium gets reduced. This phenomenon of reduction of the ionization potential, known as Ionization Potential Depression (IPD) also needs to be included to get the actual ionization scenario of any stellar environment. In Figure \ref{Fe_Saha}, the charge state distributions of Fe and Se atoms over a density -  temperature grid have been shown, as examples. Interestingly, neutral atoms are totally absent, and fully ionized charged state dominates with a few percent of H-like and He-like atoms. 

We have calculated the IPDs ($\Delta_j$) of the set of parent and daughter atoms, and also the charge states of those as a function of $n_e$ and $T$ to confirm the presence of bare atoms.

\vspace{0.2cm}



\subsubsection {{Variations of Individual Transition Rate with Temperature and Density }\label{3B-II}}

\begin{figure*}

{\includegraphics[width=85mm,height=65mm]{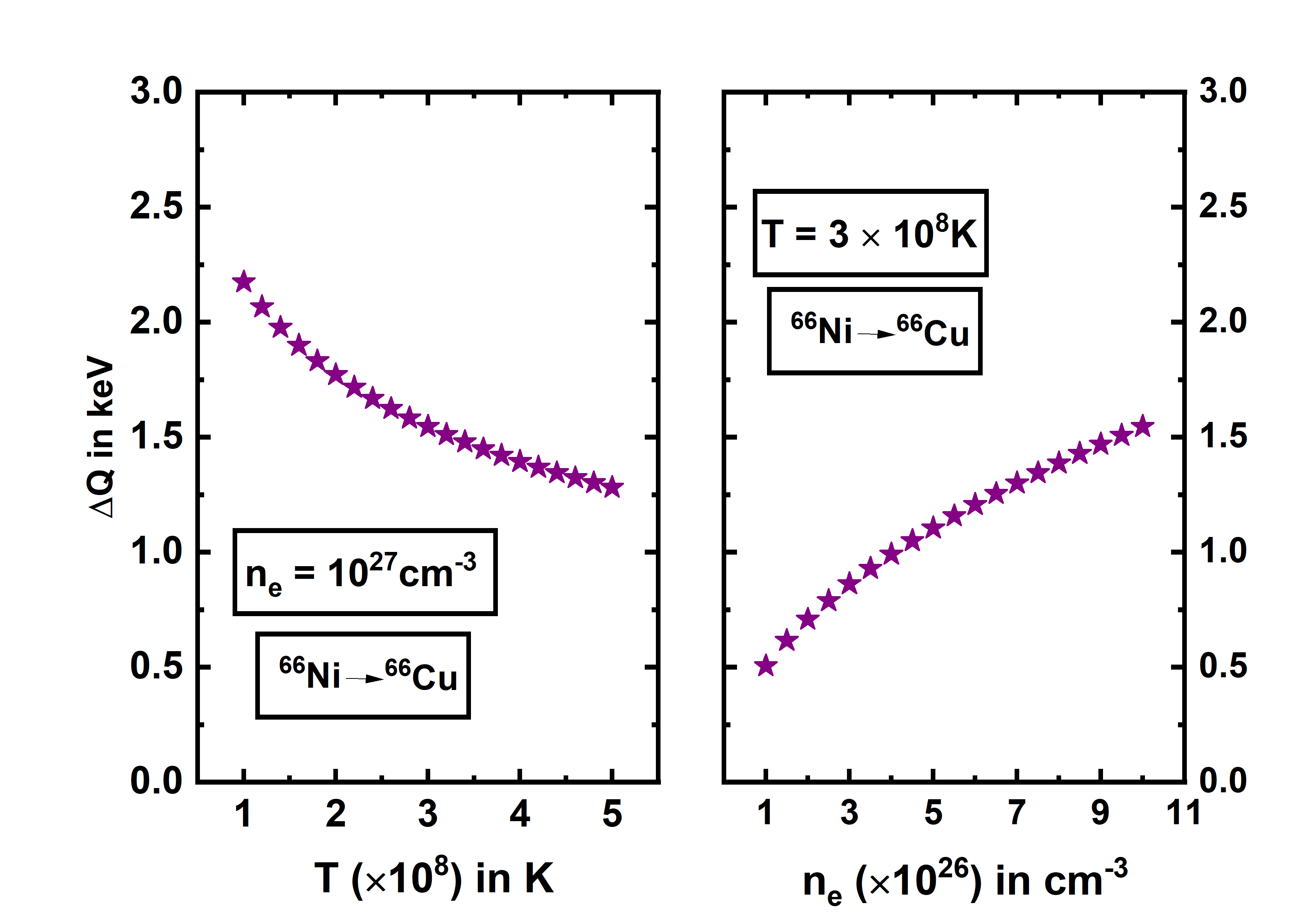}}
\includegraphics[width=85mm,height=65mm]{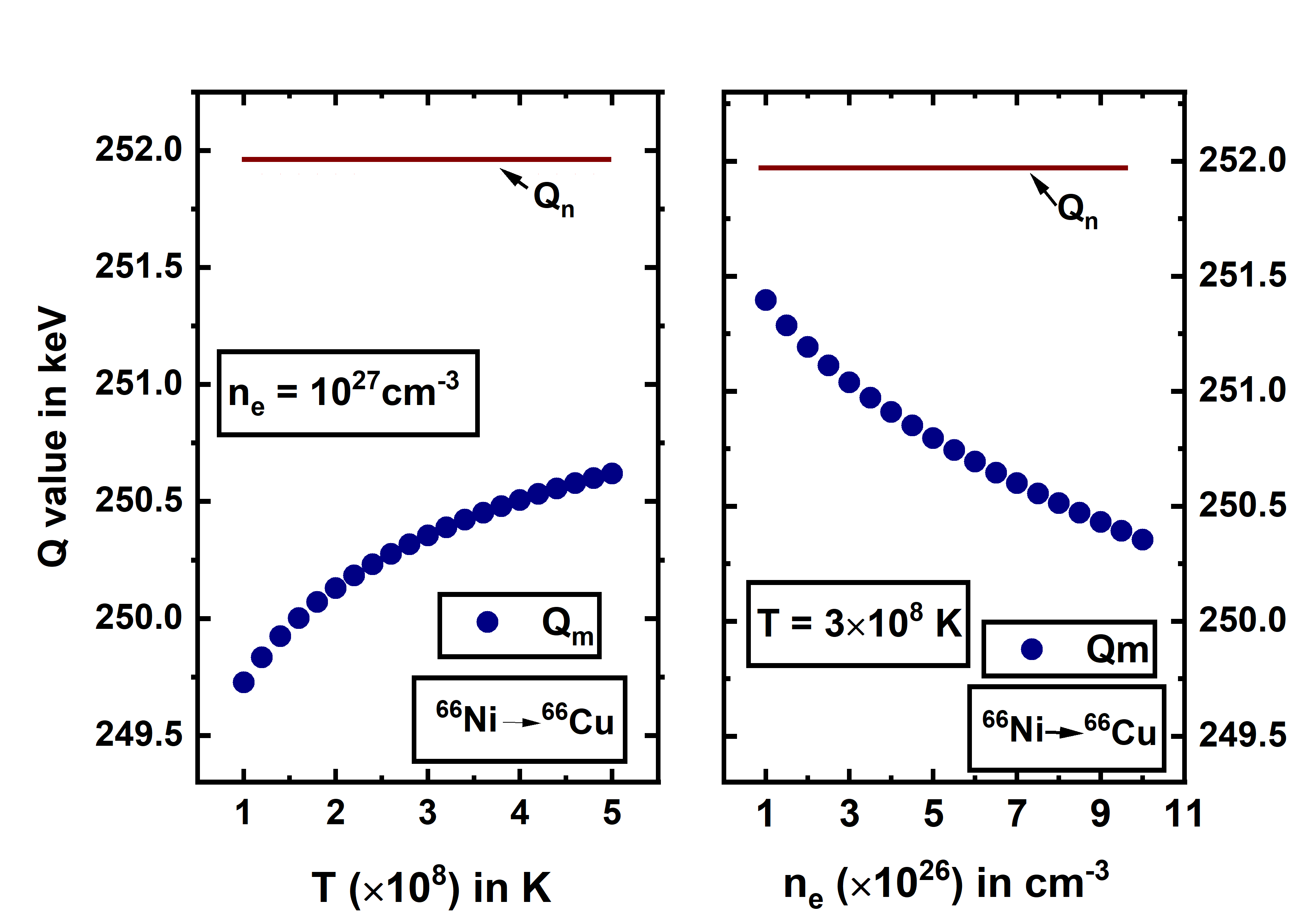}
\caption{(Color online) Left panel: Variation of IPD correction $\Delta Q = (\sum_{j_D = 0}^{Z_{D}-1} \Delta_{j}-\sum_{j_P = 0}^{Z_{P}-1} \Delta_{j})$ with temperature and free electron density; Right panel: Variation of $Q_m$ with temperature and free electron density, for the \bta decay from the ground state of $^{66}$Ni.
\label{delQ_Variation}}
\end{figure*}
  
\vspace{0.2cm}

The lowering of ionization potential (IPD) in stellar plasma \cite{Stewart_AJ_1966}, due to the effect of high temperature and free electron density, causes the neutral atom \bta decay $Q$-value $Q_n$ to be reduced by an amount $\Delta Q = (\sum_{j_D = 0}^{Z_{D}-1} \Delta_{j}-\sum_{j_P = 0}^{Z_{P}-1} \Delta_{j})$  as mentioned in Eq. \ref{Eq4}. Thus $\Delta Q$ is the IPD correction to $Q_n$. The left panel of Figure \ref{delQ_Variation} shows the variation of $\Delta Q $ with temperature and density. $\Delta Q$ decreases with temperature, and as a result, the $Q_m$ value increases. Whereas, the increase of $\Delta Q$ with free electron density reduces the $Q_m$ value. Both these variations of $Q_m$ value have been shown in the right panel of Figure \ref{delQ_Variation} for the case of ground state \bta decay of $^{66}$Ni.

\begin{figure*}

{\includegraphics[width=165mm,height=65mm]{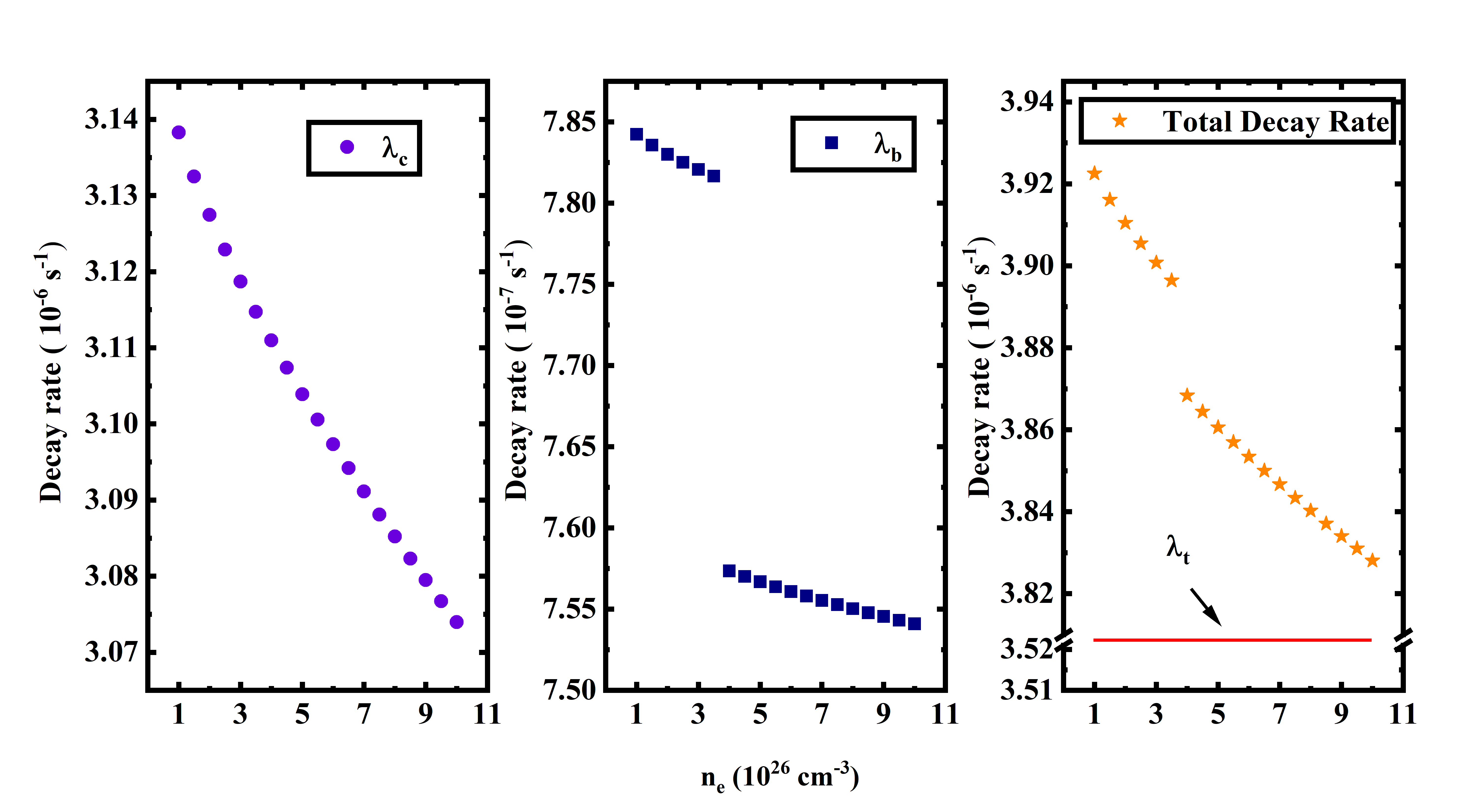}}\\
\includegraphics[width=165mm,height=65mm]{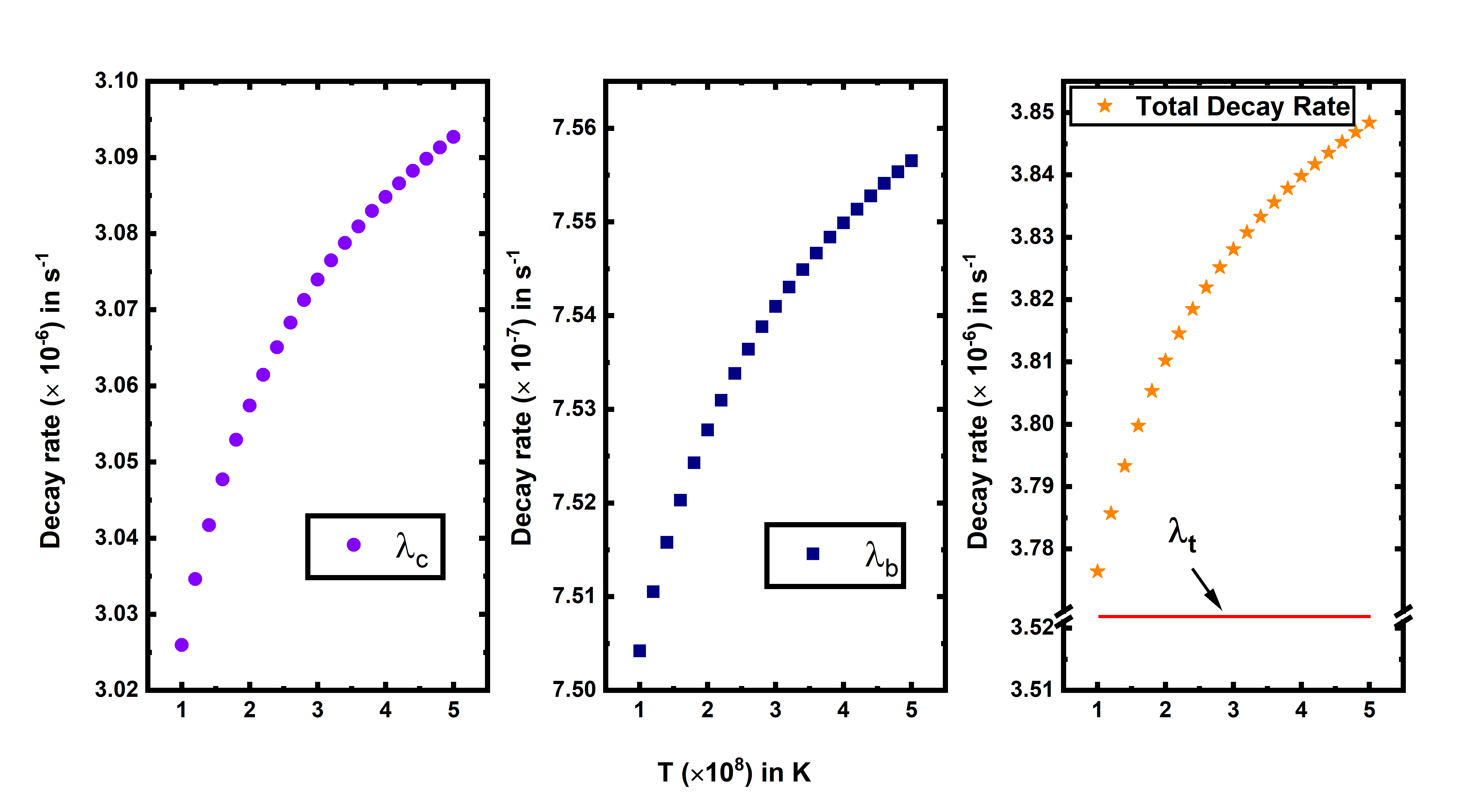}
\caption{(Color online) Upper panel: Variation of continuum decay rate, bound state decay rate and total decay rate with free electron density at $T = 3 \times 10^8$ K; Lower panel: Variation of continuum decay rate, bound state decay rate and total decay rate with temperature at $n_e = 10^{27}$ cm$^{-3}$, for the \bta decay from the ground state of $^{66}$Ni. $\lambda_t$ is the terrestrial decay rate of corresponding neutral atom. In the right most boxes of the upper and lower panels the y axis has been broken to include the terrestrial decay rate in the plot.
\label{Decay_Rate_Variation}}
\end{figure*}
  
\vspace{0.2cm}


In Figure \ref{Decay_Rate_Variation}, the variation of decay rates with temperature and density has been shown, for the case of $^{66}$Ni. It can be observed that both bound and continuum state decay rates decrease with free electron density, since the phase space factor $f^*_{a}$ of the continuum and the bound state decays are affected. The common deciding factor for this trend is the increase of IPD correction with increasing density which has been shown in Figure \ref{delQ_Variation}. Besides, in the case of decay to continuum, the factor $(1-f_{FD}(\eta,\beta))$ (Eq. \ref{Eq2}) is also responsible for the decrease of decay rate with density. On the other hand, depression of continuum results in the  disappearance of atomic bound orbits. With the increase of IPD with density (as shown in Figure  \ref{delQ_Variation}), it is possible that the atomic M, N orbits ($n>2$, n is the principal quantum number) become unbound, and hence only K, L orbits contribute to bound state decay. This in turn generates a sudden drop of bound state decay as shown in Figure  \ref{Decay_Rate_Variation}.

From the lower panel of Figure \ref{Decay_Rate_Variation}, it is to be noted that decay rates vary differently with temperature than density. With an increase in temperature, the decay rates increase for both continuum state and bound state decays. Here, again the deciding factor for this variation is the variation of IPD as shown in Figure \ref{delQ_Variation}. Moreover, temperature dependence of phase space factor, through the factor $(1-f_{FD}(\eta,\beta))$, is causing an increase in continuum state decay rate with temperature. 

In Table \ref{b_and_c_with_t}, we have shown the dependence of individual transition rates, separately for $\lambda_c$ and $\lambda_b$ at two temperatures $T = 10^8$ K and $T = 5{\times}10^8$ K for the free electron density $n_e = 10^{27} $cm$^{-3}$. Both $\lambda_b$ and $\lambda_c$ increase slightly with increase in temperature, for the set of nuclei of our interest. 

Similarly, in Table \ref{b_and_c_with_ne}, we have given the dependence of individual transition rates, separately for $\lambda_c$ and $\lambda_b$ for two densities $n_e = 10^{26} $cm$^{-3}$ and $n_e = 10^{27} $cm$^{-3}$ at a temperature  $T = 3{\times}10^8$ K. Both $\lambda_c$ and $\lambda_b$ decrease slightly with the increasing density, for all nuclei considered here.


\subsubsection {{\bta Decay from Excited Nuclear Levels}\label{3B-III}}

The equilibrium population of the thermally excited energy levels of the nuclei of interest has been calculated using Boltzmann's distribution as mentioned in Sec. \ref{sec.2B}. We have considered $\beta^-$ decay only from those levels whose population is up to $10^{-5}$ times that of the ground level.

We have illustrated the contribution of excited nuclear levels to the decay rate for various density-temperature combinations in Table \ref{Decay_from_parent}.
Due to the population of the higher energy levels of the parent, decay from these levels start to contribute as temperature rises at a fixed density. As a result, one can see how the beta branching ($I_{k(gs)}$) from the parent's ground state (gs) decreases as the temperature rises. When we choose a larger density, the situation remains almost unaffected. Additionally, the table also includes the beta branching ($I_m$) to each daughter level for the two different densities. The branching to the daughter levels decreases with an increase in temperature in the case of a transition from the parent ground state because the parent ground state losses equilibrium population to some extent. Although branching to each daughter level rises with temperature in the case of decay from excited levels.
  

\begin{figure}[H]
{\includegraphics[width=85mm,height=65mm]{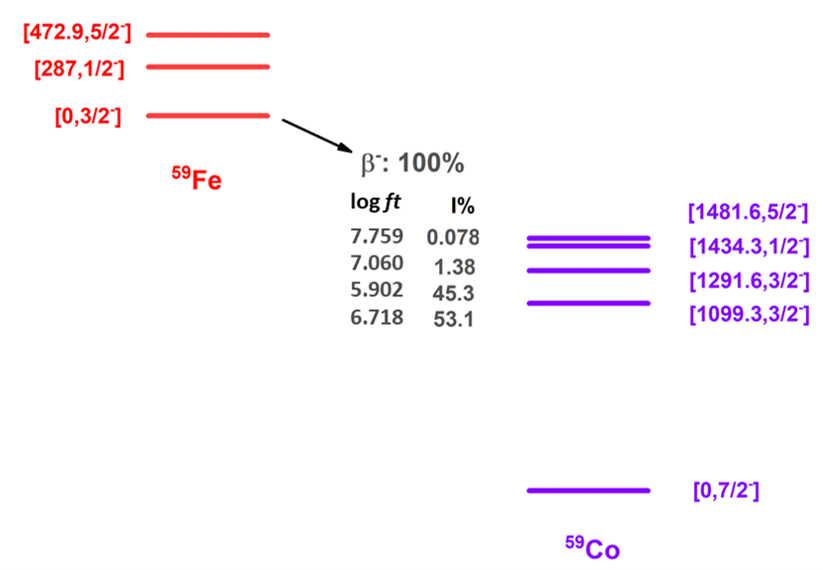}}
\caption{(Color online) \bta decay scheme for $^{59}$Fe in terrestrial laboratory. In the brackets, nuclear excitation energy ($E_{X}$) in keV and $J^{\pi}$ (spin-parity) are shown. $I$ is the \bta branching.
\label{Fe_Level_Neutral}}
\end{figure}
  
\vspace{0.2cm}

\begin{figure}[H]

{\includegraphics[width=85mm,height=65mm]{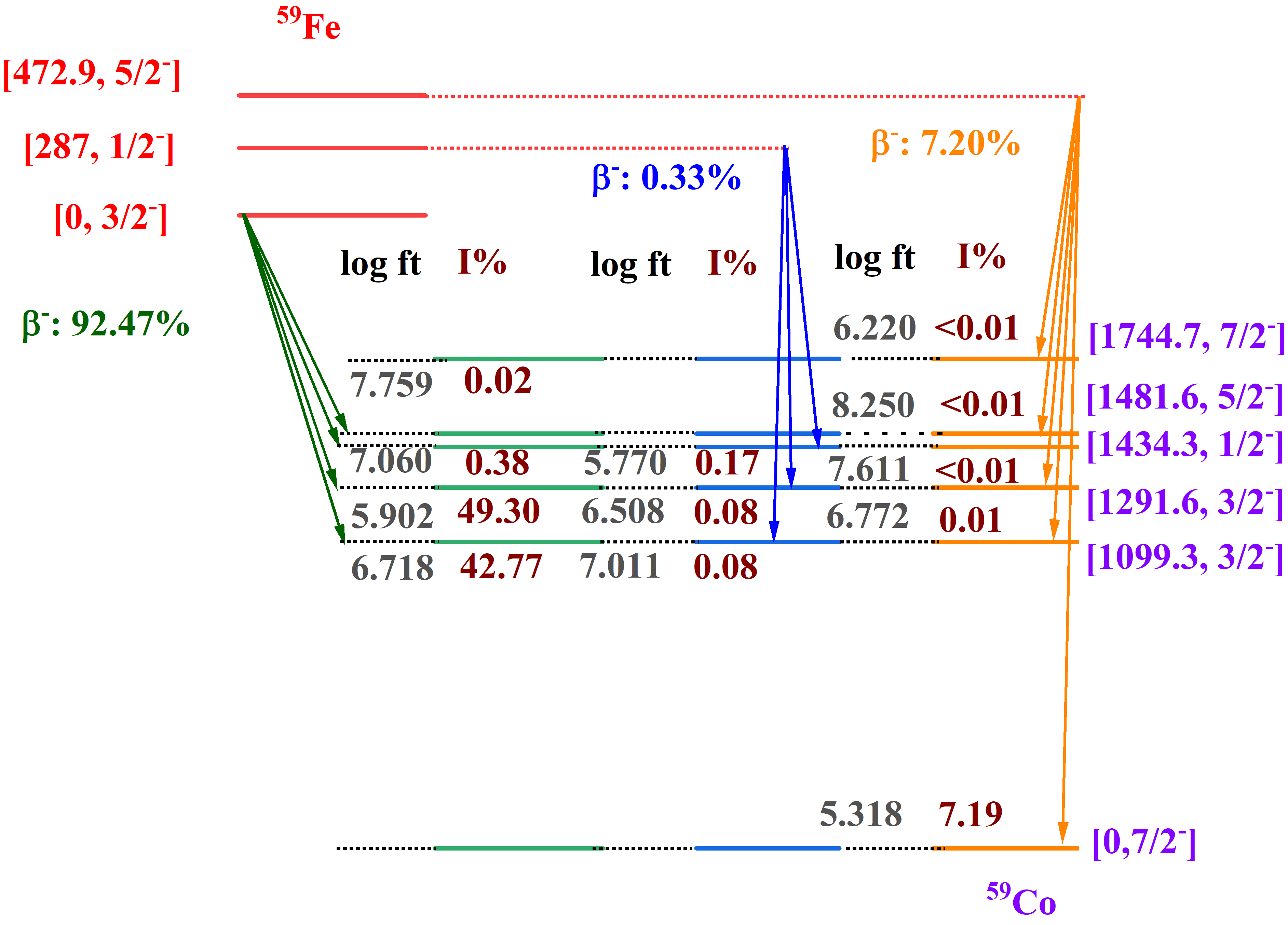}}
\caption{(Color online)  \bta decay scheme for $^{59}$Fe in stellar scenario, $T = 5{\times}10^8$ K and $n_e = 10^{26} $cm$^{-3}$. In the brackets nuclear excitation energy ($E_X$) in keV and $J^{\pi}$ (spin-parity) are shown. $I$ is the \bta branching.
\label{Fe_Level_Stellar}}
\end{figure}

We have used the example of $^{59}$Fe decaying to $^{59}$Co in a stellar situation to demonstrate the significance of \bta decay from excited levels. From Boltzmann distribution (Eq. \ref{Boltzman}), it has been found that the populations of the higher energy levels will be considerable only at $T \approx 5{\times}10^8$ K, for this particular nucleus. Terrestrially the ground state $[0.0,3/2^{-}]$ of $^{59}$Fe decays to four different levels - $[1099.3,3/2^{-}]$, $[1291.6,3/2^{-}]$, $[1434.3,1/2^{-}]$ and $[1481.6, 5/2^{-}]$ of $^{59}$Co (Figure \ref{Fe_Level_Neutral}). Branching of the decay to these four levels is almost $100\% $. But, at this temperature, the first $[287, 1/2^{-}]$ and second $[472.9,5/2^{-}]$ excited states of $^{59}$Fe get populated adequately. Now, these two levels can decay to various levels of the daughter as shown in Figure \ref{Fe_Level_Stellar}. In the total \bta decay of the parent nucleus, decay from the second excited state of the parent also contributes a substantial amount, followed by the first excited state. So we have taken into account the contribution of the excited states to bound and continuum decay for all relevant cases.     

In this paper we have shown the data (Table \ref{b_and_c_with_t}, Table \ref{b_and_c_with_ne}, Table \ref{Decay_from_parent}, Table \ref{bound_contri_a}, Table \ref{bound_contri_b}) only for those excited levels of parent which have contribution $> 0.1\%$, with respect to total decay rate. However, in Table \ref{total_decay}, we have given total decay rate taking into account all of the transitions. One may find the remaining data in the supplemental material of this paper \cite{Supplement}.

\vspace{0.2cm}



\subsubsection {{Contribution of bound state \bta decay to the total \bta decay rate}\label{3B-IV}}

\begin{figure}[H]

{\includegraphics[width=85mm,height=65mm]{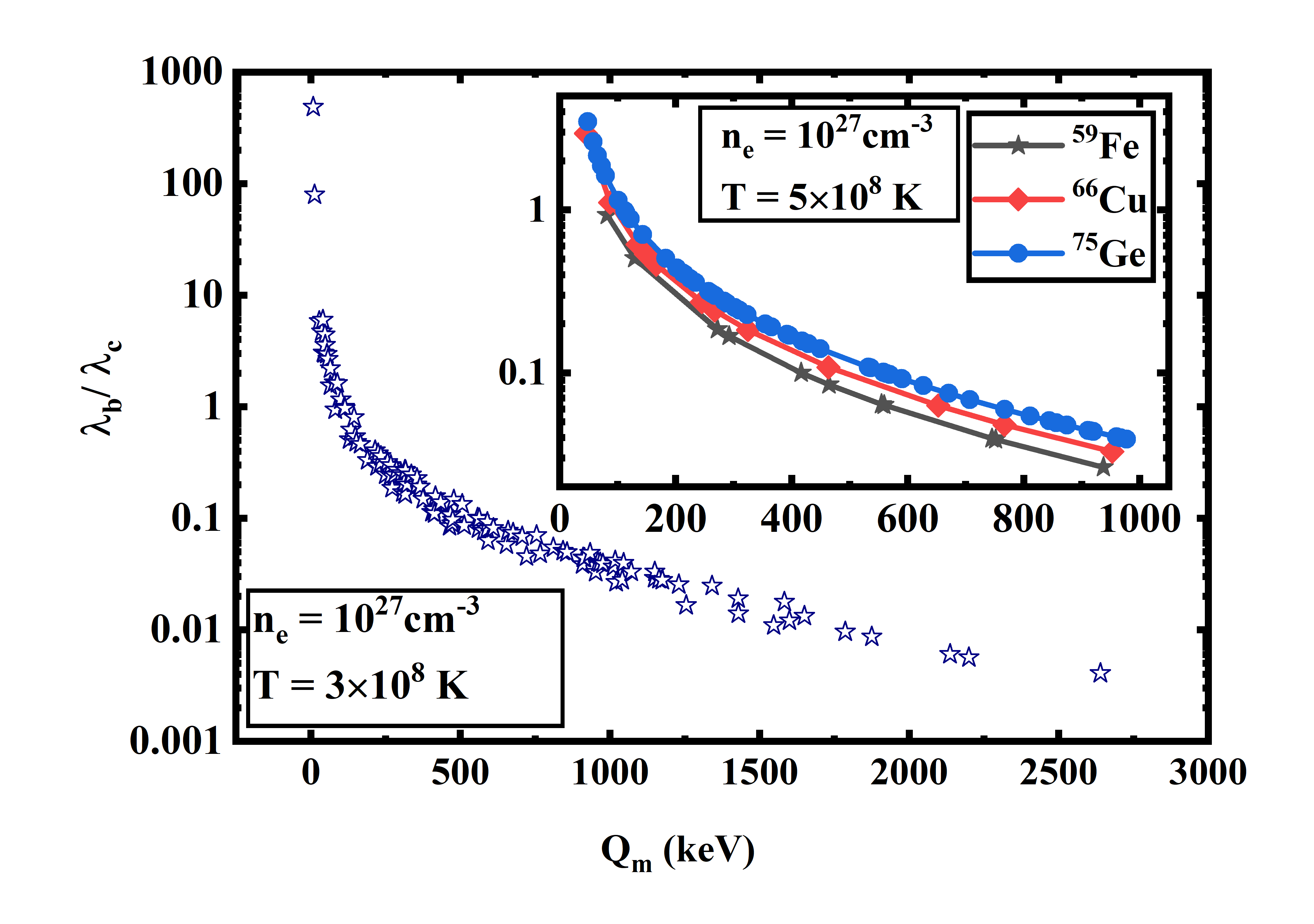}}
\caption{(Color online) Variation of the ratio of bound ($\lambda_{b}$) state to continuum ($\lambda_c$) state decay rates with $Q_m$ for a stellar density, temperature combination as shown. Spread in the ratio is due to a dependence on Z, and A of the daughter nuclei, is shown in the inner panel. See text for details.  
\label{b_to_c}}
\end{figure}

For all the terrestrially known transitions and transitions from the excited levels of the nuclei listed in Table \ref{table-logft}, we have calculated the continuum ($\lambda_c$) and bound state ($\lambda_b$) decay rates. Based on Table \ref{b_and_c_with_t} and  \ref{b_and_c_with_ne} the ratio of the bound to continuum state decay rates as a function of $Q_m$ has been shown in Figure \ref{b_to_c}. It can be clearly observed from the figure that for the transitions having high $Q_m$ values, $\lambda_c$ dominates. Whereas, in the case of the transitions having comparatively lower $Q_m$ values, $\lambda_b$ starts to compete with $\lambda_c$. From Figure \ref{b_to_c}, it can be said that for these nuclei, the transitions having $Q_m$ below $\approx$ 100 keV, the bound state decay rates dominate over continuum state decay rates. The ratio falls from  1 to about  $\approx 10^{-3}$ for 100 keV$<Q_m<2800$ keV. Moreover, it is also to be noted that the $\lambda_b/\lambda_c$ ratio is dependent on the atomic number (Z) and mass number (A) of the daughter nucleus. As the Z, A values increase, the ratio increases slightly, causing a spread in the curve as shown in Figure \ref{b_to_c}. In the inset of Figure \ref{b_to_c}, we have illustrated the Z, A dependence for a higher temperature to accommodate more transitions. This dependence is actually coming from the phase space factor of the continuum decay and through the large component of Dirac radial wave function in case of bound state decay. This points to the fact that bound state decay is more important for transitions with low $Q$-values. For the set of nuclei, the bound state decay contribution to total \bta decay ranges from about 1\% to 62\%, as shown in Table  \ref{bound_contri_a} and Table \ref{bound_contri_b}. 
For nuclei having decays from large number of excited levels, $Q$-values are generally large and contribution to the bound state decay is small. As the temperature increases at a fixed density, the contribution of the bound state decay decreases due to the following facts: i) drop of IPD at higher temperature increases the $Q_m$ value, and ii) contribution of nuclear excited levels, having higher $Q_m$ values.

From Table \ref{bound_contri_b} it can be seen that for a fixed T, bound state contribution is smaller at $n_e = 10^{27} $cm$^{-3}$ for higher $Q$-value ($Q_n>100$ keV) transitions and larger for the transitions having lower $Q$-values ($Q_n<100$ keV), compared to those at $n_e = 10^{26} $cm$^{-3}$. However, total bound state decay contribution to total decay rate decreases slightly with increasing density, at a fixed temperature for all nuclei considered here.

\subsubsection {{Total $\beta^-$ Decay Rate ($\lambda_{bare(s)}$) of Bare Atom}\label{3B-V}}

In the case of bare atoms, the total \bta decay rate is the sum of total bound state decay rate ($\lambda_b$) and total continuum state decay rate ($\lambda_c$). In Table  \ref{total_decay}, we have shown the calculated total decay rate and half-life of the nuclei. In an earlier work, Cosner and Truran \cite{Cosner_ASS_1981} calculated total \bta decay rate for the s-process nuclei with out contribution from bound state decay. Takahashi and Yokoi \cite{Takahashi_ADNDT_1987} tabulated total \bta decay rate for highly ionized s-process nuclei, with bound state decay contribution. However, they did not mention whether these rates include decay of bare atoms. In Table \ref{total_decay}, we have shown the earlier results of Ref. \cite{Cosner_ASS_1981} and Ref. \cite{Takahashi_ADNDT_1987} for comparison. Here, the ratio of the calculated half-life of neutral atom in terrestrial environment and the half-life of bare atom in stellar environment is denoted by R. The value of R shows that, most of the bare atoms are short lived compared to the corresponding neutral atoms in the terrestrial environment. This is because of the opening of the \bta decay channel to the atomic bound states from nuclear ground/isomeric states and opening of both bound and continuum state decays from the nuclear excited states. However, there are some more deciding factors, which are also reflected in a few exceptional cases, where R is almost equal to or less than one.

For the case of $^{61}$Co and $^{70}$Ga, the value of R is $\approx$ 1, which implies that for these nuclei the half-lives are almost unchanged. On the other hand, in the cases of $^{66}$Cu, $^{75}$Ge, and $^{81}$Se, R $<$ 1 in different temperature-density combinations. This indicates that the half-life of bare atom in stellar site will be larger than that of neutral atom in terrestrial condition. In these cases the competing factors are the inclusion of \bta decay from nuclear excited levels, increase in the \bta decay rate to the continuum at higher temperature due to the factor $(1-f_{FD}(\eta,\beta))$ in Eq. \ref{Eq2}, and effect of continuum depression.    

Moreover, from Table \ref{total_decay}, the general trend of R shows that, with the constant density, increasing temperature results in half-life reduction, except for $^{66}$Cu, $^{75}$Ge, and $^{81}$Se. Whereas, for a constant temperature, an increase in density results in an enhancement of the half-life of the bare atom.



\section{{Conclusion}\label{sec.4}}

In summary, in this study, at first we have calculated, using shell-model, the comparative half-lives ($ft$) for all allowed terrestrial \bta transitions in the chosen set of nuclei in the mass range A = $59 \textendash 81$. For this, we have used various Hamiltonians in fp valance space, and a single interaction in fpg space to reproduce the measured $ft$ values for different nuclei with empirically obtained GT quenching factors. Only one selected interaction has been used to calculate \logft values of all transitions in a nucleus. In most of the cases, we have found good agreement between the theoretical and the measured \logft values. The reliability of the calculated \logft for the transitions from the excited nuclear levels has been shown. We have then compared the calculated terrestrial half-lives with the measured ones. Quite good agreement for most of the cases has been found. The presence of bare atoms in the s-process density-temperature situations has been confirmed by solving Saha Ionization Equation, incorporating ionization potential depression. Then we have calculated the temperature and density dependence of the individual transition rate and have found dependence of the ratio of bound and continuum decay rates on Z and A of the daughter. The reason behind it has been explained. It is observed that $\lambda_b$ starts to compete with $\lambda_c$ and dominates over $\lambda_c$ for $Q_m$ $<$ 100 keV. We have shown the importance of bound state decay for stellar situations by providing separately the values of $\lambda_b$ and $\lambda_c$. A mild variation of individual transition rate with temperature and density has been found. Finally, total stellar \bta decay rates $\lambda_{bare(s)}$ and half-lives $T_{1/2(bare(s))}$ as a function of density and temperature have been presented. It has been noted that for bare atoms, in most of the cases, half-lives become smaller than the terrestrial half-lives of the corresponding neutral atoms. Also, it has been found that for some bare atoms, half-lives become larger than the terrestrial ones. These results may be useful for calculations of the nucleosynthesis processes.

\section*{ACKNOWLEDGMENT}
AG is appreciative of the financial assistance received from DST-INSPIRE Fellowship (IF160297). AG further expresses gratitude to SERB-DST for providing the computational facility under Government of India Project No. EMR/2016/006339. The author AG acknowledges the assistance given by Subham Burai and Sourav Paul (M.Sc. Physics Students) at various phases.

\begin{appendix} 


\section{{Shell-Model Calculations: Choices of Model Space and Hamiltonian}\label{app_shell}}

\begin{figure}
{\includegraphics[width=85mm,height=65mm]{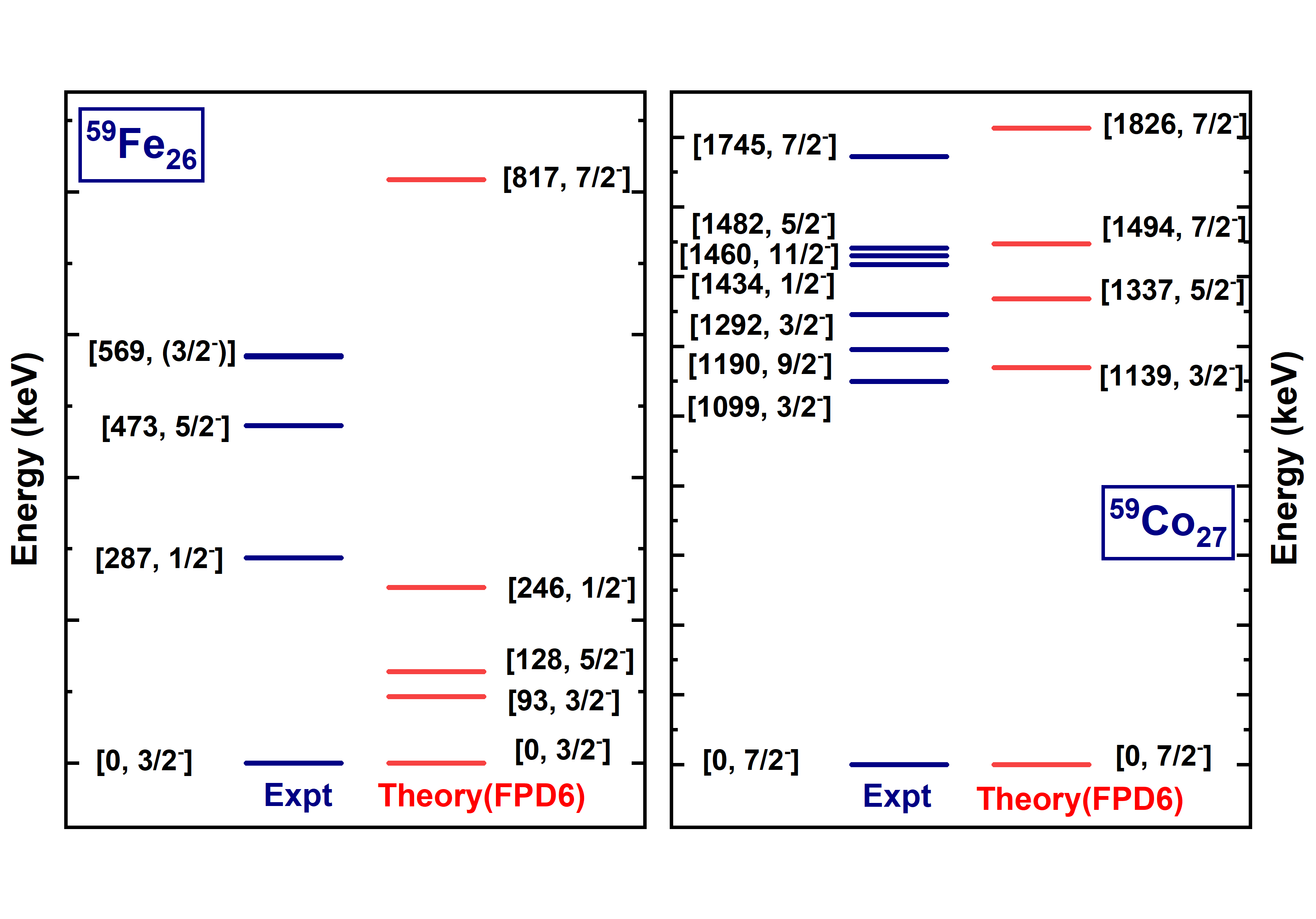}}
{\includegraphics[width=85mm,height=65mm]{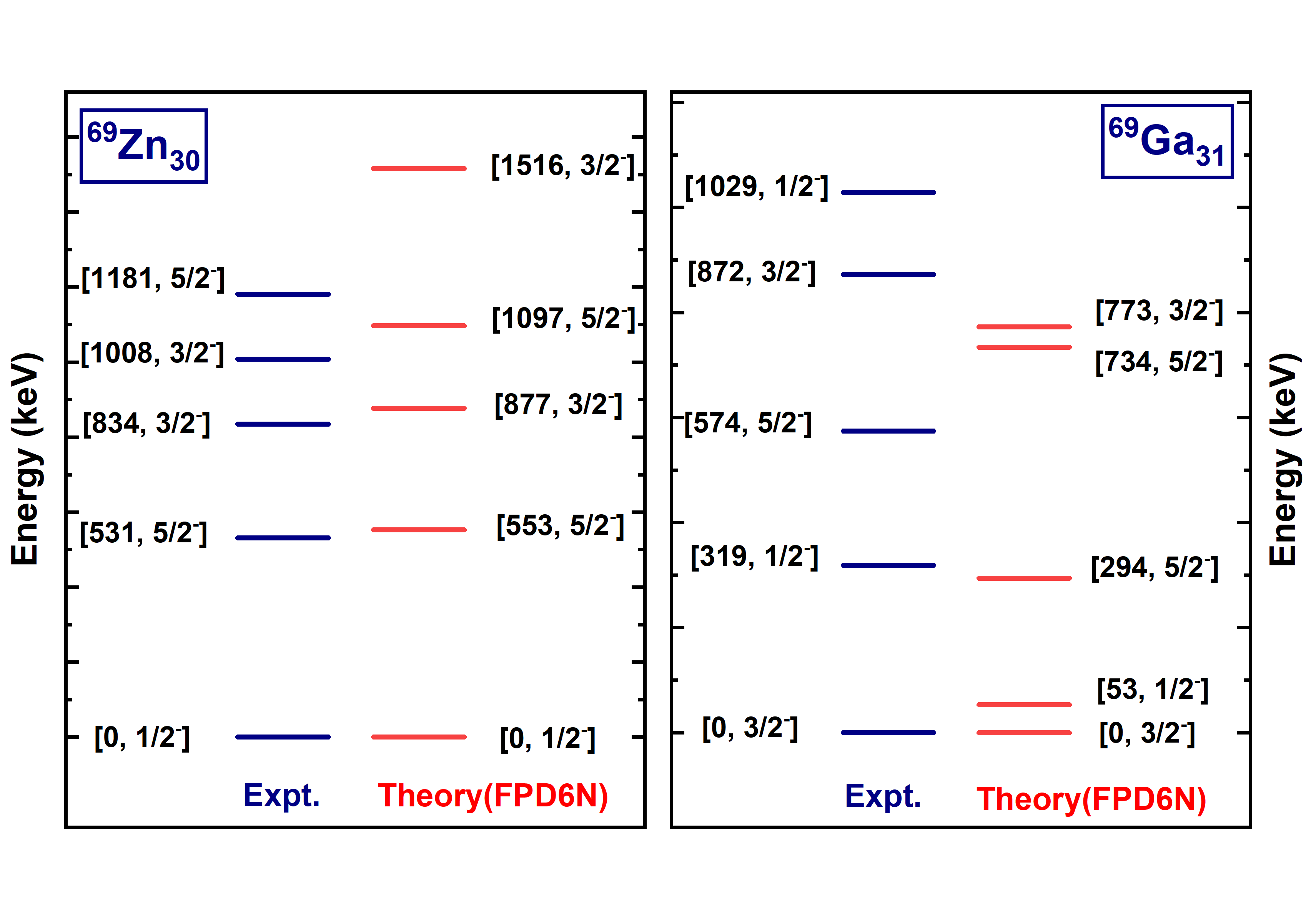}}
{\includegraphics[width=85mm,height=65mm]{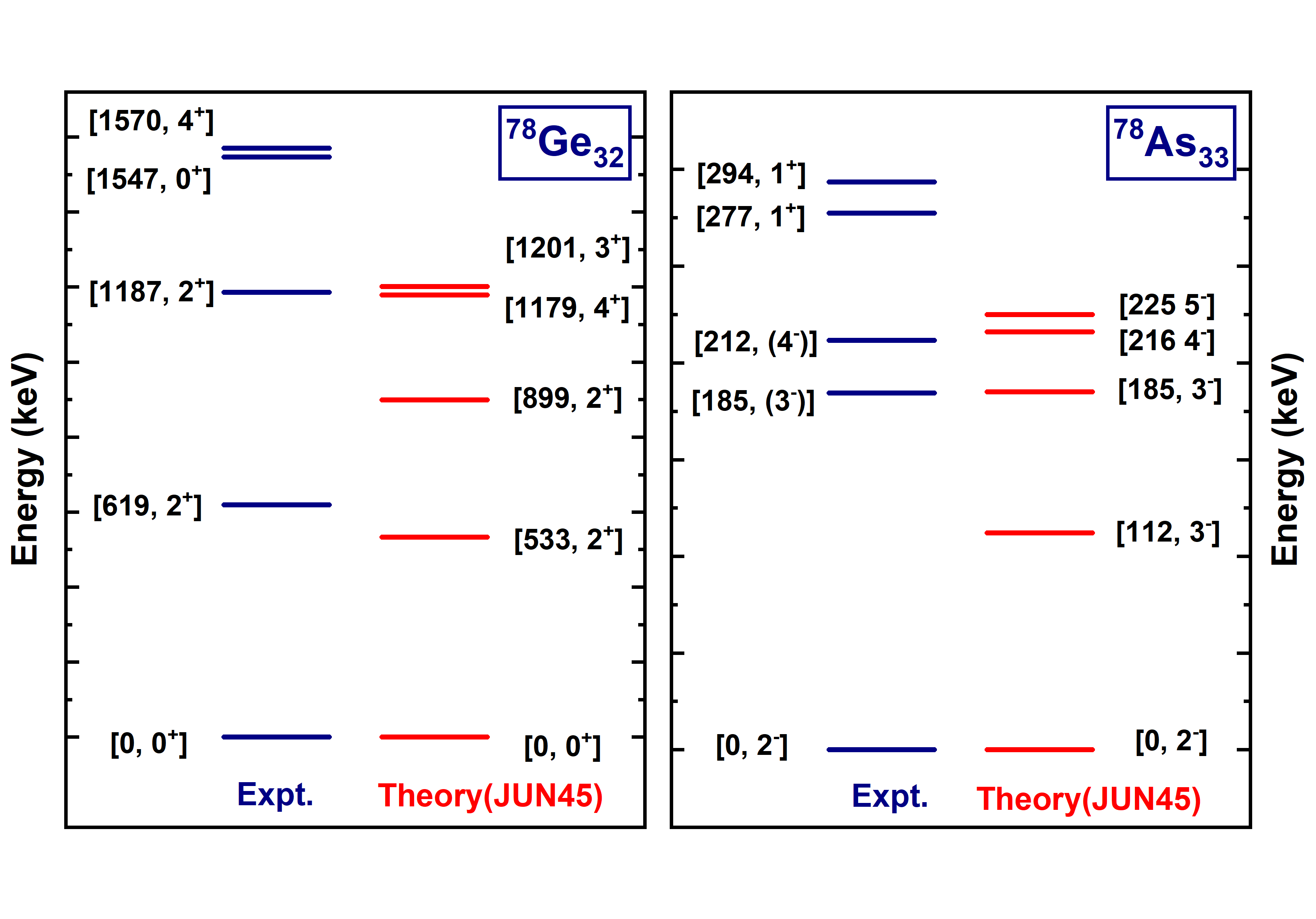}}
\vspace{0.2cm}
\renewcommand\thefigure{9}
\caption{(Color online) Comparison between theoretical and experimental level energies. Experimental results are collected from Ref. \cite{NNDC}. Experimental results are rounded off for simplicity. See text for details. 
\label{levelenergy}}
\end{figure}

We have done shell-model calculations in two valance spaces. The fp model space with $^{40}$Ca core consists of single particle orbits ($1f_{7/2} 2p_{3/2} 1f_{5/2} 2p_{1/2}$), whereas, the fpg model space with $^{56}$Ni core consists of ($2p_{3/2} 1f_{5/2} 2p_{1/2} 1g_{9/2}$) single particle orbits. Calculations have been done with various Hamiltonians available with the OXBASH and NuShellX  for each nucleus to obtain the best theoretical \logft values for terrestrial transitions. We have reported here, in Table \ref{table-logft}, only the calculated \logft that closely agrees with the corresponding experimental value and the corresponding interaction that has been used for that particular nucleus.
The description of the details of shell-model calculations with various interactions is shown in Table \ref{partition}. Some details of the particle partitions used for parent and daughter nuclei are also shown in this table. 
We have used four interaction Hamiltonians, fpd6, fpd6n, jun45 and gx1, as listed in Table   \ref{partition}, either in JT or PN formalisms. Two formalisms are equivalent. However, we have found some advantages of using PN formalism while computing with different truncations for protons and neutrons. The NuShellX admitted larger matrix dimensionalities. fpd6 interaction is older and quite tested. The interaction jun45 in the fpg valance space, has been used recently \cite{Kumar_JPG_2016} for calculating ft values in the framework of shell-model.
In the following we present in Figure \ref{levelenergy}, some of the shell-model results for energy eigenvalues of a few nuclei as representative examples.


\section {{Shell-Model Calculations: Quenching Factor}\label{app_quenching}}

\begin{figure}

{\includegraphics[width=85mm,height=65mm]{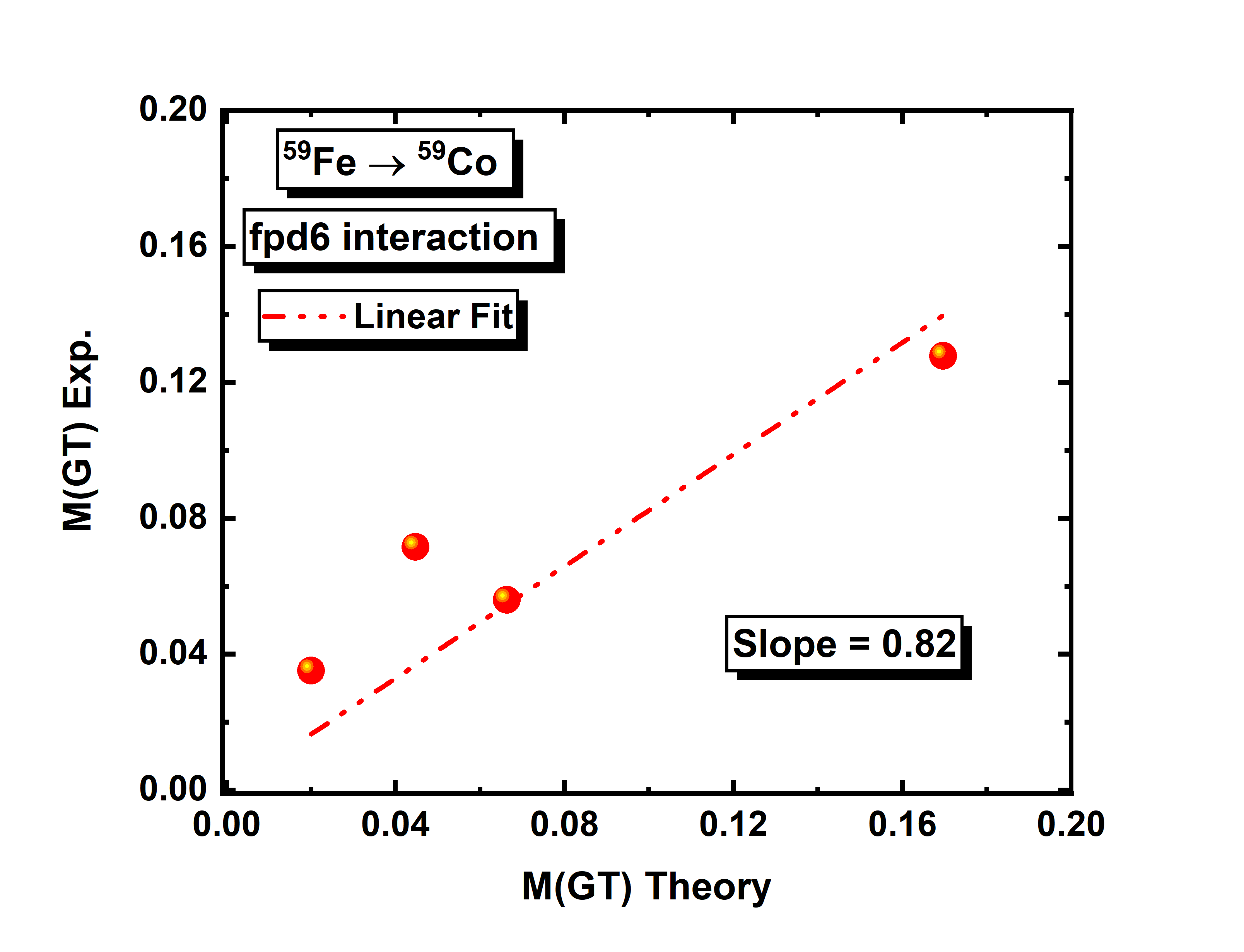}}
{\includegraphics[width=85mm,height=65mm]{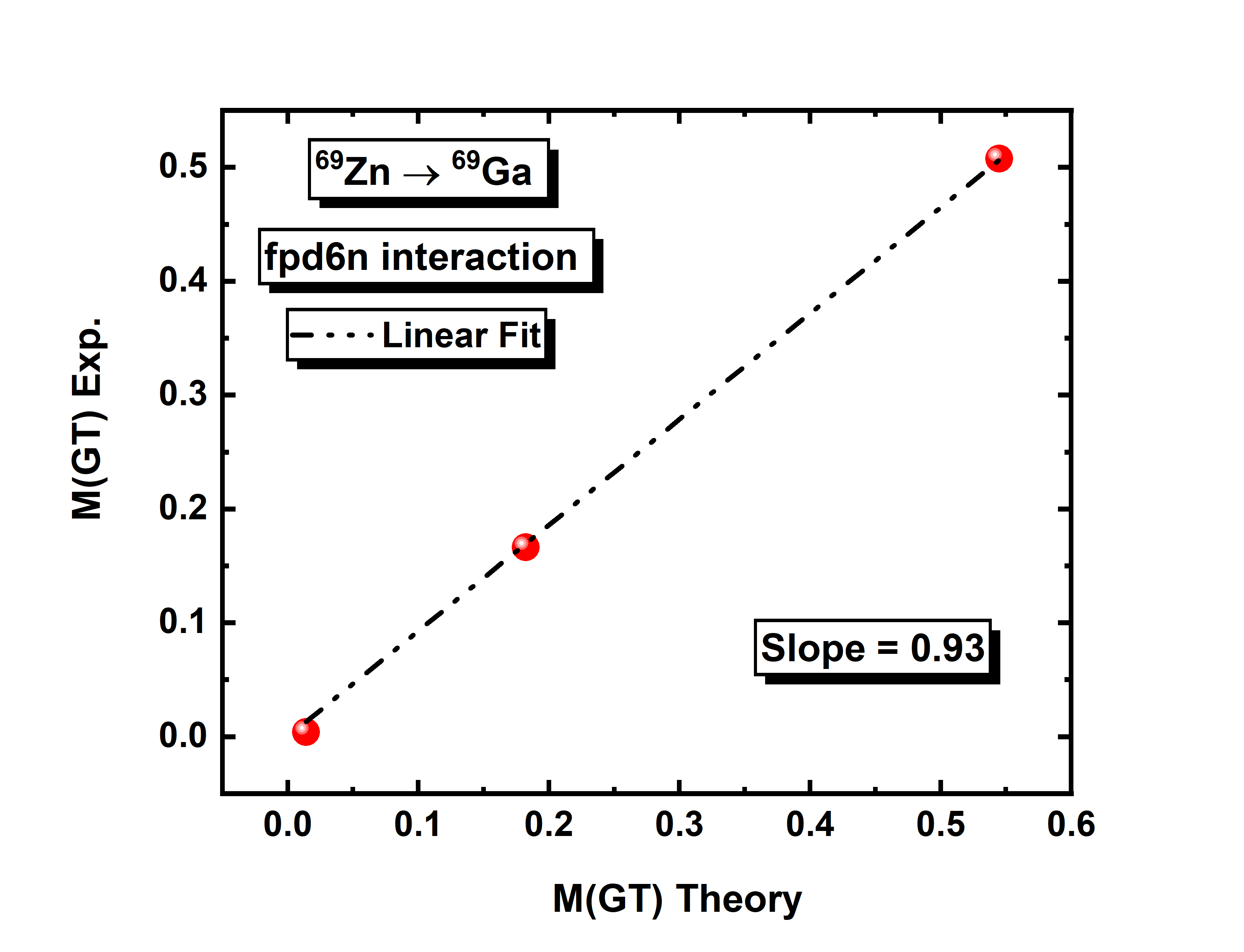}}
\renewcommand\thefigure{10}
\caption{(Color online) Theoretical vs Experimental M(GT) to obtain quenching factor. See text for details. 
\label{quenching}}
\end{figure}

The total GT strength (sum rule) measured is in general less than that predicted by shell-model calculation. Hence, theoretically obtained GT transition matrix elements are quenched by a factor q for agreement with experimental data. The quenching factor is interaction dependent. The method of obtaining the quenching factor has been discussed in section \ref{sec.3A}, and is displayed in Figure \ref{quenching} in case of multiple GT transitions of a parent level to various daughter levels. Each point in these plots represents a single transition, with theoretical M(GT) value given by x - coordinate and experimental M(GT) value \cite{NNDC} by y - coordinate. The dotted lines in these figures are the best fitted line passing through the origin. The slope of the line is the desired quenching factor. The value of the quenching factor for each nucleus is shown in Table \ref{table-logft}.

\vspace{0.2cm}



\section {{Saha Ionization equation}\label{appB}}

Under local thermodynamical equilibrium (i.e., for specific temperature, free electron density etc), Saha Ionization equation provides the fractional population of different ionized states of a particular elemental species as a function of temperature and free electron number density \cite{Takahashi_ADNDT_1987},
\begin{eqnarray}
\frac{n_{ij+1}}{n_{ij}} = \frac{g_{ij+1}}{g_{ij}} \times \left(\frac{M_{ij+1}}{M_{ij}}\right)^{3/2} \\ \nonumber
\times \text{exp} \left(-\frac{(I_{ij}- \Delta_{ij})}{k_BT}- \eta\right).
\end{eqnarray}
Where, the number density of the element i in its j - times ionized state is expressed as $n_{ij}$, atomic partition function as $g_{ij}$, mass as $M_{ij}$, ionization potential depression as $\Delta_{ij}$ (mentioned as $\Delta_{j}$ in the main text), and ionization potential of that state as $I_{ij}$. $k_B$ is the Boltzmann constant and $T$ is temperature of the stellar site. The parameter $\eta$ mentioned in the above equation is known as the degeneracy parameter which can be expressed in terms of free electron density $n_e$ as,
\begin{eqnarray}
F(\eta,b') = n_e \times \pi^2 \times \lambdabar^3.
\end{eqnarray}
Here, $\lambdabar = \hbar/m_e c$, $b' = m_e c^2 / k_BT$ and the relativistic Fermi- Dirac Integral is given by,
\begin{eqnarray}
F(\eta,b') = \int_{1}^{\infty} \frac{W \sqrt{W^2-1}} {1+\text{exp}(b'(W-1)-\eta)} dW.
\end{eqnarray}
The total mass density $\rho$ of a mixture of different elements with mass fraction of the ionized atom $x_i$ satisfy the relationship,
\begin{eqnarray}
  \rho x_i = \sum_j M_{ij}n_{ij}.
\end{eqnarray}

In equilibrium condition, the solution of Saha ionization equation must satisfy the following relation between total matter density ($\rho$) and free electron number density ($n_e$),
\begin{eqnarray}
\rho \sum_i x_i \frac{\sum_j j n_{ij}}{\sum_j M_{ij}n_{ij}}  = n_e.
\end{eqnarray}


\end{appendix}


\pagebreak[4]


\begin{table*}
\centering
\caption{Comparison of experimental \cite{NNDC} and calculated \logft. Here, $E_p$ and $E_d$ correspond to the parent and daughter level energies in keV, respectively. Errors in the experimental energy levels \cite{NNDC} are mentioned in the brackets.  $J_p^{\pi}$ and $J_d^{\pi}$ are the spin-parity of the parent and daughter levels, respectively. q is the quenching factor. The name of the interaction Hamiltonian used for each nucleus is given in Column 7. In the cases, where the spin-parity ($J^\pi$) of a level is unconfirmed in Ref.\cite{NNDC}, we have decided $J^\pi$ from SM calculation. Here, Exp. stands for experimental values \cite{NNDC} and Theo. stands for present shell-model calculation.}
\vspace{0.5cm}
\begin{tabular}{ccccccccccccccccc}
\hline
\hline
\multicolumn{10}{c}{Transition Details}   &       & \multicolumn{6}{c}{Shell-Model Results} \\   
\cline{1-10}\cline{12-17}    
\multirow{2}[4]{*}{Decay} &       & \multicolumn{3}{c}{Parent Level} &       & \multicolumn{3}{c}{Daughter Level} & Exp. &   &    &   &&            & & Theo.  \\
\cline{3-5}\cline{7-9}          &       && $J_p^{\pi}$  & $E_p$    &       && $J_d^{\pi}$  & $E_d$    & \logft &   & Interaction &  q && $E_p$     & $E_d$     & \logft \\
\hline
\hline
$^{59}$Fe $\rightarrow$ $^{59}$Co
	  &&& $3/2^{-}_{1}$  & 0.0      &&& $3/2^{-}_{1}$  & 1099.256(3) & 6.696(13) && fpd6 & 0.82  && 0     & 1139   & 6.718 \\
          &&&                &        &&& $3/2^{-}_{2}$  & 1291.605(5) & 5.979(11) &&       &         &&       & 2398   & 5.902 \\
          &&&                &        &&& $1/2^{-}_{1}$  & 1434.256(5) & 6.482(18) &&       &         &&       & 2496   & 7.060 \\
          &&&                &        &&& $5/2^{-}_{1}$  & 1481.72(12) & 7.10(4)   &&       &         &&       & 1337   & 7.759 \\
\cline{3-11} \cline{15-17}
          &&& $1/2^{-}_{1}$  & 287.023(19) &&& $3/2^{-}_{1}$  & 1099.256(3) &       &&       &         && 246    & 1139   & 7.011 \\
          &&&                &        &&& $3/2^{-}_{2}$  & 1291.605(5) &       &&       &         &&        & 2398   & 6.508 \\
          &&&                &        &&& $1/2^{-}_{1}$  & 1434.256(5) &       &&       &         &&        & 2496   & 5.770 \\
\cline{3-11} \cline{15-17}
          &&& $5/2^{-}_{1}$  & 472.87(9) &&& $7/2^{-}_{1}$  & 0.0      &       &&       &         && 128    & 0      & 5.318 \\
          &&&                &        &&& $3/2^{-}_{1}$  & 1099.256(3) &       &&       &         &&        & 1139   & 6.772 \\
          &&&                &        &&& $3/2^{-}_{2}$  & 1291.605(5) &       &&       &         &&        & 2398   & 7.611 \\
          &&&                &        &&& $5/2^{-}_{1}$  & 1481.72(12) &       &&       &         &&        & 1337   & 8.250 \\
          &&&                &        &&& $7/2^{-}_{2}$  & 1744.69(20) &       &&       &         &&        & 1494       & 6.220 \\
          \hline
$^{60}$Co $\rightarrow$ $^{60}$Ni
	  &&&   $5^{+}_{1}   $  & 0.0      &&& $4^{+}_{1}$    & 2505.753(4) & 7.512(2) && fpd6pn & 0.58 && 0  & 2242 & 7.313\\
           \cline{3-11} \cline{15-17}       
          &&& $2^{+}_{1}$    & 58.59(1)    &&& $2^{+}_{1}$    & 1332.514(4) &  7.2  &&    &&& 173    & 1326   & 7.349 \\
          
          &&&                &          &&& $2^{+}_{2}$    & 2158.632(18) &  7.4  &&    &&&      & 2095   & 7.423 \\
\cline{3-11} \cline{15-17}
          &&& $4^{+}_{1}$    & 277.20(2)    &&& $4^{+}_{1}$    & 2505.753(4) &     &&     &&& 239    & 2242  & 6.445 \\
          
          &&&                &          &&& $3^{+}_{1}$    & 2626.06(5) &     &&     &&&        & 1908 & 7.112 \\
\cline{3-11} \cline{15-17}
          &&& $3^{+}_{1}$    & 288.40(2)    &&& $2^{+}_{1}$    & 1332.514(4)   &     &&&        && 415  & 1326  & 6.892 \\
          &&&                &              &&& $2^{+}_{2}$    & 2158.632(18)  &     &&&        &&      & 2095  & 7.242 \\
          &&&                &              &&& $4^{+}_{1}$    & 2505.753(4)   &     &&&        &&      & 2242  & 6.897 \\
          &&&                &              &&& $3^{+}_{1}$    & 2626.06(5)    &     &&&        &&      & 2202  & 8.331 \\
\cline{3-11} \cline{15-17}
          &&& $5^{+}_{2}$    & 435.71(4)    &&& $4^{+}_{1}$    & 2505.753(4) &     &&&        &&  461 & 2242  & 6.996 \\
          &&&                &              &&& $4^{+}_{2}$    & 3119.87(7)  &     &&&        &&      & 2595  & 7.289 \\
\hline
$^{61}$Co $\rightarrow$ $^{61}$Ni
          & && $7/2^{-}_{1}$   &  0.0       &&& $5/2^{-}_{1}$    & 67.414(7)    & 5.240(3) && fpd6n  & 1.00      && 0    & 221   & 5.292 \\
          &&&                 &          &&& $5/2^{-}_{2}$    & 908.613(11)   &      &&   &    &&  & 783   & 5.981 \\
          &&&                 &          &&& $(7/2)^{-}_{1}$    & 917.5(7)    & 4.78(4) &&   &    &&  & 1134  & 5.788 \\
          &&&                 &          &&& $7/2^{-}_{2}$    & 1015.24(8)    &      &&   &    &&  & 1546  & 5.949 \\
          &&&                 &          &&& $5/2^{-}_{3}$    & 1132.347(18)  &      &&   &    &&  & 1118  & 5.226 \\
	 \hline  
$^{63}$Ni $\rightarrow$ $^{63}$Cu
          &&& $1/2^{-}_{1}$   &  0.0             &&& $3/2^{-}_{1}$   & 0.0    & 6.7 &&   fpd6npn  & 1.00      && 7   & 0 & 6.318 \\
\cline{3-11} \cline{15-17}
          &&& $5/2^{-}_{1}$   &  87.15(11)        &&& $3/2^{-}_{1}$    & 0.0    &     &&   &    && 0 & 0 & 5.747 \\
\cline{3-11} \cline{15-17}
          &&& $3/2^{-}_{1}$   &  155.55(15)       &&& $3/2^{-}_{1}$    & 0.0    &     &&   &    && 238 & 0 & 5.323 \\
\hline
$^{65}$Ni $\rightarrow$ $^{65}$Cu
	   &&& $5/2^{-}_{1}$       & 0.0      &&& $3/2^{-}_{1}$  & 0.0          & 6.576(2) &&  jun45  &   0.63  && 8     & 0 & 6.592 \\
          &&&                &          &&& $5/2^{-}_{1}$  & 1115.556(4)  & 6.064(6) &&   &    &&       & 1570 & 6.406 \\
          &&&                &          &&& $7/2^{-}_{1}$  & 1481.83(3)   & 4.901(4) &&   &    &&       & 1516 & 4.933 \\
          &&&                &          &&& $5/2^{-}_{2}$  & 1623.43(5)   & 6.03(1)  &&   &    &&       & 2074 & 5.609 \\
          &&&                &          &&& $3/2^{-}_{2}$  & 1725.00(5)   & 5.90(1)  &&   &    &&       & 2155 & 6.370 \\
          &&&                &          &&& $(7/2)^{-}_{2}$& 2094.34(14)  &          &&   &    &&       & 2164 & 4.385 \\
          &&&                &          &&& $(5/2)^{-}_{3}$& 2107.44(13)  &          &&   &    &&       & 2425 & 7.556 \\
\cline{3-11} \cline{15-17}
	   &&& $1/2^{-}_{1}$  & 63.37(5)  &&& $3/2^{-}_{1}$  & 0.0    & &&   &    &&  0      & 0    & 4.567 \\
          &&&                &        &&& $1/2^{-}_{1}$  & 770.64(9)  & &&   &    &&         & 931  & 6.155 \\
          &&&                &        &&& $3/2^{-}_{2}$  & 1725.00(5)  & &&   &    &&         & 2155 & 6.249 \\
          \cline{3-11} \cline{15-17}
	   &&& $3/2^{-}_{1}$  & 310.08(22)       &&& $3/2^{-}_{1}$  & 0.0     &   &&&    && 599     & 0   & 5.561 \\
          &&&                &             &&& $1/2^{-}_{1}$  & 770.64(9)   &&&    &&&         & 931  & 7.166 \\
          &&&                &             &&& $5/2^{-}_{1}$  & 1115.556(4)  &&&    &&&         & 1570 & 6.078 \\
          &&&                &             &&& $5/2^{-}_{2}$  & 1623.43(5)  &&&    &&&         & 2074 & 6.446 \\
          &&&                &             &&& $3/2^{-}_{2}$  & 1725.00(5)  &&&    &&&         & 2155 & 8.367 \\
          &&&                &             &&& $(5/2)^{-}_{3}$  & 2107.44(13)  &&&    &&&         & 2425 & 5.637 \\
          &&&                &             &&& $(1/2)^{-}_{2}$  & 2212.84(15)  &&&    &&&         & 2259 & 6.145 \\
          &&&                &             &&& $3/2^{-}_{3}$  & 2329.05(15)  &&&    &&&  & 2345 & 6.768 \\          
\hline 
\hline

    \end{tabular}
  \label{table-logft}
\end{table*}

\begin{table*}
    \centering
    ~~~~~~TABLE (I) continued
    \vspace{0.5cm}
\begin{tabular}{ccccccccccccccccc}
\hline
\hline
\multicolumn{10}{c}{Transition Details}   &       & \multicolumn{6}{c}{Shell-Model Results} \\   
\cline{1-10}\cline{12-17}    
\multirow{2}[4]{*}{Decay} &       & \multicolumn{3}{c}{Parent Level} &       & \multicolumn{3}{c}{Daughter Level} & Exp. &   &    &   &&            & & Theo.  \\
\cline{3-5}\cline{7-9}          &       && $J_p^{\pi}$  & $E_p$    &       && $J_d^{\pi}$  & $E_d$    & \logft &   & Interaction &  q && $E_p$     & $E_d$     & \logft \\
\hline
\hline

$^{66}$Ni $\rightarrow$ $^{66}$Cu
          &&& $0^{+}_{1}$   &  0.0             &&& $1^{+}_{1}$   & 0.0    &  4.3   && fpd6n & 1.0       && 0   & 0 & 4.307 \\
          \hline

$^{64}$Cu $\rightarrow$ $^{64}$Zn
          &&& $1^{+}_{1}$   &  0.0             &&& $0^{+}_{1}$   & 0.0    &  5.302(5) &&  jun45 & 1.0     && 0   & 0 & 5.368 \\
          \hline

$^{66}$Cu $\rightarrow$ $^{66}$Zn
          &&& $1^{+}_{1}$      &  0.0             &&& $0^{+}_{1}$     & 0.0          &  5.33   && jun45 &  0.51      && 67      & 0  & 5.575 \\
          &&&                  &                &&& $2^{+}_{1}$     & 1039.2279(21)     &  5.43   &&   &     &&    & 1059 & 5.203 \\
          &&&                  &                &&& $2^{+}_{2}$     & 1872.7653(24)     &  5.82(1)   &&   &     &&    & 2015 & 6.173 \\
          &&&                  &                &&& $0^{+}_{2}$     & 2372.353(4)       &  6.01(4)   &&   &     &&    & 2550 & 6.562 \\
          \cline{3-11} \cline{15-17}  
          &&& $2^{+}_{1}$      &  185.953(15)       &&& $2^{+}_{1}$     & 1039.2279(21)     &     &&   &      &&   0     & 1059  & 5.501 \\
          &&&                  &                &&& $2^{+}_{2}$     & 1872.7653(24)     &     &&   &      &&         & 2015  & 7.771 \\
          &&&                  &                &&& $2^{+}_{3}$     & 2780.157(7)       &     &&   &      &&         & 2695  & 6.440 \\
          \cline{3-11} \cline{15-17} 
          &&& $3^{+}_{1}$      &  275.030(17)           &&& $2^{+}_{1}$     & 1039.2279(21)     &     &&   &      &&  13      & 1059 & 7.925 \\
          &&&                  &                &&& $2^{+}_{2}$     & 1872.7653(24)     &     &&   &      &&          & 2015 & 8.216 \\
          &&&                  &                &&& $4^{+}_{1}$     & 2451.01(5)        &     &&   &      &&          & 2451 & 5.244 \\
          &&&                  &                &&& $4^{+}_{2}$     & 2765.56(7)        &     &&   &      &&          & 2633 & 5.920 \\
          &&&                  &                &&& $2^{+}_{3}$     & 2780.157(7)       &     &&   &      &&          & 2695 & 6.294 \\
          \cline{3-11} \cline{15-17}  
          &&& $(1^{+}_{2})$      &  385.782(10)       &&& $0^{+}_{1}$     & 0.0            &     &&   &      &&   94   & 0    & 6.085 \\
          &&&                  &                &&& $2^{+}_{1}$     & 1039.2279(21)  &     &&   &      &&        & 1059 & 5.290 \\
          &&&                  &                &&& $2^{+}_{2}$     & 1872.7653(24)  &     &&   &      &&        & 2015 & 5.955 \\
          &&&                  &                &&& $0^{+}_{2}$     & 2372.353(4)    &     &&   &      &&        & 2550 & 6.949 \\
          &&&                  &                &&& $2^{+}_{3}$     & 2780.157(7)    &     &&   &      &&        & 2695 & 6.597 \\
          &&&                  &                &&& $2^{+}_{4}$     & 2938.074(3)    &     &&   &      &&        & 2957 & 5.643 \\
          \cline{3-11} \cline{15-17}  
          &&& $2^{+}_{2}$      &  465.165(10)       &&& $2^{+}_{1}$     & 1039.2279(21)           &     &&   &     &&   61     & 1059 & 5.911 \\
          &&&                  &                &&& $2^{+}_{2}$     & 1872.7653(24)  &     &&   &     &&          & 2015 & 6.011 \\
          &&&                  &                &&& $2^{+}_{3}$     & 2780.157(7)  &     &&   &     &&          & 2695 & 6.669 \\
          &&&                  &                &&& $2^{+}_{4}$     & 2938.074(3)    &     &&   &     &&          & 2957 & 6.228 \\
          \hline

$^{67}$Cu $\rightarrow$ $^{67}$Zn
          &&& $3/2^{-}_{1}$    &  0.0             &&& $5/2^{-}_{1}$   & 0.0        &  $\approx$ 6.3   &&  fpd6n & 0.65        && 0   & 183 & 5.976 \\
          &&&                  &                &&& $1/2^{-}_{1}$   & 93.312(5)     &  $\approx$ 6.0   &&    &       &&      & 0   & 5.976 \\
          &&&                  &                &&& $3/2^{-}_{1}$   & 184.579(6)    &  $\approx$ 5.2   &&    &       &&      & 153 & 5.430 \\
          &&&                  &                &&& $3/2^{-}_{2}$   & 393.531(7)    &  $\approx$ 5.8   &&    &       &&      & 483 & 5.535 \\
          \hline
$^{69}$Zn $\rightarrow$ $^{69}$Ga
          &&& $1/2^{-}_{1}$    &  0.0             &&& $3/2^{-}_{1}$   & 0.0         &  4.48(1)   &&  {\multirow{3}[1]{*} {fpd6n}} & {\multirow{3}[1]{*} {0.93}}      && 0    & 0 & 4.482 \\
          &&&                  &                &&& $1/2^{-}_{1}$   & 318.706(21)   &  8.72(8)   &&    &     &&     & 53  & 7.654 \\
          &&&                  &                &&& $3/2^{-}_{2}$   & 871.147(22)   &  5.45(19)  &&    &     &&     & 773 & 5.434 \\
\hline
$^{72}$Zn $\rightarrow$ $^{72}$Ga
          &&& $0^{+}_{1}$    &  0.0             &&& $(0^{+}_{1})$   & 119.66(5)         &  $>$8.6   &&  {\multirow{4}[1]{*} {jun45}} & {\multirow{4}[1]{*} {0.73}}      && 0    & 253 & 7.321 \\
          &&&                  &                &&& $1^+_1, 2$   & 128.79(6)     &  7.2(3)     &&    &     &&     & 367 & 4.837 \\
          &&&                  &                &&& $1^{+}_{2}$   & 161.53(5)     &  4.468(12)   &&    &     &&     & 457 & 4.673 \\
          &&&                  &                &&& $1^{+}_{3}$   & 208.45(5)     &  4.972(17)   &&    &     &&     & 641 & 4.672 \\ 
\hline
$^{70}$Ga $\rightarrow$ $^{70}$Ge
          &&& $1^{+}_{1}$    &  0.0             &&& $0^{+}_{1}$   & 0.0         &  5.0925(18)   &&  {\multirow{3}[1]{*} {gx1}} & {\multirow{3}[1]{*} {0.52}}      && 0    & 0 & 5.113 \\
          &&&                  &                &&& $2^{+}_{1}$   & 1039.506(9) &  5.895(25)   &&    &     &&     & 1281  & 5.046 \\
          &&&                  &                &&& $0^{+}_{2}$   & 1215.621(15)&  5.431(15)   &&    &     &&     & 3190 & 5.391 \\ 
\hline                 
$^{75}$Ge $\rightarrow$ $^{75}$As
            & & & $1/2^-_1$ & 0.0     &       &       & $3/2^-_1$ & 0.0   & 5.175(7) &       & jun45      &   0.25    &       & 0     & 0     & 5.180 \\
            & & &       &       &       &       & $1/2^-_1$ & 198.6063(8) & 6.87(5)  &       &       &       &       &       & 281   & 7.160 \\
            & & &       &       &       &       & $3/2^-_2$ & 264.6581(6) & 5.63(5)  &       &       &       &       &       & 788   & 5.644 \\
            & & &       &       &       &       & $1/2^-_2$ & 468.74(17) & 6.94(5)  &       &       &       &       &       & 1757  & 7.080 \\
            & & &       &       &       &       & $1/2^-_3$ & 585(7) &       &       &       &       &       &       & 2014  & 5.476 \\
            & & &       &       &       &       & $1/2^-$, $3/2^-_3$ & 617.68(4) & 6.42(6)  &       &       &       &       &       & 1355  & 6.242 \\
            & & &       &       &       &       & $(3/2^-_4, 5/2^-)$ & 865.4(5) &       &       &       &       &       &       & 1769  & 7.336 \\
            & & &       &       &       &       & $3/2^-_5$ & 1063.3(5) &       &       &       &       &       &       & 2068  & 8.807 \\
            & & &       &       &       &       & $3/2^-_6$ & 1074.5(7) &       &       &       &       &       &       & 2292  & 7.021 \\
            & & &       &       &       &       & $1/2^-_4$, $3/2^-$ & 1127(6) &       &       &       &       &       &       & 2624  & 7.020 \\
            & & &       &       &       &       & $(1/2^-_5 to~7/2^-)$ & 1172.0(6) &       &       &       &       &       &       & 2870  & 5.975 \\
\cline{3-11} \cline{15-17} 
&       &       & $7/2^+_1$ & 139.69(3) &       &       & $9/2^+_1$ & 303.9243(8) &   &       &       &       &       & 13    & 2338  & 7.414 \\
&       &       &           &           &       &       & $5/2^+_1$ & 400.6583(6) & 6.21(13)  &       &       &       &       &       & 2551  & 7.090 \\
          &       &       &       &       &       &       &$(5/2^+_2)$ & 1080.8(8) &       &       &       &       &       &       & 2997  & 5.964 \\
          &       &       &       &       &       &       & $(5/2^+_3, 7/2^-)$ & 1100.2(6) &       &       &       &       &       &       & 3085  & 6.836 \\
          &       &       &       &       &       &       & $9/2^+_2$ & 1261(5) &       &       &       &       &       &       & 2572  & 6.556 \\
          &       &       &       &       &       &       & $5/2^+_4$ & 1302.3(7) &       &       &       &       &       &       & 3116  & 6.152 \\
                    \cline{3-11} \cline{15-17} 
 &       &       & $5/2^+_1$ & 192.19(6) &       &       & $5/2^+_1$ & 400.6583(6) &       &       &       &       &       & 9     & 2551  & 6.584 \\
          &       &       &       &       &       &       & $(5/2^+_2)$ & 1080.8(8) &       &       &       &       &       &       & 2997  & 5.809 \\
          &       &       &       &       &       &       & $(5/2^+_3, 7/2^-)$ & 1100.2(6) &       &       &       &       &       &       & 3085  & 6.030 \\
          &       &       &       &       &       &       & $5/2^+_4$ & 1302.3(7) &       &       &       &       &       &       & 3116  & 5.852 \\       
\hline 
\hline
\end{tabular}
\label{table-logft-b}
\end{table*}

\begin{table*}
\centering
~~~~~~TABLE (I) continued
\vspace{0.5cm}
\begin{tabular}{ccccccccccccccccc}
\hline
\hline
\multicolumn{10}{c}{Transition Details}   &       & \multicolumn{6}{c}{Shell-Model Results} \\   
\cline{1-10}\cline{12-17}    
\multirow{2}[4]{*}{Decay} &       & \multicolumn{3}{c}{Parent Level} &       & \multicolumn{3}{c}{Daughter Level} & Exp. &   &    &   &&            & & Theo.  \\
\cline{3-5}\cline{7-9}          &       && $J_p^{\pi}$  & $E_p$    &       && $J_d^{\pi}$  & $E_d$    & \logft &   & Interaction &  q && $E_p$     & $E_d$     & \logft \\
\hline
\hline

 &       &       & $9/2^+_1$ & 199.89(11) &       &       & $9/2^+_1$ & 303.9243(8) &       &       &       &       &       & 183   & 2338  & 6.901 \\
          &       &       &       &       &       &       & $9/2^+_2$ & 1261(5) &       &       &       &       &       &       & 2572  & 6.979 \\ 
\cline{3-11} \cline{15-17}        
&       &       & $3/2^-_1$ & 253.15(6) &       &       & $3/2^-_1$ & 0.0     &       &       &       &       &       & 434   & 0     & 5.725 \\
          &       &       &       &       &       &       & $1/2^-_1$ & 198.6063(8) &       &       &       &       &       &       & 281   & 5.796 \\
          &       &       &       &       &       &       & $3/2^-_2$ & 264.6581(6) &       &       &       &       &       &       & 788   & 6.262 \\
          &       &       &       &       &       &       & $5/2^-_1$ & 279.5428(8) &       &       &       &       &       &       & 600   & 7.960 \\
          &       &       &       &       &       &       & $1/2^-_2$ & 468.74(17) &       &       &       &       &       &       & 1757  & 7.883 \\
          &       &       &       &       &       &       & $5/2^-_2$ & 572.41(3) &       &       &       &       &       &       & 841   & 6.913 \\
          &       &       &       &       &       &       & $1/2^-_3$ & 585(7) &       &       &       &       &       &       & 2014  & 7.289 \\
          &       &       &       &       &       &       & $1/2^-$, $3/2^-_3$ & 617.68(4) &       &       &       &       &       &       & 1355  & 6.583 \\
          &       &       &       &       &       &       & $(3/2^-_4, 5/2^-)$ & 865.4(5) &       &       &       &       &       &       & 1769  & 6.504 \\
          &       &       &       &       &       &       & $3/2^-_5$ & 1063.3(5) &       &       &       &       &       &       & 2068  & 6.084 \\
          &       &       &       &       &       &       & $3/2^-_6$ & 1074.5(7) &       &       &       &       &       &       & 2292  & 7.186 \\
          &       &       &       &       &       &       & $1/2^-_4, 3/2^-$ & 1127(6) &       &       &       &       &       &       & 2652  & 6.757 \\
          &       &       &       &       &       &       & $(1/2^-_5 to~ 7/2^-)$ & 1172.0(6) &       &       &       &       &       &       & 2624  & 6.757 \\
          &       &       &       &       &       &       & $3/2^-_7$ & 1203.5(6) &       &       &       &       &       &       & 2770  & 6.534 \\
          &       &       &       &       &       &       & $3/2^-_8$ & 1349.4(6) &       &       &       &       &       &       & 2984  & 6.799 \\
          &       &       &       &       &       &       & $(3/2^-)_9$ & 1370.8(7) &       &       &       &       &       &       & 3052  & 6.080 \\
          &       &       &       &       &       &       & $(5/2^-)_3$ & 1420.2(5) &       &       &       &       &       &       & 1047 & 6.055 \\ 
\cline{3-11} \cline{15-17}      
 &       &       & $5/2^-_1$ & 316.81(7) &       &       & $3/2^-_1$ & 0.0     &       &       &       &       &       & 48    & 0     & 6.529 \\
          &       &       &       &       &       &       & $3/2^-_2$ & 264.6581(6) &       &       &       &       &       &       & 788   & 6.556 \\
          &       &       &       &       &       &       & $5/2^-_1$ & 279.5428(8) &       &       &       &       &       &       & 599   & 6.030 \\
          &       &       &       &       &       &       & $5/2^-_2$ & 572.41(3) &       &       &       &       &       &       & 841   & 6.312 \\
          &       &       &       &       &       &       & $1/2^-, 3/2^-_3$ & 617.68(4) &       &       &       &       &       &       & 1355  & 7.048 \\
          &       &       &       &       &       &       & $7/2^-_1$ & 821.62(15) &       &       &       &       &       &       & 576   & 5.264 \\
          &       &       &       &       &       &       & $(3/2^-_4, 5/2^-)$ & 865.4(5) &       &       &       &       &       &       & 1769  & 7.401 \\
          &       &       &       &       &       &       & $7/2^-_2$ & 1043.4(6) &       &       &       &       &       &       & 1467  & 6.075 \\
          &       &       &       &       &       &       & $3/2^-_5$ & 1063.3(5) &       &       &       &       &       &       & 2068  & 7.248 \\
          &       &       &       &       &       &       & $3/2^-_6$ & 1074.5(7) &       &       &       &       &       &       & 2292  & 9.850 \\
          &       &       &       &       &       &       & $(7/2^-)_3$ & 1096.3(7) &       &       &       &       &       &       & 1743  & 7.251 \\
          &       &       &       &       &       &       & $(5/2^+, 7/2^-_4)$ & 1100.2(6) &       &       &       &       &       &       & 1992  & 6.745 \\
          &       &       &       &       &       &       & $3/2^-_7$ & 1203.5(6) &       &       &       &       &       &       & 2652  & 7.349 \\
          &       &       &       &       &       &       & $7/2^-_5$ & 1309.5(4) &       &       &       &       &       &       & 2493  & 6.283 \\
          &       &       &       &       &       &       & $3/2^-_8$ & 1349.4(6) &       &       &       &       &       &       & 2770  & 8.363 \\
          &       &       &       &       &       &       & $(3/2^-)_9$ & 1370.8(7) &       &       &       &       &       &       & 2984  & 7.270 \\
          &       &       &       &       &       &       & $(5/2^-)_3$ & 1420.2(5) &       &       &       &       &       &       & 1046  & 5.915 \\
          &       &       &       &       &       &       & $3/2^-_{10}$ & 1430.5(6) &       &       &       &       &       &       &  3052     & 7.015 \\
\cline{3-11} \cline{15-17}
        &       &       & $5/2^-_2$ & 457.07(7) &       &       & $3/2^-_1$ & 0.0     &       &       &       &       &       & 374   & 0     & 7.128 \\
          &       &       &       &       &       &       & $3/2^-_2$ & 264.6581(6) &       &       &       &       &       &       & 788   & 8.497 \\
          &       &       &       &       &       &       & $5/2^-_1$ & 279.5428(8) &       &       &       &       &       &       & 599   & 7.989 \\
          &       &       &       &       &       &       & $5/2^-_2$ & 572.41(3) &       &       &       &       &       &       & 841   & 6.645 \\
          &       &       &       &       &       &       & $1/2^-, 3/2^-_3$ & 617.68(4) &       &       &       &       &       &       & 1355  & 8.312 \\
          &       &       &       &       &       &       & $7/2^-_1$ & 821.62(15) &       &       &       &       &       &       & 576   & 6.053 \\
          &       &       &       &       &       &       & $(3/2^-_4, 5/2^-)$ & 865.4(5) &       &       &       &       &       &       & 1769  & 9.618 \\
          &       &       &       &       &       &       & $7/2^-_2$ & 1043.4(6) &       &       &       &       &       &       & 1467  & 7.110 \\
          &       &       &       &       &       &       & $3/2^-_5$ & 1063.3(5) &       &       &       &       &       &       & 2068  & 7.444 \\
          &       &       &       &       &       &       & $3/2^-_6$ & 1074.5(7) &       &       &       &       &       &       & 2292  & 8.669 \\
          &       &       &       &       &       &       & $(7/2^-)_3$ & 1096.3(7) &       &       &       &       &       &       & 1743  & 6.731 \\
          &       &       &       &       &       &       & $(5/2^+, 7/2^-_4)$ & 1100.2(6) &       &       &       &       &       &       & 1992  & 6.836 \\
          &       &       &       &       &       &       & $3/2^-_7$ & 1203.5(6) &       &       &       &       &       &       & 2652  & 7.522\\
          &       &       &       &       &       &       & $7/2^-_5$ & 1309.5(4) &       &       &       &       &       &       & 2493  & 8.629 \\
          &       &       &       &       &       &       & $3/2^-_8$ & 1349.4(6) &       &       &       &       &       &       & 2770  & 8.833 \\
          &       &       &       &       &       &       & $(3/2^-)_{9}$ & 1370.8(7) &       &       &       &       &       &       & 2984  & 8.667 \\
          &       &       &       &       &       &       & $(5/2^-)_3$ & 1420.2(5) &       &       &       &       &       &       & 1046  & 6.777 \\
          &       &       &       &       &       &       & $3/2^-_{10}$ & 1430.5(6) &       &       &       &       &       &       &  3052     & 7.934 \\
          &       &       &       &       &       &       & $1/2^-, 3/2^-_{11}$ & 1606.3(5) &       &       &       &       &       &       &  3097     & 7.789 \\   
\hline
$^{78}$Ge $\rightarrow$ $^{78}$As
          &&& $0^{+}_{1}$    &  0.0             &&& $1^{+}_{1}$   & 277.3(3)         &  4.264(25)   &&  {\multirow{2}[1]{*} {jun45}} & {\multirow{2}[1]{*} {1.00}}      && 0    & 316 & 4.646 \\
          &&&                  &                &&& $1^{+}_{2}$   & 293.9(5)     &  5.61(12)   &&    &     &&     & 518 & 5.046 \\ 
                    &&&                  &                &&& $1^{+}_{3}$   & 536(4)     &  ---   &&    &     &&     & 593 & 3.875 \\              
                                                         
\hline
\hline
\end{tabular}
\label{table-logft-c}
\end{table*}

\begin{table*}
\centering
~~~~~~TABLE (I) continued
\vspace{0.5cm}
\begin{tabular}{ccccccccccccccccc}
\hline
\hline
\multicolumn{10}{c}{Transition Details}   &       & \multicolumn{6}{c}{Shell-Model Results} \\   
\cline{1-10}\cline{12-17}    
\multirow{2}[4]{*}{Decay} &       & \multicolumn{3}{c}{Parent Level} &       & \multicolumn{3}{c}{Daughter Level} & Exp. &   &    &   &&            & & Theo.  \\
\cline{3-5}\cline{7-9}          &       && $J_p^{\pi}$  & $E_p$    &       && $J_d^{\pi}$  & $E_d$    & \logft &   & Interaction &  q && $E_p$     & $E_d$     & \logft \\
\hline
\hline
$^{81}$Se $\rightarrow$ $^{81}$Br &       &       & $1/2^-_1$ & 0.0     & & & $3/2^-_1$ & 0.0     & 5.010(4)  &       &  jun45  &  0.62   &  29    & 0    & 0 & 5.047 \\
          &       &       &       &       &       &       & $1/2^-_1$, $3/2^-$ & 538.20(8) & 7.77(7)  &       &       &       &       &       & 351   & 7.957 \\
          &       &       &       &       &       &       & $3/2^-_2$ & 566.04(5) & 6.36(5)  &       &       &       &       &       & 671   & 7.297 \\
          &       &       &       &       &       &       & $(3/2^-)_3$ & 649.90(8) & 7.83(6)  &       &       &       &       &       & 793   & 6.901 \\
          &       &       &       &       &       &       & $3/2^-_4$ & 828.29(5) & 6.18(5)  &       &       &       &       &       & 1162  & 5.870 \\
          &       &       &       &       &       &       & $(1/2)^-_2$ & 1105.3(6) &       &       &       &       &       &       & 908   & 6.625 \\
          &       &       &       &       &       &       & $(3/2^-_5, 5/2, 7/2^-)$ & 1266.8(6) &       &       &       &       &       &       & 1865  & 6.998 \\
          &       &       &       &       &       &       & $(3/2^-)_6$ & 1535.9(7) &       &       &       &       &       &       & 1921  & 6.849 \\
          &       &       &       &       &       &       & $(3/2^-)_7$ & 1543.2(5) &       &       &       &       &       &       & 1991  & 6.391 \\
          
\cline{3-11} \cline{15-17} 
          &       &       & $7/2^+_1$ & 103.00(6) &       &       & $9/2^+_1$ & 536.20(9) & 8.25(22)      &       &       &       &       & 0     & 2042  & 8.068 \\
          &       &       &       &       &       &       & $7/2^+_1$ & 1371.5(13) &       &       &       &       &       &       & 2390  & 5.528 \\
          &       &       &       &       &       &       & $(9/2^+)_2$ & 1541.6(13) &       &       &       &       &       &       & 2643  & 5.050 \\
          
\cline{3-11} \cline{15-17}  

         &       &       & $9/2^+_1$ & 294.30(17) &       &       & $9/2^+_1$ & 536.20(9) &       &       &       &       &       & 171   & 2042  & 7.074 \\
          &       &       &       &       &       &       & $7/2^+_1$ & 1371.5(13) &       &       &       &       &       &       & 2390  & 6.427 \\
          &       &       &       &       &       &       & $(11/2^+)_1$ & 1522.3(8) &       &       &       &       &       &       & 2454  & 6.900 \\
          &       &       &       &       &       &       & $(9/2^+)_2$ & 1541.6(13) &       &       &       &       &       &       & 2643  & 5.113 \\
          &       &       &       &       &       &       & $(7/2^+)_2$ & 1788.9(13) &       &       &       &       &       &       & 2481  & 5.804 \\
\cline{3-11} \cline{15-17}

      &       &       & $3/2^-_1$ & 467.74(8) &       &       & $3/2^-_1$ & 0.0     &       &       &       &       &       & 638   & 0 & 5.400 \\
          &       &       &       &       &       &       & $5/2^-_1$ & 275.985(12) &       &       &       &       &       &       & 541   & 6.257 \\
          &       &       &       &       &       &       & $1/2^-_1, 3/2^-$ & 538.20(8) &       &       &       &       &       &       & 351   & 7.837 \\
          &       &       &       &       &       &       & $3/2^-_2$ & 566.04(5) &       &       &       &       &       &       & 671   & 6.138 \\
          &       &       &       &       &       &       & $(3/2)^-_3$ & 649.90(8) &       &       &       &       &       &       & 793   & 8.761 \\
          &       &       &       &       &       &       & $(5/2^-)_2$ & 767.04(9) &       &       &       &       &       &       & 538   & 5.530 \\
          &       &       &       &       &       &       & $3/2^-_4$ & 828.29(5) &       &       &       &       &       &       & 1162  & 6.937 \\
          &       &       &       &       &       &       & $5/2^{(-)}_3$ & 1023.7(4) &       &       &       &       &       &       & 845   & 6.439 \\
          &       &       &       &       &       &       & $(1/2)^-_2$ & 1105.3(6) &       &       &       &       &       &       & 908   & 6.664 \\
          &       &       &       &       &       &       & $5/2^-_4, 7/2^-$ & 1189.9(21) &       &       &       &       &       &       & 993   & 7.169 \\
          &       &       &       &       &       &       & $(3/2^-_5, 5/2,7/2^-)$ & 1266.8(6) &       &       &       &       &       &       & 1865  & 6.181 \\
          &       &       &       &       &       &       & $(5/2)^-_5$ & 1323.0(4) &       &       &       &       &       &       & 1578  & 6.654 \\
          &       &       &       &       &       &       & $(3/2^-_6, 1/2^-)$ & 1512.9(10) &       &       &       &       &       &       & 1921  & 5.938 \\
          &       &       &       &       &       &       & $(3/2^-)_7$ & 1535.9(7) &       &       &       &       &       &       & 1991  & 5.989 \\
          &       &       &       &       &       &       & $(3/2^-)_8$ & 1543.2(5) &       &       &       &       &       &       & 2267  & 5.749 \\
          &       &       &       &       &       &       & $(5/2^-)_6$ & 1798.9(10) &       &       &       &       &       &       & 1902  & 6.388 \\
          &       &       &       &       &       &       & $(3/2^-)_9$ & 1866.4(10) &       &       &       &       &       &       & 2399  & 6.561 \\
          &       &       &       &       &       &       & $(3/2^-_{10}, 5/2, 7/2^-)$ & 1885.2(7) &       &       &       &       &       &       & 2507  & 5.769 \\
\cline{3-11} \cline{15-17} 

         &       &       & $(5/2)^-_1$ & 491.06(9) &       &       & $3/2 ^-_1$ & 0.0     &       &       &       &       &       & 479   & 0     & 8.745 \\
          &       &       &       &       &       &       & $5/2 ^-_1$ & 275.985(12) &       &       &       &       &       &       & 541   & 5.346 \\
          &       &       &       &       &       &       & $3/2 ^-_2$ & 566.04(5) &       &       &       &       &       &       & 671   & 7.250 \\
          &       &       &       &       &       &       & $(3/2)^-_3$ & 649.90(8) &       &       &       &       &       &       & 793   & 7.514 \\
          &       &       &       &       &       &       & $(5/2)^-_2$ & 767.04 &       &       &       &       &       &       & 538   & 6.188 \\
          &       &       &       &       &       &       & $3/2^-_4$ & 828.29(5) &       &       &       &       &       &       & 1162  & 7.817 \\

          &       &       &       &       &       &       & $7/2^-_1$ & 836.78(9) &       &       &       &       &       &       & 834   & 6.138 \\
          &       &       &       &       &       &       & $5/2^{(-)}_3$ & 1023.7(4) &       &       &       &       &       &       & 845   & 7.221 \\
          &       &       &       &       &       &       & $5/2^-_4,7/2^-$ & 1189.9(21) &       &       &       &       &       &       & 993   & 6.752 \\
          &       &       &       &       &       &       & $(3/2^-_5, 5/2, 7/2^-)$ & 1266.8(6) &       &       &       &       &       &       & 1865  & 8.336 \\
          &       &       &       &       &       &       & $(5/2)^-_5$ & 1323.0(4)  &       &       &       &       &       &       & 1578  & 6.263 \\
          &       &       &       &       &       &       & $(7/2^-)_2$ & 1481.7(6) &       &       &       &       &       &       & 986   & 5.337 \\
          &       &       &       &       &       &       & $(3/2^-_6, 1/2^-)$ & 1512.9(10) &       &       &       &       &       &       & 1921  & 6.905 \\
          &       &       &       &       &       &       & $(3/2^-)_7$ & 1535.9(7) &       &       &       &       &       &       & 1991  & 7.989 \\
          &       &       &       &       &       &       & $(3/2^-)_8$ & 1543.2(5) &       &       &       &       &       &       & 2267  & 6.980 \\
          &       &       &       &       &       &       & $(7/2^-)_3$ & 1681.2(8) &       &       &       &       &       &       & 1549  & 6.134 \\
          &       &       &       &       &       &       & $(5/2^-)_6$ & 1798.9(10) &       &       &       &       &       &       & 1902  & 7.432 \\
          &       &       &       &       &       &       & $(3/2^-)_9$ & 1866.4(10) &       &       &       &       &       &       & 2399  & 7.836 \\
          &       &       &       &       &       &       & $(3/2^-_{10},5/2,7/2^-)$ & 1885.2(7) &       &       &       &       &       &       & 2507  & 7.206 \\
          &       &       &       &       &       &       & $7/2^{(-)}_4$ & 1995.9(8) &       &       &       &       &       &       & 1742  & 4.583 \\
          &       &       &       &       &       &       & $1/2^-, 3/2^-_{11}$ & 2056.0(21) &       &       &       &       &       &       & 2538  &  7.015 \\ 
\hline
\hline
\end{tabular}
\label{table-logft-c}
\end{table*}    

\begin{table*}
\centering
\caption{A comparison of terrestrial half-lives ($T_{1/2(t)}$). $J_p^{\pi}$ and $J_d^{\pi}$ are the spin-parity of the parent and daughter levels, respectively. Experimental \bta branching ($I_{m(exp)}$) from a parent level given in Ref. \cite{NNDC} is normalized to 100\%. Theoretical \bta branching ($I_m$) calculations take into account only \bta decay from the parent levels.}

\vspace{1cm}
\begin{tabular}{cccccccccc}
\hline
\hline
\multicolumn{8}{c}{Transition Details \cite{NNDC}}                 & \multicolumn{2}{c}{Shell-Model Results} \\
\cline{1-7} \cline{9-10}
\multicolumn{2}{c}{Decay} & \multicolumn{3}{c}{$J^\pi_p$ $\rightarrow$ $J^\pi_d$} & Exp. \bta & Exp. & & Theo. \bta  & Theo. \\
         &&&  & & $I_m(\%)$  & $T_{1/2(t)}$ & & $I_m(\%)$ & $T_{1/2(t)}$  \\
    \cline{3-4}
    
\hline
\hline
\multicolumn{2}{c}{$^{59}$Fe $\rightarrow$ $^{59}$Co} & \multicolumn{3}{c} {$3/2^{-}_{1}$ $\rightarrow$ $3/2^{-}_{1}$} & 53.1  & 44.490 d && 48.10 & 41.21 d \\
          &       & \multicolumn{3}{c} {$3/2^{-}_{1}$ $\rightarrow$ $3/2^{-}_{2}$}& 45.3  &         && 51.55 &  \\
          &       &  \multicolumn{3}{c} {$3/2^{-}_{1}$ $\rightarrow$ $1/2^{-}_{1}$} & 1.31  &       && 0.33  &  \\
          &       &  \multicolumn{3}{c} {$3/2^{-}_{1}$ $\rightarrow$ $5/2^{-}_{1}$} & 0.078 &       && 0.02  &  \\
\hline
\multicolumn{2}{c}{$^{60}$Co $\rightarrow$ $^{60}$Ni} & \multicolumn{3}{c} {$5^{+}_1$ $\rightarrow$ $4^{+}_1$ } & 100.0 & 1925.28 d  &&  100.00     &  1198.71 d \\
\cline{3-10} 
          &       & \multicolumn{3}{c} {$^{a,b}$ $2^{+}_1$    $\rightarrow$ $2^{+}_1$}  & 96    & 2.91 d && 95.68      & 5.41 d  \\
          &       & \multicolumn{3}{c} {$^{a,b}$  $2^{+}_1$    $\rightarrow$ $2^{+}_2$} & 4     &       &&  4.32     &  \\
    \hline
    \multicolumn{2}{c}{$^{61}$Co $\rightarrow$ $^{61}$Ni} & \multicolumn{3}{c} {$7/2^{-}_1$ $\rightarrow$  $5/2^{-}_1$ }  & 95.6  & 1.649 hr && 98.86 & 1.87 hr \\
          &       & \multicolumn{3}{c} {$7/2^{-}_1$  $\rightarrow$     $5/2^{-}_2$} & ---   &       && 0.35  &  \\
          &       & \multicolumn{3}{c} {$7/2^{-}_1$  $\rightarrow$     $7/2^{-}_1$} & 4.4   &       && 0.50  &  \\
          &       & \multicolumn{3}{c} {$7/2^{-}_1$  $\rightarrow$     $7/2^{-}_2$} & ---   &       && 0.14  &  \\
          &       & \multicolumn{3}{c} {$7/2^{-}_1$  $\rightarrow$     $5/2^{-}_3$} & ---   &       && 0.15  &  \\
    \hline
    \multicolumn{2}{c} {$^{63}$Ni $\rightarrow$ $^{63}$Cu} & \multicolumn{3}{c} {$1/2^{-}_1$ $\rightarrow$ $3/2^{-}_1$ }     & 100.0   & 101.2 yr &&  100.00   & 44.69 yr \\
    \hline
    \multicolumn{2}{c} {$^{65}$Ni $\rightarrow$ $^{65}$Cu} & \multicolumn{3}{c} {$5/2^{-}_1$  $\rightarrow$ $3/2^{-}_1$ }     & 60.0    & 2.51 hr && 63.08 & 2.67 hr \\
          &       & \multicolumn{3}{c}    {$5/2^{-}_1$ $\rightarrow$ $5/2^{-}_1$   } & 10.18 &       && 5.07  &  \\
          &       & \multicolumn{3}{c}    {$5/2^{-}_1$ $\rightarrow$ $7/2^{-}_1$   } & 28.4  &       && 29.07 &  \\
          &       & \multicolumn{3}{c}    {$5/2^{-}_1$ $\rightarrow$ $5/2^{-}_2$   } & 0.89  &       && 2.58  &  \\
          &       & \multicolumn{3}{c}    {$5/2^{-}_1$ $\rightarrow$ $3/2^{-}_2$   } & 0.555 &       && 0.21  &  \\
          &       & \multicolumn{3}{c}    {$5/2^{-}_1$ $\rightarrow$ $7/2^{-}_2$   } & ---   &       && 0.01  & \\
          &       & \multicolumn{3}{c}    {$5/2^{-}_1$ $\rightarrow$ $5/2^{-}_3$   } & ---   &       && $<$0.01  &  \\         
    \hline
    \multicolumn{2}{c} {$^{66}$Ni $\rightarrow$ $^{66}$Cu} & \multicolumn{3}{c} { $0^{+}_1$ $\rightarrow$ $1^{+}_1$}    & 100   & 54.6 hr && 100    & 57.96 hr \\
    \hline
    \multicolumn{2}{c} {$^{64}$Cu $\rightarrow$ $^{64}$Zn} & \multicolumn{3}{c} {$1^{+}_1$ $\rightarrow$ $0^{+}_1$}     & 100   & 32.9886 hr &&  100  & 37.27 hr \\
    \hline
    \multicolumn{2}{c} {$^{66}$Cu $\rightarrow$ $^{66}$Zn} & \multicolumn{3}{c} {$1^{+}_1$ $\rightarrow$ $0^{+}_1$} & 90.77 & 5.120 m && 77.12 & 7.42 m \\
          &       &  \multicolumn{3}{c}    { $1^{+}_1$ $\rightarrow$  $2^{+}_1$    } & 9.01   &       && 22.73 &  \\
          &       &  \multicolumn{3}{c}    { $1^{+}_1$ $\rightarrow$  $2^{+}_2$    } & 0.220  &       && 0.14  &  \\
          &       &  \multicolumn{3}{c}    { $1^{+}_1$ $\rightarrow$  $0^{+}_2$    } & 0.0037 &       && 0.00  &  \\
    \hline
    \multicolumn{2}{c} {$^{67}$Cu $\rightarrow$ $^{67}$Zn} & \multicolumn{3}{c}  {$3/2^{-}_1$ $\rightarrow$ $5/2^-_1$ }     & $\approx$ 20    & 61.83 hr && 40.64 & 68.61 hr \\
          &       & \multicolumn{3}{c}  { $3/2^{-}_1$ $\rightarrow$ $1/2^{-}_1$  }  & $\approx$ 22    &       && 21.50 &  \\
          &       &  \multicolumn{3}{c} { $3/2^{-}_1$ $\rightarrow$ $3/2^{-}_1$  } & $\approx$ 57    &       && 35.87 &  \\
          &       &   \multicolumn{3}{c} { $3/2^{-}_1$ $\rightarrow$ $3/2^{-}_2$ } & $\approx$ 1.1   &       && 1.99  &  \\
    \hline
    \multicolumn{2}{c} {$^{69}$Zn $\rightarrow$ $^{69}$Ga} & \multicolumn{3}{c} { $1/2^{-}_1$ $\rightarrow$ $3/2^{-}_1$  }     & 99.9986 & 56.4 m && 99.99 & 53.77 m \\
          &       & \multicolumn{3}{c} {$1/2^{-}_1$ $\rightarrow$ $1/2^{-}_1$  } & 0.0012 &       && 0.01  &  \\
          &       & \multicolumn{3}{c} {$1/2^{-}_1$ $\rightarrow$  $3/2^{-}_2$ } & 0.00025 &       && $\approx$ 0.00  &  \\
    \hline
    \multicolumn{2}{c} {$^{72}$Zn $\rightarrow$ $^{72}$Ga} & \multicolumn{3}{c} {$0^{+}_1$ $\rightarrow$ $0^{+}_{1}$} & $<$0.01 & 46.5 hr && ---  & 54.56 hr \\
          &       &  \multicolumn{3}{c} {$0^{+}_1$ $\rightarrow$ $1^{+}_{1}$} & 0.21  &       && --- &  \\
          &       &  \multicolumn{3}{c} {$0^{+}_1$ $\rightarrow$ $1^{+}_{2}$} & 85.1  &       && 64.42 &  \\
          &       &  \multicolumn{3}{c} {$0^{+}_1$ $\rightarrow$ $1^{+}_{3}$} & 14.7  &       && 35.58 &  \\
    \hline
    \multicolumn{2}{c} {$^{70}$Ga $\rightarrow$ $^{70}$Ge} & \multicolumn{3}{c} {$1^{+}_1$ $\rightarrow$ $0^{+}_{1}$ }    & 98.91 & 21.23 m && 97.01 & 21.09 m \\
          &       & \multicolumn{3}{c}  {$1^{+}_1$ $\rightarrow$ $2^{+}_{1}$ } & 0.36  &       && 2.63 &  \\
          &       &  \multicolumn{3}{c} {$1^{+}_1$ $\rightarrow$ $0^{+}_{2}$} & 0.32  &       && 0.36  &  \\
\hline
    \multicolumn{2}{c} {$^{75}$Ge $\rightarrow$ $^{75}$As} & \multicolumn{3}{c} {$1/2^{-}_1$ $\rightarrow$ $3/2^{-}_{1}$ }     & 87.1  & 82.78 m && 84.52 & 78.94 m \\
          &       & \multicolumn{3}{c} {$1/2^{-}_1$ $\rightarrow$ $1/2^{-}_{1}$ } & 0.86  &       && 0.44  &  \\
          &       & \multicolumn{3}{c} {$1/2^{-}_1$ $\rightarrow$ $3/2^{-}_{2}$ } & 11.5  &       && 11.04  &  \\
          &       & \multicolumn{3}{c} {$1/2^{-}_1$ $\rightarrow$ $1/2^{-}_{2}$ } & 0.225 &       && 0.16  &  \\
          &       & \multicolumn{3}{c} {$1/2^{-}_1$ $\rightarrow$ $1/2^{-}_{3}$ } & ---   &       && 3.36  &  \\
          &       & \multicolumn{3}{c} {$1/2^{-}_1$ $\rightarrow$ $3/2^{-}_{3}$ } & 0.32  &       && 0.29  &  \\
          &       & \multicolumn{3}{c} {$1/2^{-}_1$ $\rightarrow$ $3/2^{-}_{4}$ } & ---   &       && 0.01  &  \\
          &       & \multicolumn{3}{c} {$1/2^{-}_1$ $\rightarrow$ $3/2^{-}_{5}$ } & ---   &       && $<0.01$  &  \\
          &       & \multicolumn{3}{c} {$1/2^{-}_1$ $\rightarrow$ $3/2^{-}_{6}$ } & ---   &       && $<0.01$  &  \\
          &       & \multicolumn{3}{c} {$1/2^{-}_1$ $\rightarrow$ $1/2^{-}_{4}$ } &  ---  &       && $<0.01$  &  \\
          &       & \multicolumn{3}{c} {$1/2^{-}_1$ $\rightarrow$ $1/2^{-}_{5}$ } &  ---  &       && $<0.01$  &  \\
    \cline{3-10}         
     &       & \multicolumn{3}{c} {$^{a,b}$ $7/2^{+}_{1}$ $\rightarrow$ $5/2^{+}_1$ } & 100  & 44.17 hr &&  100.00  & 326.17 hr \\        
   \hline
   \hline
    \end{tabular}%
  \label{table-halflife}
\end{table*}



\begin{table*}
\centering
~~~~~~~~~TABLE (II) continued

\vspace{1cm}
\begin{tabular}{cccccccccc}
\hline
\hline
\multicolumn{8}{c}{Transition Details \cite{NNDC}}                 & \multicolumn{2}{c}{Shell-Model Results} \\
\cline{1-7} \cline{9-10}
\multicolumn{2}{c}{Decay} & \multicolumn{3}{c}{$J^\pi_p$ $\rightarrow$ $J^\pi_d$} & Exp. \bta & Exp. & & Theo. \bta  & Theo. \\
         &&&  & & $I_m(\%)$  & $T_{1/2(t)}$ & & $I_m(\%)$ & $T_{1/2(t)}$  \\
    \cline{3-4}
    
\hline
\hline

    \multicolumn{2}{c} {$^{78}$Ge $\rightarrow$ $^{78}$As} & \multicolumn{3}{c} {$0^{+}_1$ $\rightarrow$ $1^{+}_{1}$ } & 96    & 88 m && 40.64 & 86.59 m \\
          &       & \multicolumn{3}{c} {$0^{+}_1$ $\rightarrow$ $1^{+}_{2}$ } & 4     &       && 14.81  &  \\
          &       & \multicolumn{3}{c} {$0^{+}_1$ $\rightarrow$ $1^{+}_{3}$ } & ---     &       && 44.54  &  \\
    \hline
    \multicolumn{2}{c} {$^{81}$Se $\rightarrow$ $^{81}$Br} & \multicolumn{3}{c} {$1/2^{-}_1$ $\rightarrow$ $3/2^{-}_{1}$ }     & 98.73 & 18.45 m && 98.78 & 19.39 m \\
          &       & \multicolumn{3}{c} {$1/2^{-}_1$ $\rightarrow$ $1/2^{-}_{1}$ } & 0.034 &       && 0.02  &  \\
          &       & \multicolumn{3}{c} {$1/2^{-}_1$ $\rightarrow$ $3/2^{-}_{2}$ } & 0.79  &       && 0.10  &  \\
          &       & \multicolumn{3}{c} {$1/2^{-}_1$ $\rightarrow$ $3/2^{-}_{3}$ } & 0.0196 &       && 0.18  &  \\
          &       & \multicolumn{3}{c} {$1/2^{-}_1$ $\rightarrow$ $3/2^{-}_{4}$ } & 0.40   &       && 0.89  &  \\
          &       & \multicolumn{3}{c} {$1/2^{-}_1$ $\rightarrow$ $1/2^{-}_{2}$ } & --- &       && 0.03  &  \\
          &       & \multicolumn{3}{c} {$1/2^{-}_1$ $\rightarrow$ $3/2^{-}_{5}$ } & --- &       && $\approx$0.01  &  \\
          &       & \multicolumn{3}{c} {$1/2^{-}_1$ $\rightarrow$ $3/2^{-}_{6}$ } & --- &       && $<$0.01  &  \\
          &       & \multicolumn{3}{c} {$1/2^{-}_1$ $\rightarrow$ $3/2^{-}_{7}$ } & --- &       && $<$0.01  &  \\
\cline{3-10}
          &       & \multicolumn{3}{c} {$^{a,b}$ $7/2^{+}_1$ $\rightarrow$ $9/2^{+}_{1}$ } & 100 &  78.00 d   && 100  &  71.43 d\\
   \hline
   \hline
  \multicolumn{10}{c}{}\\
 \multicolumn{10}{l} {$^a$ isomer}.\\
\multicolumn{10}{c}{$^b$ Total \bta branching normalized to 100\%. For example, the $7/2^+_1$ isomer in}\\
\multicolumn{10}{l}{\hspace{0.3cm} $^{75}$Ge has only 0.03\% \bta decay. We have taken this  0.03\% as 100\%.}

    \end{tabular}%

\end{table*}%

%

\begin{table*}
\centering
\renewcommand\thetable{III.A}
\caption{Bound state decay rate ($\lambda_b$ in s$^{-1}$) and continuum state decay rate ($\lambda_c$ in s$^{-1}$) variation with temperature at a free electron density $n_e = 1{\times}10^{27}$ cm$^{-3}$ for each transition. There are blank spaces in column 4 and column 5 because those excited states are not populated at the corresponding temperature.}
\vspace{1cm}
\begin{tabular}{cLLcccccc}
\hline\hline
\multicolumn{3}{c}{Transition Details} &       & \multicolumn{5}{c}{$n_e = 1{\times}10^{27}$ cm$^{-3}$} \\
\cline{1-3} \cline{5-9}
          &       &       &       & \multicolumn{2}{c}{T = 1${\times}10^{8}$ K} &       & \multicolumn{2}{c}{T = 5${\times}10^{8}$ K} \\
\cline{5-6} \cline{8-9}  
 Decay & J^{\pi}_p & J^{\pi}_d &   & $\lambda_{c}$ & $\lambda_{b}$ &   & $\lambda_{c}$ & $\lambda_{b}$ \\
\hline
$^{59}$Fe  $\rightarrow$ $^{59}$Co & 3/2^-_1 & 3/2^-_1 &       & 8.93${\times}10^{-8}$ & 7.62${\times}10^{-9}$ &       & 9.02${\times}10^{-8}$ & 7.65${\times}10^{-9}$ \\
          &       & 3/2^-_2 &       & 9.27${\times}10^{-8}$ & 1.74${\times}10^{-8}$ &       & 9.45${\times}10^{-8}$ & 1.75${\times}10^{-8}$ \\
          &       & 1/2^-_1 &       & 5.46${\times}10^{-10}$ & 2.85${\times}10^{-10}$ &       & 5.70${\times}10^{-10}$ & 2.88${\times}10^{-10}$ \\
          &       & 5/2^-_1 &       & 2.45${\times}10^{-11}$ & 2.39${\times}10^{-11}$ &       & 2.63${\times}10^{-11}$ & 2.44${\times}10^{-11}$ \\
\cline{2-9}          
          & 1/2^-_1 & 3/2^-_1 &       &       &       &       & 2.60${\times}10^{-7}$ & 1.01${\times}10^{-8}$ \\
          &         & 3/2^-_2 &       &       &       &       & 2.82${\times}10^{-7}$ & 1.79${\times}10^{-8}$ \\
          &         & 1/2^-_1 &       &       &       &       & 5.47${\times}10^{-7}$ & 5.47${\times}10^{-8}$ \\
\cline{2-9}       
          & 5/2^-_1 & 7/2^-_1 &       &       &       &       & 6.47${\times}10^{-4}$ & 3.62${\times}10^{-6}$ \\
          &         & 3/2^-_1 &       &       &       &       & 1.03${\times}10^{-6}$ & 2.72${\times}10^{-8}$ \\
          &         & 3/2^-_2 &       &       &       &       & 6.32${\times}10^{-8}$ & 2.49${\times}10^{-9}$ \\
          &         & 5/2^-_1 &       &       &       &       & 4.98${\times}10^{-9}$ & 3.19${\times}10^{-10}$ \\
          &         & 7/2^-_2 &       &       &       &       & 5.75${\times}10^{-8}$ & 9.65${\times}10^{-9}$ \\
\hline
$^{60}$Co  $\rightarrow$ $^{60}$Ni & 5^+_1 & 4^+_1 &       & 6.24${\times}10^{-9}$ & 1.03${\times}10^{-9}$ &       & 6.34${\times}10^{-9}$ & 1.03${\times}10^{-9}$ \\
\cline{2-9}        
          & 2^+_1 & 2^+_1 &       & 2.02${\times}10^{-6}$ & 2.21${\times}10^{-8}$ &       & 2.02${\times}10^{-6}$ & 2.21${\times}10^{-8}$ \\
          &       & 2^+_2 &       & 8.97${\times}10^{-8}$ & 4.08${\times}10^{-9}$ &       & 9.03${\times}10^{-8}$ & 4.09${\times}10^{-9}$ \\
\cline{2-9}   
          & 4^+_1 & 4^+_1 &       &       &       &       & 4.21${\times}10^{-7}$ & 2.64${\times}10^{-8}$ \\
          &       & 3^+_1 &       &       &       &       & 4.05${\times}10^{-8}$ & 3.63${\times}10^{-9}$ \\
\cline{2-9} 
          & 3^+_1 & 2^+_1 &       &       &       &       & 1.02${\times}10^{-5}$ & 8.35${\times}10^{-8}$ \\
          &       & 2^+_2 &       &       &       &       & 3.85${\times}10^{-7}$ & 1.08${\times}10^{-8}$ \\
          &       & 4^+_1 &       &       &       &       & 1.59${\times}10^{-7}$ & 9.67${\times}10^{-9}$ \\
          &       & 3^+_1 &       &       &       &       & 3.87${\times}10^{-9}$ & 3.35${\times}10^{-10}$ \\
\cline{2-9}  
          & 5^+_2 & 4^+_1 &       &       &       &       & 2.80${\times}10^{-7}$ & 1.19${\times}10^{-8}$ \\
          &       & 4^+_2 &       &       &       &       & 4.28${\times}10^{-10}$ & 2.18${\times}10^{-10}$ \\
\hline
$^{61}$Co  $\rightarrow$ $^{61}$Ni & 7/2^-_1  & 5/2^-_1 &       & 9.98${\times}10^{-5}$ & 1.66${\times}10^{-6}$ &       & 1.00${\times}10^{-4}$ & 1.66${\times}10^{-6}$ \\
          &       & 5/2^-_2 &       & 3.40${\times}10^{-7}$ & 3.76${\times}10^{-8}$ &       & 3.44${\times}10^{-7}$ & 3.78${\times}10^{-8}$ \\
          &       & 7/2^-_1 &       & 4.92${\times}10^{-7}$ & 5.62${\times}10^{-8}$ &       & 4.98${\times}10^{-7}$ & 5.64${\times}10^{-8}$ \\
          &       & 7/2^-_2 &       & 1.31${\times}10^{-7}$ & 2.25${\times}10^{-8}$ &       & 1.34${\times}10^{-7}$ & 2.26${\times}10^{-8}$ \\
          &       & 5/2^-_3 &       & 1.39${\times}10^{-7}$ & 4.66${\times}10^{-8}$ &       & 1.43${\times}10^{-7}$ & 4.70${\times}10^{-8}$ \\
\hline
$^{63}$Ni  $\rightarrow$ $^{63}$Cu & 1/2^-_1  & 3/2^-_1  &       & 3.50${\times}10^{-10}$ & 5.72${\times}10^{-10}$ &       & 3.84${\times}10^{-10}$ & 5.86${\times}10^{-10}$ \\
\cline{2-9}         
         & 5/2^-_1  & 3/2^-_1  &       & 2.13${\times}10^{-8}$ & 1.04${\times}10^{-8}$ &       & 2.21${\times}10^{-8}$ & 1.06${\times}10^{-8}$ \\
\cline{2-9}     
         & 3/2^-_1  & 3/2^-_1  &       &       &       &       & 1.97${\times}10^{-7}$ & 5.72${\times}10^{-8}$ \\
\hline  
$^{65}$Ni  $\rightarrow$ $^{65}$Cu 
          & 5/2^-_1 & 3/2^-_1 &       & 4.50${\times}10^{-5}$ & 2.71${\times}10^{-7}$ &       & 4.51${\times}10^{-5}$ & 2.72${\times}10^{-7}$ \\
          &       & 5/2^-_1 &       & 3.58${\times}10^{-6}$ & 9.57${\times}10^{-8}$ &       & 3.59${\times}10^{-6}$ & 9.58${\times}10^{-8}$ \\
          &       & 7/2^-_1 &       & 2.02${\times}10^{-5}$ & 1.18${\times}10^{-6}$ &       & 2.04${\times}10^{-5}$ & 1.18${\times}10^{-6}$ \\
          &       & 5/2^-_2 &       & 1.78${\times}10^{-6}$ & 1.53${\times}10^{-7}$ &       & 1.80${\times}10^{-6}$ & 1.54${\times}10^{-7}$ \\
          &       & 3/2^-_2 &       & 1.42${\times}10^{-7}$ & 1.72${\times}10^{-8}$ &       & 1.44${\times}10^{-7}$ & 1.73${\times}10^{-8}$ \\
          &       & 7/2^-_2  &       & 2.38${\times}10^{-9}$ & 7.90${\times}10^{-9}$ &       & 2.75${\times}10^{-9}$ & 8.19${\times}10^{-9}$ \\
          &       & 5/2^-_3  &       & 1.21${\times}10^{-12}$ & 8.00${\times}10^{-12}$ &       & 1.49${\times}10^{-12}$ & 8.41${\times}10^{-12}$ \\
\cline{2-9}        
          & 1/2^-_1  & 3/2^-_1 &     & 5.39${\times}10^{-3}$ & 3.05${\times}10^{-5}$ &       & 5.40${\times}10^{-3}$ & 3.05${\times}10^{-5}$ \\
          &       & 1/2^-_1 &       & 2.38${\times}10^{-5}$ & 3.33${\times}10^{-7}$ &       & 2.38${\times}10^{-5}$ & 3.33${\times}10^{-7}$ \\
          &       & 3/2^-_2 &       & 3.10${\times}10^{-7}$ & 3.01${\times}10^{-8}$ &       & 3.13${\times}10^{-7}$ & 3.02${\times}10^{-8}$ \\
\hline
$^{66}$Ni  $\rightarrow$ $^{66}$Cu & 0^+_1 & 1^+_1 &       & 3.03${\times}10^{-6}$ & 7.50${\times}10^{-7}$ &       & 3.09${\times}10^{-6}$ & 7.55${\times}10^{-7}$ \\
\hline
$^{64}$Cu  $\rightarrow$ $^{64}$Zn & 1^+_1 & 0^+_1 &       & 4.95${\times}10^{-6}$ & 3.81${\times}10^{-7}$ &       & 5.00${\times}10^{-6}$ & 3.83${\times}10^{-7}$ \\
\hline
\hline
\end{tabular}
\label{b_and_c_with_t}
\end{table*}

\begin{table*}
\centering
~~~~~~TABLE (III.A) continued

\vspace{0.5cm}
\begin{tabular}{cLLcccccc}
\hline
\hline
\multicolumn{3}{c}{Transition Details} &       & \multicolumn{5}{c}{$n_e = 1{\times}10^{27}$ cm$^{-3}$} \\
\cline{1-3} \cline{5-9}
          &       &       &       & \multicolumn{2}{c}{T = 1${\times}10^{8}$ K} &       & \multicolumn{2}{c}{T = 5${\times}10^{8}$ K} \\
\cline{5-6} \cline{8-9}  

 Decay & J^{\pi}_p & J^{\pi}_d &   & $\lambda_{c}$ & $\lambda_{b}$ &   & $\lambda_{c}$ & $\lambda_{b}$ \\
\hline
$^{66}$Cu $\rightarrow$ $^{66}$Zn & 1^+_1 & 0^+_1 &       & 1.19${\times}10^{-3}$ & 4.85${\times}10^{-6}$ &       & 1.19${\times}10^{-3}$ & 4.85${\times}10^{-6}$ \\
          &       & 2^+_1 &       & 3.48${\times}10^{-4}$ & 4.21${\times}10^{-6}$ &       & 3.49${\times}10^{-4}$ & 4.21${\times}10^{-6}$ \\
          &       & 2^+_2 &       & 2.17${\times}10^{-6}$ & 1.04${\times}10^{-7}$ &       & 2.18${\times}10^{-6}$ & 1.05${\times}10^{-7}$ \\
          &       & 0^+_2 &       & 2.19${\times}10^{-8}$ & 5.35${\times}10^{-9}$ &       & 2.23${\times}10^{-8}$ & 5.38${\times}10^{-9}$ \\

\hline
$^{67}$Cu  $\rightarrow$ $^{67}$Zn & 3/2^-_1  & 5/2^-_1 &       & 1.09${\times}10^{-6}$ & 8.83${\times}10^{-8}$ &       & 1.10${\times}10^{-6}$ & 8.86${\times}10^{-8}$ \\
          &       & 1/2^-_1 &       & 5.72${\times}10^{-7}$ & 6.16${\times}10^{-8}$ &       & 5.79${\times}10^{-7}$ & 6.19${\times}10^{-8}$ \\
          &       & 3/2^-_1 &       & 9.45${\times}10^{-7}$ & 1.41${\times}10^{-7}$ &       & 9.58${\times}10^{-7}$ & 1.42${\times}10^{-7}$ \\
          &       & 3/2^-_2 &       & 4.85${\times}10^{-8}$ & 2.28${\times}10^{-8}$ &       & 5.02${\times}10^{-8}$ & 2.31${\times}10^{-8}$ \\
\hline
$^{69}$Zn  $\rightarrow$ $^{69}$Ga & 1/2^-_1 & 3/2^-_1 &       & 2.09${\times}10^{-4}$ & 8.07${\times}10^{-6}$ &       & 2.10${\times}10^{-4}$ & 8.09${\times}10^{-6}$ \\
          &       & 1/2^-_1 &       & 2.88${\times}10^{-8}$ & 2.31${\times}10^{-9}$ &       & 2.91${\times}10^{-8}$ & 2.32${\times}10^{-9}$ \\
          &       & 3/2^-_2 &       & 4.07${\times}10^{-10}$ & 2.11${\times}10^{-9}$ &       & 4.83${\times}10^{-10}$ & 2.20${\times}10^{-9}$ \\
\hline
$^{72}$Zn  $\rightarrow$ $^{72}$Ga & 0^+_1 & 0^+_1 &       & 7.50${\times}10^{-9}$ & 1.34${\times}10^{-9}$ &       & 7.63${\times}10^{-9}$ & 1.52${\times}10^{-9}$ \\
          &       & 1^+_1   &       & 2.07${\times}10^{-6}$ & 4.35${\times}10^{-7}$ &       & 2.11${\times}10^{-6}$ & 4.37${\times}10^{-7}$ \\
          &       & 1^+_2 &       & 2.08${\times}10^{-6}$ & 5.12${\times}10^{-7}$ &       & 2.12${\times}10^{-6}$ & 5.15${\times}10^{-7}$ \\
          &       & 1^+_3 &       & 1.13${\times}10^{-6}$ & 3.60${\times}10^{-7}$ &       & 1.16${\times}10^{-6}$ & 3.62${\times}10^{-7}$ \\
\hline
$^{70}$Ga  $\rightarrow$ $^{70}$Ge & 1^+_1 & 0^+_1 &       & 5.22${\times}10^{-4}$ & 6.93${\times}10^{-6}$ &       & 5.24${\times}10^{-4}$ & 6.94${\times}10^{-6}$ \\
          &       & 2^+_1 &       & 1.38${\times}10^{-5}$ & 1.13${\times}10^{-6}$ &       & 1.39${\times}10^{-5}$ & 1.13${\times}10^{-6}$ \\
          &       & 0^+_2 &       & 1.87${\times}10^{-6}$ & 2.60${\times}10^{-7}$ &       & 1.89${\times}10^{-6}$ & 2.61${\times}10^{-7}$ \\
\hline
$^{75}$Ge  $\rightarrow$ $^{75}$As & 1/2^-_1 & 3/2^-_1 &       & 1.21${\times}10^{-4}$ & 3.38${\times}10^{-6}$ &       & 1.22${\times}10^{-4}$ & 3.39${\times}10^{-6}$ \\
          &       & 1/2^-_1 &       & 6.25${\times}10^{-7}$ & 2.45${\times}10^{-8}$ &       & 6.28${\times}10^{-7}$ & 2.46${\times}10^{-8}$ \\
          &       & 3/2^-_2 &       & 1.57${\times}10^{-5}$ & 7.00${\times}10^{-7}$ &       & 1.58${\times}10^{-5}$ & 7.02${\times}10^{-7}$ \\
          &       & 1/2^-_2 &       & 2.26${\times}10^{-7}$ & 1.55${\times}10^{-8}$ &       & 2.27${\times}10^{-7}$ & 1.56${\times}10^{-8}$ \\
          &       & 1/2^-_3 &       & 4.74${\times}10^{-6}$ & 4.38${\times}10^{-7}$ &       & 4.78${\times}10^{-6}$ & 4.39${\times}10^{-7}$ \\
          &       & 3/2^-_3 &       & 6.63${\times}10^{-7}$ & 6.71${\times}10^{-8}$ &       & 6.69${\times}10^{-7}$ & 6.73${\times}10^{-8}$ \\
          &       & 3/2^-_4 &       & 6.99${\times}10^{-9}$ & 1.71${\times}10^{-9}$ &       & 7.12${\times}10^{-9}$ & 1.72${\times}10^{-9}$ \\
          &       & 3/2^-_5 &       & 8.09${\times}10^{-12}$ & 8.30${\times}10^{-12}$ &       & 8.56${\times}10^{-12}$ & 8.44${\times}10^{-12}$ \\
          &       & 3/2^-_6 &       & 3.49${\times}10^{-10}$ & 4.18${\times}10^{-10}$ &       & 3.72${\times}10^{-10}$ & 4.25${\times}10^{-10}$ \\
          &       & 1/2^-_4 &       & 2.96${\times}10^{-11}$ & 1.14${\times}10^{-10}$ &       & 3.39${\times}10^{-11}$ & 1.18${\times}10^{-10}$ \\
          &       & 1/2^-_5 &       &  ---                   & 6.13${\times}10^{-11}$ &       &  ---                   & 7.09${\times}10^{-11}$ \\
\cline{2-9}   
          & 7/2^+_1 & 9/2^+_1 &       &       &       &       & 3.99${\times}10^{-7}$ & 1.47${\times}10^{-8}$ \\     
          &         & 5/2^+_1 &       &       &       &       & 5.75${\times}10^{-7}$ & 2.53${\times}10^{-8}$ \\
          &         & 5/2^+_2 &       &       &       &       & 6.59${\times}10^{-8}$ & 2.36${\times}10^{-8}$ \\
          &         & 5/2^+_3 &       &       &       &       & 6.66${\times}10^{-9}$ & 2.68${\times}10^{-9}$ \\
          &         & 9/2^+_2 &       &       &       &       & 1.42${\times}10^{-10}$ & 4.13${\times}10^{-10}$ \\
          &         & 5/2^+_4 &       &       &       &       & 1.84${\times}10^{-12}$ & 1.33${\times}10^{-10}$ \\
\hline
\hline
\end{tabular}
\end{table*}

\begin{table*}
\centering
~~~~~~TABLE (III.A) continued

\vspace{0.5cm}
\begin{tabular}{cLLcccccc}
\hline
\hline
\multicolumn{3}{c}{Transition Details} &       & \multicolumn{5}{c}{$n_e = 1{\times}10^{27}$ cm$^{-3}$} \\
\cline{1-3} \cline{5-9}
          &       &       &       & \multicolumn{2}{c}{T = 1${\times}10^{8}$ K} &       & \multicolumn{2}{c}{T = 5${\times}10^{8}$ K} \\
\cline{5-6} \cline{8-9}   

 Decay & J^{\pi}_p & J^{\pi}_d &   & $\lambda_{c}$ & $\lambda_{b}$ &   & $\lambda_{c}$ & $\lambda_{b}$ \\
\hline

          & 3/2^-_1 & 3/2^-_1 &       &       &       &       & 7.46${\times}10^{-5}$ & 1.42${\times}10^{-6}$ \\
          &       & 1/2^-_1 &       &       &       &       & 3.52${\times}10^{-5}$ & 8.97${\times}10^{-7}$ \\
          &       & 3/2^-_2 &       &       &       &       & 9.71${\times}10^{-6}$ & 2.75${\times}10^{-7}$ \\
          &       & 5/2^-_1 &       &       &       &       & 1.85${\times}10^{-7}$ & 5.37${\times}10^{-9}$ \\
          &       & 1/2^-_2 &       &       &       &       & 1.11${\times}10^{-7}$ & 4.49${\times}10^{-9}$ \\
          &       & 5/2^-_2 &       &       &       &       & 6.76${\times}10^{-7}$ & 3.34${\times}10^{-8}$ \\
          &       & 1/2^-_3 &       &       &       &       & 2.69${\times}10^{-7}$ & 1.36${\times}10^{-8}$ \\
          &       & 3/2^-_3 &       &       &       &       & 1.18${\times}10^{-6}$ & 6.42${\times}10^{-8}$ \\
          &       & 3/2^-_4 &       &       &       &       & 3.79${\times}10^{-7}$ & 3.75${\times}10^{-8}$ \\
          &       & 3/2^-_5 &       &       &       &       & 2.21${\times}10^{-7}$ & 4.23${\times}10^{-8}$ \\
          &       & 3/2^-_6 &       &       &       &       & 1.57${\times}10^{-8}$ & 3.15${\times}10^{-9}$ \\
          &       & 1/2^-_4 &       &       &       &       & 2.46${\times}10^{-8}$ & 6.18${\times}10^{-9}$ \\
          &       & 1/2^-_5 &       &       &       &       & 1.43${\times}10^{-8}$ & 4.53${\times}10^{-9}$ \\
          &       & 3/2^-_7 &       &       &       &       & 1.55${\times}10^{-8}$ & 5.88${\times}10^{-9}$ \\
          &       & 3/2^-_8 &       &       &       &       & 2.83${\times}10^{-10}$ & 4.58${\times}10^{-10}$ \\
          &       & 3/2^-_9 &       &       &       &       & 5.29${\times}10^{-10}$ & 1.38${\times}10^{-9}$ \\
          &       & 5/2^-_3 &       &       &       &       & 2.73${\times}10^{-13}$ & 1.08${\times}10^{-10}$ \\

\hline
$^{78}$Ge $\rightarrow$ $^{78}$As  & 0^+_1 & 1^+_1 &       & 5.21${\times}10^{-5}$ & 3.86${\times}10^{-6}$ &       & 5.25${\times}10^{-5}$ & 3.87${\times}10^{-6}$ \\
          &       & 1^+_2 &       & 1.89${\times}10^{-5}$ & 1.46${\times}10^{-6}$ &       & 1.91${\times}10^{-5}$ & 1.47${\times}10^{-6}$ \\
          &       & 1^+_3 &       & 5.57${\times}10^{-5}$ & 8.82${\times}10^{-6}$ &       & 5.65${\times}10^{-5}$ & 8.86${\times}10^{-6}$ \\
\hline
$^{81}$Se $\rightarrow$ $^{81}$Br & 1/2^-_1 & 3/2^-_1 &       & 5.77${\times}10^{-4}$ & 1.03${\times}10^{-5}$ &       & 5.79${\times}10^{-4}$ & 1.03${\times}10^{-5}$ \\
          &       & 1/2^-_1 &       & 1.40${\times}10^{-7}$ & 5.56${\times}10^{-9}$ &       & 1.41${\times}10^{-7}$ & 5.57${\times}10^{-9}$ \\
          &       & 3/2^-_2 &       & 5.77${\times}10^{-7}$ & 2.41${\times}10^{-8}$ &       & 5.80${\times}10^{-7}$ & 2.41${\times}10^{-8}$ \\
          &       & 3/2^-_3 &       & 1.04${\times}10^{-6}$ & 5.05${\times}10^{-8}$ &       & 1.04${\times}10^{-6}$ & 5.06${\times}10^{-8}$ \\
          &       & 3/2^-_4 &       & 5.07${\times}10^{-6}$ & 3.57${\times}10^{-7}$ &       & 5.11${\times}10^{-6}$ & 3.58${\times}10^{-7}$ \\
          &       & 1/2^-_2 &       & 1.74${\times}10^{-7}$ & 2.55${\times}10^{-8}$ &       & 1.76${\times}10^{-7}$ & 2.56${\times}10^{-8}$ \\
          &       & 3/2^-_5 &       & 1.80${\times}10^{-8}$ & 4.85${\times}10^{-9}$ &       & 1.83${\times}10^{-8}$ & 4.88${\times}10^{-9}$ \\
          &       & 3/2^-_6 &       & 4.41${\times}10^{-11}$ & 2.12${\times}10^{-10}$ &       & 5.09${\times}10^{-11}$ & 2.20${\times}10^{-10}$ \\
          &       & 3/2^-_7 &       & 7.02${\times}10^{-11}$ & 4.64${\times}10^{-10}$ &       & 8.31${\times}10^{-11}$ & 4.84${\times}10^{-10}$ \\
\hline
\hline
\end{tabular}
\end{table*}

\begin{table*}
\centering
\renewcommand\thetable{III.B}
\caption{Bound state decay rate ($\lambda_b$ in s$^{-1}$) and continuum state decay rate ($\lambda_c$ in s$^{-1}$) variation with free electron density at a temperature $T = 3{\times}10^{8}$ K for each transition. This table does not include the levels that are not populated at this temperature.}
\vspace{0.5cm}
\begin{tabular}{cLLcccccc}
\hline\hline
\multicolumn{3}{c}{Transition Details} &       & \multicolumn{5}{c}{$T = 3{\times}10^{8}$ K} \\
\cline{1-3} \cline{5-9}
          &       &     &       & \multicolumn{2}{c}{$n_e = 1{\times}10^{26}$ cm$^{-3}$} &       & \multicolumn{2}{c}{$n_e = 1{\times}10^{27}$ cm$^{-3}$}\\
\cline{5-6} \cline{8-9}   

 Decay & J^{\pi}_p & J^{\pi}_d &   & $\lambda_{c}$ & $\lambda_{b}$ &   & $\lambda_{c}$ & $\lambda_{b}$ \\
\hline
$^{59}$Fe  $\rightarrow$ $^{59}$Co & 3/2^-_1 & {3/2^-_1} &       & 9.09${\times}10^{-8}$ & 7.92${\times}10^{-9}$ &       & 8.99${\times}10^{-8}$ & 7.64${\times}10^{-9}$ \\
          &       & {3/2^-_2} &       & 9.57${\times}10^{-8}$ & 1.81${\times}10^{-8}$ &       & 9.40${\times}10^{-8}$ & 1.75${\times}10^{-8}$ \\
          &       & {1/2^-_1} &       & 5.85${\times}10^{-10}$ & 3.00${\times}10^{-10}$ &       & 5.64${\times}10^{-10}$ & 2.87${\times}10^{-10}$ \\
          &       & {5/2^-_1} &       & 2.73${\times}10^{-11}$ & 2.55${\times}10^{-11}$ &       & 2.58${\times}10^{-11}$ & 2.43${\times}10^{-11}$ \\
\hline
$^{60}$Co  $\rightarrow$ $^{60}$Ni & 5^+_1 & {4^+_1} &       & 6.41${\times}10^{-9}$ & 1.07${\times}10^{-9}$ &       & 6.31${\times}10^{-9}$ & 1.03${\times}10^{-9}$ \\
\cline{2-9}          
          & 2^+_1 & {2^+_1} &       & 2.03${\times}10^{-6}$ & 2.29${\times}10^{-8}$ &       & 2.02${\times}10^{-6}$ &  2.21${\times}10^{-8}$ \\
          &       & {2^+_2} &       & 9.07${\times}10^{-8}$ & 4.24${\times}10^{-9}$ &       & 9.01${\times}10^{-8}$ & 4.09${\times}10^{-9}$ \\
\cline{2-9}         
          & 4^+_1 & {4^+_1} &       & 4.23${\times}10^{-7}$ & 2.73${\times}10^{-8}$ &       & 4.20${\times}10^{-7}$ & 2.64${\times}10^{-8}$ \\
          &       & {3^+_1} &       & 4.08${\times}10^{-8}$ & 3.76${\times}10^{-9}$ &       & 4.03${\times}10^{-8}$ & 3.63${\times}10^{-9}$ \\
\hline
$^{61}$Co  $\rightarrow$ $^{61}$Ni & 7/2^-_1  & {5/2^-_1} &       & 1.00${\times}10^{-4}$ & 1.72${\times}10^{-6}$ &       & 1.00${\times}10^{-4}$ & 1.66${\times}10^{-6}$ \\
          &       & {5/2^-_2} &       & 3.47${\times}10^{-7}$ & 3.91${\times}10^{-8}$ &       & 3.43${\times}10^{-7}$ & 3.77${\times}10^{-8}$ \\
          &       & {7/2^-_1} &       & 5.02${\times}10^{-7}$ & 5.85${\times}10^{-8}$ &       & 4.96${\times}10^{-7}$ & 5.64${\times}10^{-8}$ \\
          &       & {7/2^-_2} &       & 1.35${\times}10^{-7}$ & 2.35${\times}10^{-8}$ &       & 1.33${\times}10^{-7}$ & 2.26${\times}10^{-8}$ \\
          &       & {5/2^-_3} &       & 1.46${\times}10^{-7}$ & 4.88${\times}10^{-8}$ &       & 1.43${\times}10^{-7}$ & 4.69${\times}10^{-8}$ \\
    \hline
$^{63}$Ni  $\rightarrow$ $^{63}$Cu & 1/2^-_1  & {3/2^-_1} &       & 4.03${\times}10^{-10}$ & 6.14${\times}10^{-10}$ &       & 3.76${\times}10^{-10}$ & 5.82${\times}10^{-10}$ \\
\cline{2-9}          & 5/2^-_1  & {3/2^-_1} &       & 2.26${\times}10^{-8}$ & 1.10${\times}10^{-8}$ &       & 2.19${\times}10^{-8}$ & 1.05${\times}10^{-8}$ \\
\cline{2-9}          & 3/2^-_1  & {3/2^-_1} &       & 2.00${\times}10^{-7}$ & 5.94${\times}10^{-8}$ &       & 1.96${\times}10^{-7}$ & 5.70${\times}10^{-8}$ \\
\hline
$^{65}$Ni  $\rightarrow$ $^{65}$Cu & 5/2^-_1  & 3/2^-_1 &       & 4.52${\times}10^{-5}$ & 2.81${\times}10^{-7}$ &       & 4.51${\times}10^{-5}$ & 2.71${\times}10^{-7}$ \\
          &       & {5/2^-_1} &       & 3.60${\times}10^{-6}$ & 9.92${\times}10^{-8}$ &       & 3.59${\times}10^{-6}$ & 9.58${\times}10^{-8}$ \\
          &       & {7/2^-_1} &       & 2.05${\times}10^{-5}$ & 1.22${\times}10^{-6}$ &       & 2.03${\times}10^{-5}$ & 1.18${\times}10^{-6}$ \\
          &       & {5/2^-_2} &       & 1.81${\times}10^{-6}$ & 1.59${\times}10^{-7}$ &       & 1.79${\times}10^{-6}$ & 1.54${\times}10^{-7}$ \\
          &       & {3/2^-_2} &       & 1.45${\times}10^{-7}$ & 1.79${\times}10^{-8}$ &       & 1.43${\times}10^{-7}$ & 1.72${\times}10^{-8}$ \\
          &       & 7/2^-_2 &       & 2.95${\times}10^{-9}$ & 8.64${\times}10^{-9}$ &       & 2.66${\times}10^{-9}$ & 8.11${\times}10^{-9}$ \\
          &       & 5/2^-_3 &       & 1.65${\times}10^{-12}$ & 8.93${\times}10^{-12}$ &       & 1.42${\times}10^{-12}$ & 8.28${\times}10^{-12}$ \\
\cline{2-9}          
          & 1/2^-_1  & {3/2^-_1} &       & 5.41${\times}10^{-3}$ & 3.15${\times}10^{-5}$ &       & 5.40${\times}10^{-3}$ & 3.05${\times}10^{-5}$ \\
          &       & {1/2^-_1} &       & 2.39${\times}10^{-5}$ & 3.45${\times}10^{-7}$ &       & 2.38${\times}10^{-5}$ & 3.33${\times}10^{-7}$ \\
          &       & {3/2^-_2} &       & 3.16${\times}10^{-7}$ & 3.13${\times}10^{-8}$ &       & 3.12${\times}10^{-7}$ & 3.02${\times}10^{-8}$ \\
\hline
$^{66}$Ni  $\rightarrow$ $^{66}$Cu & 0^+_1 & {1^+_1} &       & 3.14${\times}10^{-6}$ & 7.84${\times}10^{-7}$ &       & 3.07${\times}10^{-6}$ & 7.54${\times}10^{-7}$ \\
    \hline
$^{64}$Cu  $\rightarrow$ $^{64}$Zn & 1^+_1 & {0^+_1} &       & 5.03${\times}10^{-6}$ & 3.96${\times}10^{-7}$ &       & 4.98${\times}10^{-6}$ & 3.82${\times}10^{-7}$ \\
    \hline
$^{66}$Cu $\rightarrow$ $^{66}$Zn & 1^+_1 & {0^+_1} &       & 1.19${\times}10^{-3}$ & 5.02${\times}10^{-6}$ &       & 1.19${\times}10^{-3}$ & 4.85${\times}10^{-6}$ \\
          &       & {2^+_1} &       & 3.50${\times}10^{-4}$ & 4.36${\times}10^{-6}$ &       & 3.49${\times}10^{-4}$ & 4.21${\times}10^{-6}$ \\
          &       & {2^+_2} &       & 2.19${\times}10^{-6}$ & 1.08${\times}10^{-7}$ &       & 2.18${\times}10^{-6}$ & 1.05${\times}10^{-7}$ \\
          &       & {0^+_2} &       & 2.26${\times}10^{-8}$ & 5.58${\times}10^{-9}$ &       & 2.22${\times}10^{-8}$ & 5.37${\times}10^{-9}$ \\
\hline
$^{67}$Cu  $\rightarrow$ $^{67}$Zn & 3/2^-_1  & {5/2^-_1} &       & 1.11${\times}10^{-6}$ & 9.18${\times}10^{-8}$ &       & 1.10${\times}10^{-6}$ & 8.85${\times}10^{-8}$ \\
          &       & {1/2^-_1} &       & 5.83${\times}10^{-7}$ & 6.41${\times}10^{-8}$ &       & 5.77${\times}10^{-7}$ & 6.18${\times}10^{-8}$ \\
          &       & {3/2^-_1} &       & 9.68${\times}10^{-7}$ & 1.47${\times}10^{-7}$ &       & 9.54${\times}10^{-7}$ & 1.42${\times}10^{-7}$ \\
          &       & {3/2^-_2} &       & 5.13${\times}10^{-8}$ & 2.40${\times}10^{-8}$ &       & 4.97${\times}10^{-8}$ & 2.30${\times}10^{-8}$ \\
\hline
$^{69}$Zn  $\rightarrow$ $^{69}$Ga & 1/2^-_1 & {3/2^-_1} &       & 2.11${\times}10^{-4}$ & 8.38${\times}10^{-6}$ &       & 2.10${\times}10^{-4}$ & 8.09${\times}10^{-6}$ \\
          &       & {1/2^-_1} &       & 2.92${\times}10^{-8}$ & 2.40${\times}10^{-9}$ &       & 2.90${\times}10^{-8}$ & 2.32${\times}10^{-9}$ \\
          &       & {3/2^-_2} &       & 5.26${\times}10^{-10}$ & 2.33${\times}10^{-9}$ &       & 4.64${\times}10^{-10}$ & 2.18${\times}10^{-9}$ \\
\hline
$^{72}$Zn  $\rightarrow$ $^{72}$Ga & 0^+_1 & {0^+_1} &       & 7.72${\times}10^{-9}$ & 1.58${\times}10^{-9}$ &       & 7.59${\times}10^{-9}$ & 1.52${\times}10^{-9}$ \\
          &       & {1^+_1} &       & 2.13${\times}10^{-6}$ & 4.54${\times}10^{-7}$ &       & 2.10${\times}10^{-6}$ & 4.37${\times}10^{-7}$ \\
          &       & {1^+_2} &       & 2.15${\times}10^{-6}$ & 5.35${\times}10^{-7}$ &       & 2.11${\times}10^{-6}$ & 5.14${\times}10^{-7}$ \\
          &       & {1^+_3} &       & 1.18${\times}10^{-6}$ & 3.76${\times}10^{-7}$ &       & 1.15${\times}10^{-6}$ & 3.61${\times}10^{-7}$ \\
\hline
$^{70}$Ga  $\rightarrow$ $^{70}$Ge & 1^+_1 & {0^+_1} &       & 5.25${\times}10^{-4}$ & 7.18${\times}10^{-6}$ &       & 5.23${\times}10^{-4}$ & 6.94${\times}10^{-6}$ \\
          &       & {2^+_1} &       & 1.40${\times}10^{-5}$ & 1.17${\times}10^{-6}$ &       & 1.39${\times}10^{-5}$ & 1.13${\times}10^{-6}$ \\
          &       & {0^+_2} &       & 1.91${\times}10^{-6}$ & 2.71${\times}10^{-7}$ &       & 1.89${\times}10^{-6}$ & 2.61${\times}10^{-7}$ \\
\hline
\hline
\end{tabular}
\label{b_and_c_with_ne}
\end{table*}

\begin{table*}
\centering
~~~~~~TABLE (III.B) continued

\vspace{0.3cm}
\begin{tabular}{cLLcccccc}
\hline\hline
\multicolumn{3}{c}{Transition Details} &       & \multicolumn{5}{c}{$T = 3{\times}10^{8}$ K} \\
\cline{1-3} \cline{5-9}
          &       &     &       & \multicolumn{2}{c}{$n_e = 1{\times}10^{26}$ cm$^{-3}$} &       & \multicolumn{2}{c}{$n_e = 1{\times}10^{27}$ cm$^{-3}$}\\
\cline{5-6} \cline{8-9}   

 Decay & J^{\pi}_p & J^{\pi}_d &   & $\lambda_{c}$ & $\lambda_{b}$ &   & $\lambda_{c}$ & $\lambda_{b}$ \\
\hline

$^{75}$Ge  $\rightarrow$ $^{75}$As & 1/2^-_1 & {3/2^-_1} &       & 1.22${\times}10^{-4}$ & 3.51${\times}10^{-6}$ &       & 1.22${\times}10^{-4}$ & 3.38${\times}10^{-6}$ \\
          &       & {1/2^-_1} &       & 6.31${\times}10^{-7}$ & 2.54${\times}10^{-8}$ &       & 6.27${\times}10^{-7}$ & 2.46${\times}10^{-8}$ \\
          &       & {3/2^-_2} &       & 1.59${\times}10^{-5}$ & 7.27${\times}10^{-7}$ &       & 1.58${\times}10^{-5}$ & 7.01${\times}10^{-7}$ \\
          &       & {1/2^-_2} &       & 2.29${\times}10^{-7}$ & 1.61${\times}10^{-8}$ &       & 2.27${\times}10^{-7}$ & 1.56${\times}10^{-8}$ \\
          &       & {1/2^-_3} &       & 4.81${\times}10^{-7}$ & 4.55${\times}10^{-7}$ &       & 4.77${\times}10^{-6}$ & 4.39${\times}10^{-7}$ \\
          &       & 3/2^-_3   &       & 6.74${\times}10^{-7}$ & 6.97${\times}10^{-8}$ &       & 6.67${\times}10^{-7}$ & 6.72${\times}10^{-8}$ \\
          &       & {3/2^-_4} &       & 7.21${\times}10^{-9}$ & 1.78${\times}10^{-9}$ &       & 7.08${\times}10^{-9}$ & 1.72${\times}10^{-9}$ \\
          &       & {3/2^-_5} &       & 8.85${\times}10^{-12}$ & 8.80${\times}10^{-12}$ &       & 8.43${\times}10^{-12}$ & 8.40${\times}10^{-12}$ \\
          &       & {3/2^-_6} &       & 3.86${\times}10^{-10}$ & 4.44${\times}10^{-10}$ &       & 3.66${\times}10^{-10}$ & 4.23${\times}10^{-10}$ \\
          &       & 1/2^-_4  &       & 3.63${\times}10^{-11}$ & 1.24${\times}10^{-10}$ &       & 3.28${\times}10^{-11}$ & 1.17${\times}10^{-10}$ \\
          &       & {1/2^-_5} &       & 5.35${\times}10^{-17}$ & 8.02${\times}10^{-11}$ &       & ---      & 6.80${\times}10^{-11}$ \\
\cline{2-9} 
& 7/2^+_1 & {9/2^+_1} &       & 4.01${\times}10^{-7}$ & 1.52${\times}10^{-8}$ &       & 3.98${\times}10^{-7}$ & 1.47${\times}10^{-8}$ \\
 &        & {5/2^+_1} &       & 5.78${\times}10^{-7}$ & 2.62${\times}10^{-8}$ &       & 5.75${\times}10^{-7}$ & 2.53${\times}10^{-8}$ \\
          &       & {5/2^+_2} &       & 6.71${\times}10^{-8}$ & 2.45${\times}10^{-8}$ &       & 6.54${\times}10^{-8}$ & 2.35${\times}10^{-8}$ \\
          &       & {5/2^+_3} &       & 6.78${\times}10^{-9}$ & 2.79${\times}10^{-9}$ &       & 6.60${\times}10^{-9}$ & 2.68${\times}10^{-9}$ \\
          &       & {9/2^+_2} &       & 1.52${\times}10^{-10}$ & 4.34${\times}10^{-10}$ &       & 1.38${\times}10^{-10}$ & 4.09${\times}10^{-10}$ \\
          &       & {5/2^+_4} &       & 2.49${\times}10^{-12}$ & 1.44${\times}10^{-10}$ &       & 1.63${\times}10^{-12}$ & 1.29${\times}10^{-10}$ \\
\cline{2-9}       

          & 3/2^-_1 & {3/2^-_1} &       & 7.48${\times}10^{-5}$ & 1.47${\times}10^{-6}$ &       & 7.45${\times}10^{-5}$ & 1.42${\times}10^{-6}$ \\
          &       & {1/2^-_1} &       & 3.53${\times}10^{-5}$ & 9.29${\times}10^{-7}$ &       & 3.51${\times}10^{-5}$ & 8.97${\times}10^{-7}$ \\
          &       & {3/2^-_2} &       & 9.74${\times}10^{-6}$ & 2.85${\times}10^{-7}$ &       & 9.70${\times}10^{-6}$ & 2.75${\times}10^{-7}$ \\
          &       & {5/2^-_1} &       & 1.86${\times}10^{-7}$ & 5.56${\times}10^{-9}$ &       & 1.85${\times}10^{-7}$ & 5.37${\times}10^{-9}$ \\
          &       & {1/2^-_2} &       & 1.12${\times}10^{-7}$ & 4.65${\times}10^{-9}$ &       & 1.11${\times}10^{-7}$ & 4.49${\times}10^{-9}$ \\
          &       & {5/2^-_2} &       & 6.80${\times}10^{-7}$ & 3.46${\times}10^{-8}$ &       & 6.75${\times}10^{-7}$ & 3.34${\times}10^{-8}$ \\
          &       & {1/2^-_3} &       & 2.71${\times}10^{-7}$ & 1.41${\times}10^{-8}$ &       & 2.69${\times}10^{-7}$ & 1.36${\times}10^{-8}$ \\
          &       &  3/2^-_3  &       & 1.19${\times}10^{-6}$ & 6.65${\times}10^{-8}$ &       & 1.18${\times}10^{-6}$ & 6.41${\times}10^{-8}$ \\
          &       & 3/2^-_4   &       & 3.82${\times}10^{-7}$ & 3.89${\times}10^{-8}$ &       & 3.78${\times}10^{-7}$ & 3.75${\times}10^{-8}$ \\
          &       & {3/2^-_5} &       & 2.24${\times}10^{-7}$ & 4.38${\times}10^{-8}$ &       & 2.20${\times}10^{-7}$ & 4.22${\times}10^{-8}$ \\
          &       & {3/2^-_6} &       & 1.59${\times}10^{-8}$ & 3.26${\times}10^{-9}$ &       & 1.57${\times}10^{-8}$ & 3.14${\times}10^{-9}$ \\
          &       & {1/2^-_4} &       & 2.49${\times}10^{-8}$ & 6.42${\times}10^{-9}$ &       & 2.45${\times}10^{-8}$ & 6.17${\times}10^{-9}$ \\
          &       & {1/2^-_5} &       & 1.46${\times}10^{-8}$ & 4.70${\times}10^{-9}$ &       & 1.43${\times}10^{-8}$ & 4.52${\times}10^{-9}$ \\
          &       & {3/2^-_7} &       & 1.58${\times}10^{-8}$ & 6.11${\times}10^{-9}$ &       & 1.54${\times}10^{-8}$ & 5.86${\times}10^{-9}$ \\
          &       & {3/2^-_8} &       & 2.96${\times}10^{-10}$ & 4.79${\times}10^{-10}$ &       & 2.77${\times}10^{-10}$ & 4.55${\times}10^{-10}$ \\
          &       & {3/2^-_9}  &       & 5.61${\times}10^{-10}$ & 1.45${\times}10^{-9}$ &       & 5.14${\times}10^{-10}$ & 1.37${\times}10^{-9}$ \\
          &       & {5/2^-_3} &       & 4.86${\times}10^{-13}$ & 1.19${\times}10^{-10}$ &       & 2.16${\times}10^{-13}$ & 1.05${\times}10^{-10}$ \\
\hline
$^{78}$Ge $\rightarrow$ $^{78}$As  & 0^+_1 & {1^+_1} &       & 5.28${\times}10^{-5}$ & 4.01${\times}10^{-6}$ &       & 5.24${\times}10^{-5}$ & 3.87${\times}10^{-6}$ \\
          &       & {1^+_2} &       & 1.92${\times}10^{-5}$ & 1.52${\times}10^{-6}$ &       & 1.91${\times}10^{-5}$ & 1.47${\times}10^{-6}$ \\
          &       & {1^+_3} &       & 5.70${\times}10^{-5}$ & 9.19${\times}10^{-6}$ &       & 5.62${\times}10^{-5}$ & 8.85${\times}10^{-6}$ \\
\hline
$^{81}$Se $\rightarrow$ $^{81}$Br   & 1/2^-_1 & {3/2^-_1} &       & 5.80${\times}10^{-4}$ & 1.07${\times}10^{-5}$ &       & 5.78${\times}10^{-4}$ & 1.03${\times}10^{-5}$ \\
          &       & 1/2^-_1   &       & 1.41${\times}10^{-7}$ & 5.77${\times}10^{-9}$ &       & 1.40${\times}10^{-7}$ & 5.56${\times}10^{-9}$ \\
          &       & {3/2^-_2}  &       & 5.82${\times}10^{-7}$ & 2.50${\times}10^{-8}$ &       & 5.79${\times}10^{-7}$ & 2.41${\times}10^{-8}$ \\
          &       & {3/2^-_3} &       & 1.05${\times}10^{-6}$ & 5.25${\times}10^{-8}$ &       & 1.04${\times}10^{-6}$ & 5.06${\times}10^{-8}$ \\
          &       & {3/2^-_4} &       & 5.14${\times}10^{-6}$ & 3.71${\times}10^{-7}$ &       & 5.10${\times}10^{-6}$ & 3.58${\times}10^{-7}$ \\
          &       & {1/2^-_2} &       & 1.78${\times}10^{-7}$ & 2.66${\times}10^{-8}$ &       & 1.75${\times}10^{-7}$ & 2.56${\times}10^{-8}$ \\
          &       & {3/2^-_5} &       & 1.85${\times}10^{-8}$ & 5.07${\times}10^{-9}$ &       & 1.82${\times}10^{-8}$ & 4.88${\times}10^{-9}$ \\
          &       & {3/2^-_6} &       & 5.50${\times}10^{-11}$ & 2.31${\times}10^{-10}$ &       & 4.92${\times}10^{-11}$ & 2.17${\times}10^{-10}$ \\
          &       & {3/2^-_7} &       & 9.09${\times}10^{-11}$ & 5.11${\times}10^{-10}$ &       & 7.99${\times}10^{-11}$ & 4.78${\times}10^{-10}$ \\
\hline
\hline

\end{tabular}
\end{table*}


\begin{table*}
\centering
\renewcommand\thetable{IV}
\caption{Calculated bare atom \bta decay rate ($\lambda_k$, where $\lambda_k = (\lambda_b + \lambda_c) \times n_{ik}$; in s$^{-1}$) and branching ($I_k$) from each parent level to daughter level m with branching ($I_m$) for various density - temperature combinations. Here, Column 2: spin-parity of parent energy levels; Column 7 and Column 12: spin-parity of daughter energy levels. Temperature (T) is given in Column 3, Column 8 and Column 13; where $T_1 = 1{\times}10^{8}$ K, $T_3 = 3{\times}10^{8}$ K and $ T_5= 5{\times}10^{8}$ K, respectively. Column 4: equilibrium population of the k - th nuclear state of the i - th nucleus ($n_{ik}$) at different temperatures. Column 5 and Column 10: Total \bta branching ($I_k$) from each parent level at different density - temperature combinations. Column 6 and Column 11: stellar \bta decay rate ($\lambda_k$) of bare atom i from its k - th nuclear level. Column 9 and Column 14: \bta branching ($I_m$) to each daughter level m.}
\begin{tabular}{cLLccccLLccccLLc}
 &       &   &    &       &       &       &       &       &       &       &       &       &       &       &  \\
\hline
\hline
 &       &   &    &       & \multicolumn{5}{c}{$n_e = 1{\times}10^{26}$ cm$^{-3}$} &       & \multicolumn{5}{c}{$n_e = 1{\times}10^{27}$ cm$^{-3}$} \\
\cline{6-10} \cline{12-16}
{Decay} & J^{\pi}_p & T   & $n_{ik}$ &       & $I_k(\%)$ & $\lambda_k$ & J^{\pi}_d &  T     & $I_m(\%)$ &       & $I_k(\%)$ &     $\lambda_k$ & J^{\pi}_d &  T     & $I_m(\%)$ \\
\hline
\hline
$^{59}$Fe $\rightarrow$ $^{59}$Co & 3/2^-_1 & T_1    & 1.00$\times 10^{0}$ &       & 100.00 & {2.13$\times 10^{-7}$} & 3/2^-_1 & T_1    & 46.29 &       & 100.00 & 2.08$\times 10^{-7}$ & 3/2^-_1 & T_1    & 46.62 \\
          &       & T_3    & 1.00$\times 10^{0}$ &       & 100.00 & 2.14$\times 10^{-7}$ &       & T_3    & 46.26 &       & 100.00 & 2.10$\times 10^{-7}$ &       & T_3    & 46.48 \\
          &       & T_5    & 9.99$\times 10^{-1}$ &       & 92.47 & 2.14$\times 10^{-7}$ &       & T_5    & 42.77 &       & 92.40 & 2.11$\times 10^{-7}$ &       & T_5    & 42.90 \\
\cline{8-10} \cline{14-16}
          &       &       &       &       &       &       & 3/2^-_2 & T_1    & 53.27 &       &       &       & 3/2^-_2 & T_1    & 52.96 \\
          &       &       &       &       &       &       &       & T_3    & 53.30 &       &       &       &       & T_3    & 53.09 \\
          &       &       &       &       &       &       &       & T_5    & 49.30 &       &       &       &       & T_5    & 49.09 \\
\cline{8-10} \cline{14-16}
          &       &       &       &       &       &       & 1/2^-_1 & T_1    & 0.41  &       &       &       & 1/2^-_1 & T_1    & 0.40 \\
          &       &       &       &       &       &       &       & T_3    & 0.41  &       &       &       &       & T_3    & 0.41 \\
          &       &       &       &       &       &       &       & T_5    & 0.38  &       &       &       &       & T_5    & 0.38 \\
\cline{8-10} \cline{14-16}
          &       &       &       &       &       &       & 5/2^-_1 & T_1    & 0.02  &       &       &       & 5/2^-_1 & T_1    & 0.02 \\
          &       &       &       &       &       &       &       & T_3    & 0.02  &       &       &       &       & T_3    & 0.02 \\
          &       &       &       &       &       &       &       & T_5    & 0.02  &       &       &       &       & T_5    & 0.02 \\
\cline{2-16}       
   & 1/2^-_1 & T_1    & 0.00$\times 10^{0}$ &       & 0.00  & 0.00$\times 10^{0}$ & 3/2^-_1 & T_1    & 0.00  &       & 0.00  & 0.00$\times 10^{0}$ & 3/2^-_1 & T_1    & 0.00 \\
          &       & T_3    & 0.00$\times 10^{0}$ &       & 0.00  & 0.00$\times 10^{0}$ &       & T_3    & 0.00  &       & 0.00  & 0.00$\times 10^{0}$ &       & T_3    & 0.00 \\
          &       & T_5    & 6.38$\times 10^{-4}$ &       & 0.33  & 7.55$\times 10^{-10}$ &       & T_5    & 0.08  &       & 0.33  & 7.48$\times 10^{-10}$ &       & T_5    & 0.08 \\
\cline{8-10} \cline{14-16}
          &       &       &       &       &       &       & 3/2^-_2 & T_1    & 0.00  &       &       &       & 3/2^-_2 & T_1    & 0.00 \\
          &       &       &       &       &       &       &       & T_3    & 0.00  &       &       &       &       & T_3    & 0.00 \\
          &       &       &       &       &       &       &       & T_5    & 0.08  &       &       &       &       & T_5    & 0.08 \\
\cline{8-10} \cline{14-16}
          &       &       &       &       &       &       & 1/2^-_1 & T_1    & 0.00  &       &       &       & 1/2^-_1 & T_1    & 0.00 \\
          &       &       &       &       &       &       &       & T_3    & 0.00  &       &       &       &       & T_3    & 0.00 \\
          &       &       &       &       &       &       &       & T_5    & 0.17  &       &       &       &       & T_5    & 0.17 \\
\cline{2-16}   
       & 5/2^-_1 & T_1    & 0.00$\times 10^{0}$ &       & 0.00  & 0.00$\times 10^{0}$ & 7/2^-_1 & T_1    & 0.00  &       & 0.00  & 0.00$\times 10^{0}$ & 7/2^-_1 & T_1    & 0.00 \\
          &       & T_3    & 0.00$\times 10^{0}$ &       & 0.00  & 0.00$\times 10^{0}$ &       & T_3    & 0.00  &       & 0.00  & 0.00$\times 10^{0}$ &       & T_3    & 0.00 \\
          &       & T_5    & 2.55$\times 10^{-5}$ &       & 7.20  & 1.66$\times 10^{-8}$ &       & T_5    & 7.19  &       & 7.29  & 1.66$\times 10^{-8}$ &       & T_5    & 7.27 \\
\cline{8-10} \cline{14-16}
          &       &       &       &       &       &       & 3/2^-_1 & T_1    & 0.00  &       &       &       & 3/2^-_1 & T_1    & 0.00 \\
          &       &       &       &       &       &       &       & T_3    & 0.00  &       &       &       &       & T_3    & 0.00 \\
          &       &       &       &       &       &       &       & T_5    & 0.01  &       &       &       &       & T_5    & 0.01 \\
\hline
    $^{60}$Co  $\rightarrow$ $^{60}$Ni & 5^+_1 & T_1    & 1.00${\times}10^{0}$ &       & 97.18 & 4.54${\times}10^{-9}$ & 4^+_1 & T_1    & 97.18 &       & 99.68 & 4.44${\times}10^{-9}$ & 4^+_1 & T_1    & 99.68 \\
          &       & T_3    & 9.56${\times}10^{-1}$ &       &  26.56     & 4.35${\times}10^{-9}$ &       & T_3    & 26.56      &       & 26.40 & 4.29${\times}10^{-9}$ &       & T_3    & 26.40 \\
          &       & T_5    & 8.95${\times}10^{-1}$ &       & 11.91 & 4.07${\times}10^{-9}$ &       & T_5    & 11.91 &       & 11.86 & 4.03${\times}10^{-9}$ &       & T_5    & 11.86 \\
\cline{2-16}   
       & 2^+_1 & T_1    & 4.87${\times}10^{-4}$ &       & 2.82  & 1.32${\times}10^{-10}$ & 2^+_1 & T_1    & 0.31  &       & 0.32  & 1.43${\times}10^{-11}$ & 2^+_1 & T_1    & 0.03 \\
          &       & T_3    & 4.45${\times}10^{-2}$ &       &    73.43   & 1.20${\times}10^{-8}$      &       & T_3    & 7.97      &       & 73.59 & 1.20${\times}10^{-8}$ &       & T_3    & 8.02 \\
          &       & T_5    & 1.04${\times}10^{-1}$ &       & 81.96 & 2.80${\times}10^{-8}$ &       & T_5    & 8.90  &       & 82.07 & 2.79${\times}10^{-8}$ &       & T_5    & 8.93 \\
\cline{8-10} \cline{14-16}
          &       &       &       &       &       &       & 2^+_2 & T_1    & 2.51  &       &       &       & 2^+_2 & T_1    & 0.29 \\
          &       &       &       &       &       &       &       & T_3    & 65.45      &       &       &       &       & T_3    & 65.57 \\
          &       &       &       &       &       &       &       & T_5    & 73.06 &       &       &       &       & T_5    & 73.14 \\
\cline{2-16}        
  & 4^+_1 & T_1    & 0.00${\times}10^{0}$ &       & 0.00  & 0.00${\times}10^{0}$ & 4^+_1 & T_1    & 0.00  &       & 0.00  & 0.00${\times}10^{0}$ & 4^+_1 & T_1    & 0.00 \\
          &       & T_3    & 1.72${\times}10^{-5}$ &       &  0.01     &   1.63${\times}10^{-12}$    &       & T_3    &  0.01     &       & 0.01  & 1.62${\times}10^{-12}$ &       & T_3    & 0.01 \\
          &       & T_5    & 1.17${\times}10^{-3}$ &       & 0.33  & 1.12${\times}10^{-10}$ &       & T_5    & 0.32  &       & 0.33  & 1.11${\times}10^{-10}$ &       & T_5    & 0.32 \\
\cline{8-10} \cline{14-16}
          &       &       &       &       &       &       & 3^+_1 & T_1    & 0.00  &       &       &       & 3^+_1 & T_1    & 0.00 \\
          &       &       &       &       &       &       &       & T_3    & 0.00      &       &       &       &       & T_3    & 0.00 \\
          &       &       &       &       &       &       &       & T_5    & 0.01  &       &       &       &       & T_5    & 0.01 \\
\hline
\hline
    \end{tabular}
  \label{Decay_from_parent}
\end{table*}
\begin{table*}
\centering
~~~~~~TABLE (IV) continued
\vspace{0.5cm}
\begin{tabular}{cLLccccLLccccLLc}
\hline
\hline
& 3^+_1 & T_1    & 0.00${\times}10^{0}$ &       & 0.00  & 0.00${\times}10^{0}$ & 2^+_1 & T_1    & 0.00  &       & 0.00  & 0.00${\times}10^{0}$ & 2^+_1 & T_1    & 0.00 \\
          &       & T_3    & 0.00${\times}10^{0}$ &      &   0.00     & 0.00${\times}10^{0}$  &       & T_3    & 0.00  &       & 0.00  & 0.00${\times}10^{0}$ &       & T_3    & 0.00 \\
          &       & T_5    & 7.03${\times}10^{-4}$ &       & 5.73  & 1.96${\times}10^{-9}$ &       & T_5    & 0.38  &       & 5.74  & 1.95${\times}10^{-9}$ &       & T_5    & 0.38 \\
\cline{8-10} \cline{14-16}
          &       &       &       &       &       &       & 2^+_2 & T_1    & 0.00  &       &       &       & 2^+_2 & T_1    & 0.00 \\
          &       &       &       &       &       &       &       & T_3    & 0.00  &       &       &       &       & T_3    & 0.00 \\
          &       &       &       &       &       &       &       & T_5    & 4.92  &       &       &       &       & T_5    & 4.92 \\
\cline{8-10} \cline{14-16}
          &       &       &       &       &       &       & 4^+_1 & T_1    & 0.00  &       &       &       & 4^+_1 & T_1    & 0.00 \\
          &       &       &       &       &       &       &       & T_3    & 0.00  &       &       &       &       & T_3    & 0.00 \\
          &       &       &       &       &       &       &       & T_5    & 0.43  &       &       &       &       & T_5    & 0.43 \\
\cline{8-10} \cline{14-16}
          &       &       &       &       &       &       & 3^+_1 & T_1    & 0.00  &       &       &       & 3^+_1 & T_1    & 0.00 \\
          &       &       &       &       &       &       &       & T_3    & 0.00  &       &       &       &       & T_3    & 0.00 \\
          &       &       &       &       &       &       &       & T_5    & 0.01  &       &       &       &       & T_5    & 0.01 \\
\hline
$^{61}$Co $\rightarrow$ $^{61}$Ni & 7/2^-_1  & T_1    & 1.00$\times 10^{0}$ &       & 100.00 & 1.02$\times 10^{-4}$ & 5/2^-_1 & T_1    & 98.74 &       & 100.00 & 1.03$\times 10^{-4}$ & 5/2^-_1 & T_1    & 98.77 \\
          &       & T_3    & 1.00$\times 10^{0}$ &       & 100.00 & 1.02$\times 10^{-4}$ &       & T_3    & 98.74 &       & 100.00 & 1.03$\times 10^{-4}$ &       & T_3    & 98.76 \\
          &       & T_5    & 1.00$\times 10^{0}$ &       & 100.00 & 1.02$\times 10^{-4}$ &       & T_5    & 98.74 &       & 100.00 & 1.03$\times 10^{-4}$ &       & T_5    & 98.76 \\
\cline{8-10} \cline{14-16}          
          &       &       &       &       &       &       & 5/2^-_2 & T_1    & 0.37  &       &       &       & 5/2^-_2 & T_1    & 0.37 \\
          &       &       &       &       &       &       &       & T_3    & 0.37  &       &       &       &       & T_3    & 0.37 \\
          &       &       &       &       &       &       &       & T_5    & 0.37  &       &       &       &       & T_5    & 0.37 \\
\cline{8-10} \cline{14-16}
          &       &       &       &       &       &       & 7/2^-_1 & T_1    & 0.54  &       &       &       & 7/2^-_1 & T_1    & 0.53 \\
          &       &       &       &       &       &       &       & T_3    & 0.54  &       &       &       &       & T_3    & 0.54 \\
          &       &       &       &       &       &       &       & T_5    & 0.54  &       &       &       &       & T_5    & 0.54 \\
\cline{8-10} \cline{14-16}
          &       &       &       &       &       &       & 7/2^-_2 & T_1    & 0.15  &       &       &       & 7/2^-_2 & T_1    & 0.15 \\
          &       &       &       &       &       &       &       & T_3    & 0.15  &       &       &       &       & T_3    & 0.15 \\
          &       &       &       &       &       &       &       & T_5    & 0.15  &       &       &       &       & T_5    & 0.15 \\
\cline{8-10} \cline{14-16}
          &       &       &       &       &       &       & 5/2^-_3 & T_1    & 0.19  &       &       &       & 5/2^-_3 & T_1    & 0.18 \\
          &       &       &       &       &       &       &       & T_3    & 0.19  &       &       &       &       & T_3    & 0.18 \\
          &       &       &       &       &       &       &       & T_5    & 0.19  &       &       &       &       & T_5    & 0.19 \\
\hline
$^{63}$Ni $\rightarrow$ $^{63}$Cu & 1/2^-_1  & T_1    & 1.00$\times 10^{0}$ &       & 99.60 & {1.01$\times 10^{-9}$} & 3/2^-_1  & T_1    & 99.60 &       & 99.59 & 9.21$\times 10^{-10}$ & 3/2^-_1  & T_1    & 99.59 \\
          &       & T_3    & 9.03$\times 10^{-1}$ &       & 17.76 & 9.18$\times 10^{-10}$ &       & T_3    & 17.76 &       & 17.37 & 8.64$\times 10^{-10}$ &       & T_3    & 17.37 \\
          &       & T_5    & 6.90$\times 10^{-1}$ &       & 3.60  & 7.04$\times 10^{-10}$ &       & T_5    & 3.60  &       & 3.51  & 6.75$\times 10^{-10}$ &       & T_5    & 3.51 \\
\cline{2-16}        
  & 5/2^-_1  & T_1    & 1.20$\times 10^{-4}$ &       & 0.40  & {4.02$\times 10^{-12}$} & 3/2^-_1  & T_1    & 0.40  &       & 0.41  & 3.82$\times 10^{-12}$ & 3/2^-_1  & T_1    & 0.41 \\
          &       & T_3    & 9.27$\times 10^{-2}$ &       & 60.23 & 3.11$\times 10^{-9}$ &       & T_3    & 60.23 &       & 60.36 & 3.00$\times 10^{-9}$ &       & T_3    & 60.36 \\
          &       & T_5    & 2.73$\times 10^{-1}$ &       & 46.95 & 9.19$\times 10^{-9}$ &       & T_5    & 46.95 &       & 46.81 & 8.92$\times 10^{-9}$ &       & T_5    & 46.81 \\
\cline{2-16}        
  & 3/2^-_1  & T_1    & 0.00$\times 10^{0}$ &       & 0.00  & 0.00$\times 10^{0}$ & 3/2^-_1  & T1    & 0.00  &       & 0.00  & 0.00$\times 10^{0}$ & 3/2^-_1  & T_1    & 0.00 \\
          &       & T_3    & 4.38$\times 10^{-3}$ &       & 22.02 & 1.14$\times 10^{-9}$ &       & T_3    & 22.02 &       & 22.27 & 1.11$\times 10^{-9}$ &       & T_3    & 22.27 \\
          &       & T_5    & 3.72$\times 10^{-2}$ &       & 49.45 & 9.68$\times 10^{-9}$ &       & T_5    & 49.45 &       & 49.67 & 9.46$\times 10^{-9}$ &       & T_5    & 49.67 \\
\hline
$^{65}$Ni $\rightarrow$ $^{65}$Cu & 5/2^-_1  & T1    & 1.00$\times 10^{0}$ &       & 98.44 & {7.29$\times 10^{-5}$} & 3/2^-_1 & T_1    & 61.29 &       & 98.43 & 7.24$\times 10^{-5}$ & 3/2^-_1 & T_1    & 61.50 \\
          &       & T_3    & 9.72$\times 10^{-1}$ &       & 31.79 & 7.10$\times 10^{-5}$ &       & T_3    & 19.79 &       & 31.74 & 7.06$\times 10^{-5}$ &       & T_3    & 19.80 \\
          &       & T_5    & 9.29$\times 10^{-1}$ &       & 14.85 & 6.69$\times 10^{-5}$ &       & T_5    & 9.25  &       & 14.83 & 6.75$\times 10^{-5}$ &       & T_5    & 9.25 \\
\cline{8-10} \cline{14-16}
          &       &       &       &       &       &       & 5/2^-_1 & T_1    & 4.99  &       &       &       & 5/2^-_1 & T_1    & 4.99 \\
          &       &       &       &       &       &       &       & T_3    & 1.61  &       &       &       &       & T_3    & 1.61 \\
          &       &       &       &       &       &       &       & T_5    & 0.75  &       &       &       &       & T_5    & 0.75 \\
 \cline{8-10} \cline{14-16}
          &       &       &       &       &       &       & 7/2^-_1 & T_1    & 29.27 &       &       &       & 7/2^-_1 & T_1    & 29.09 \\
          &       &       &       &       &       &       &       & T_3    & 9.46  &       &       &       &       & T_3    & 9.40 \\
          &       &       &       &       &       &       &       & T_5    & 4.41  &       &       &       &       & T_5    & 4.40 \\
\cline{8-10} \cline{14-16}
          &       &       &       &       &       &       & 5/2^-_2 & T_1    & 2.65  &       &       &       & 5/2^-_2 & T_1    & 2.62 \\
          &       &       &       &       &       &       &       & T_3    & 0.86  &       &       &       &       & T_3    & 0.85 \\
          &       &       &       &       &       &       &       & T_5    & 0.40  &       &       &       &       & T_5    & 0.40 \\
\cline{8-10} \cline{14-16}
          &       &       &       &       &       &       & 3/2^-_2 & T_1    & 0.22  &       &       &       & 3/2^-_2 & T_1    & 0.22 \\
          &       &       &       &       &       &       &       & T_3    & 0.07  &       &       &       &       & T_3    & 0.07 \\
          &       &       &       &       &       &       &       & T_5    & 0.03  &       &       &       &       & T_5    & 0.03 \\
\cline{8-10} \cline{14-16}
          &       &       &       &       &       &       & 7/2^-_2 & T_1    & 0.02  &       &       &       & 7/2^-_2 & T_1    & 0.01 \\
          &       &       &       &       &       &       &       & T_3    & 0.01  &       &       &       &       & T_3    & 0.00 \\
          &       &       &       &       &       &       &       & T_5    & 0.00  &       &       &       &       & T_5    & 0.00 \\

\hline
\hline   
\end{tabular}
\label{tab:addlabel}
\end{table*}

\begin{table*}
\centering
~~~~~~TABLE (IV) continued
\vspace{0.5cm}
\begin{tabular}{cLLccccLLccccLLc}
\hline 
\hline
     & 1/2^-_1  & T_1    & 2.12$\times 10^{-4}$ &       & 1.56  & {1.16$\times 10^{-6}$} & 3/2^-1 & T_1    & 1.56  &       & 1.57  & 1.15$\times 10^{-6}$ & 3/2^-1 & T_1    & 1.56 \\
          &       & T_3    & 2.79$\times 10^{-2}$ &       & 68.21 & 1.52$\times 10^{-4}$ &       & T_3    & 67.90 &       & 68.26 & 1.52$\times 10^{-4}$ &       & T_3    & 67.95 \\
          &       & T_5    & 7.10$\times 10^{-2}$ &       & 85.06 & 3.83$\times 10^{-4}$ &       & T_5    & 84.68 &       & 85.08 & 3.87$\times 10^{-4}$ &       & T_5    & 84.69 \\
\cline{8-10} \cline{14-16}
          &       &       &       &       &       &       & 1/2^-_1 & T_1    & 0.01  &       &       &       & 1/2^-_1 & T_1    & 0.01 \\
          &       &       &       &       &       &       &       & T_3    & 0.30  &       &       &       &       & T_3    & 0.30 \\
          &       &       &       &       &       &       &       & T_5    & 0.38  &       &       &       &       & T_5    & 0.38 \\
\cline{8-10} \cline{14-16}
          &       &       &       &       &       &       & 3/2^-_2 & T_1    & 0.00  &       &       &       & 3/2^-_2 & T_1    & 0.00 \\
          &       &       &       &       &       &       &       & T_3    & 0.00  &       &       &       &       & T_3    & 0.00 \\
          &       &       &       &       &       &       &       & T_5    & 0.01  &       &       &       &       & T_5    & 0.01 \\
\hline
$^{66}$Ni $\rightarrow$ $^{66}$Cu & 0^+_1 & T_1    & 1.00$\times 10^{0}$ &       & 100.00 & 3.91$\times 10^{-6}$ & 0^+_1 & T_1    & 100.00 &       & 100.00 & 3.78$\times 10^{-6}$ & 0^+_1 & T_1    & 100.00 \\
          &       & T_3    & 1.00$\times 10^{0}$ &       & 100.00 & 3.92$\times 10^{-6}$ &       & T_3    & 100.00 &       & 100.00 & 3.83$\times 10^{-6}$ &       & T_3    & 100.00 \\
          &       & T_5    & 1.00$\times 10^{0}$ &       & 100.00 & 3.93$\times 10^{-6}$ &       & T_5    & 100.00 &       & 100.00 & 3.85$\times 10^{-6}$ &       & T_5    & 100.00 \\
\hline
$^{64}$Cu $\rightarrow$ $^{64}$Zn & 1^+_1 & T_1    & 1.00$\times 10^{0}$ &       & 100.00 & 5.42$\times 10^{-6}$ & 0^+_1 & T_1    & 100.00 &       & 100.00 & 5.34$\times 10^{-6}$ & 0^+_1 & T_1    & 100.00 \\
          &       & T_3    & 1.00$\times 10^{0}$ &       & 100.00 & 5.43$\times 10^{-6}$ &       & T_3    & 100.00 &       & 100.00 & 5.37$\times 10^{-6}$ &       & T_3    & 100.00 \\
          &       & T_5    & 1.00$\times 10^{0}$ &       & 100.00 & 5.43$\times 10^{-6}$ &       & T_5    & 100.00 &       & 100.00 & 5.38$\times 10^{-6}$ &       & T_5    & 100.00 \\
\hline
$^{66}$Cu $\rightarrow$ $^{66}$Zn & 1^+_1 & T_1    & 1.00$\times 10^{0}$ &       & 100.00 & {1.55$\times 10^{-3}$} & 0^+_1 & T_1    & 77.03 &       & 100.00 & 1.55$\times 10^{-3}$ & 0^+_1 & T_1    & 77.07 \\
          &       & T_3    & 9.99$\times 10^{-1}$ &       & 99.98 & 1.55$\times 10^{-3}$ &       & T_3    & 77.01 &       & 99.98 & 1.55$\times 10^{-3}$ &       & T_3    & 77.04 \\
          &       & T_5    & 9.74$\times 10^{-1}$ &       & 99.60 & 1.51$\times 10^{-3}$ &       & T_5    & 76.72 &       & 99.59 & 1.51$\times 10^{-3}$ &       & T_5    & 76.74 \\
\cline{8-10} \cline{14-16}
          &       &       &       &       &       &       & 2^+_1 & T_1    & 22.82 &       &       &       & 2^+_1 & T_1    & 22.78 \\
          &       &       &       &       &       &       &       & T_3    & 22.81 &       &       &       &       & T_3    & 22.79 \\
          &       &       &       &       &       &       &       & T_5    & 22.73 &       &       &       &       & T_5    & 22.71 \\
\cline{8-10} \cline{14-16}
          &       &       &       &       &       &       & 2^+_2 & T_1    & 0.15  &       &       &       & 2^+_2 & T_1    & 0.15 \\
          &       &       &       &       &       &       &       & T_3    & 0.15  &       &       &       &       & T_3    & 0.15 \\
          &       &       &       &       &       &       &       & T_5    & 0.15  &       &       &       &       & T_5    & 0.15 \\

\hline
$^{67}$Cu  $\rightarrow$ $^{67}$Zn & 3/2^-_1  &  T_1     & 1.00${\times}10^{0}$ && 100.00  & 2.98${\times}10^{-6}$   &  5/2^-_1 &  T_1 & 39.50       &       & 100.00 & 2.94${\times}10^{-6}$ & 5/2^-_1 & T_1    & 39.64 \\
          &       &  T_3     & 1.00${\times}10^{0}$ &       & 100.00      &  2.99${\times}10^{-6}$     &       & T_3    &  39.48  &       & 100.00 & 2.96${\times}10^{-6}$ &       & T_3    & 39.57 \\
          &       &  T_5     & 1.00${\times}10^{0}$ &       & 100.00     &   2.99${\times}10^{-6}$     &       & T_5    &  39.48  &       & 100.00 & 2.97${\times}10^{-6}$ &       & T_5    & 39.54 \\
\cline{8-10} \cline{14-16}
          &       &       &       &       &       &       &  1/2^-_1  &  T_1    &  21.32  &       &       &       & 1/2^-_1 & T_1    & 21.34 \\
          &       &       &       &       &       &       &           &  T_3    & 21.31   &       &       &       &       & T_3    & 21.33 \\
          &       &       &       &       &       &       &           &  T_5    & 21.31   &       &       &       &       & T_5    & 21.32 \\
\cline{8-10} \cline{14-16}
          &       &       &       &       &       &       &    3/2^-_1   &  T_1   & 36.70   &       &       &       & 3/2^-_1 & T_1    & 36.60 \\
          &       &       &       &       &       &       &              &  T_3   &  36.72  &       &       &       &       & T_3    & 36.65 \\
          &       &       &       &       &       &       &              &  T_5   & 36.72  &       &       &       &       & T_5    & 36.67 \\
\cline{8-10} \cline{14-16}
          &       &       &       &       &       &       &  3/2^-_2   & T_1      & 2.48   &       &       &       & 3/2^-_2 & T_1    & 2.42 \\
          &       &       &       &       &       &       &             &  T_3    &  2.49  &       &       &       &       & T_3    & 2.45 \\
          &       &       &       &       &       &       &             &  T_5    &  2.49  &       &       &       &       & T_5    & 2.46 \\
\hline
$^{69}$Zn $\rightarrow$ $^{69}$Ga & 1/2^-_1 & T_1    & 1.00$\times 10^{0}$ &       & 100.00 & {2.19$\times 10^{-4}$} & 3/2^-_1 & T_1    & 99.98 &       & 100.00 & 2.17$\times 10^{-4}$ & 3/2^-_1 & T_1    & 99.98 \\
          &       & T_3    & 1.00$\times 10^{0}$ &       & 100.00 & 2.19$\times 10^{-4}$ &       & T_3    & 99.98 &       & 100.00 & 2.18$\times 10^{-4}$ &       & T_3    & 99.98 \\
          &       & T_5    & 1.00$\times 10^{0}$ &       & 100.00 & 2.19$\times 10^{-4}$ &       & T_5    & 99.98 &       & 100.00 & 2.18$\times 10^{-4}$ &       & T_5    & 99.98 \\
\cline{8-10} \cline{14-16}
          &       &       &       &       &       &       & 1/2^-_1 & T_1    & 0.01  &       &       &       & 1/2^-_1 & T_1    & 0.01 \\
          &       &       &       &       &       &       &       & T_3    & 0.01  &       &       &       &       & T_3    & 0.01 \\
          &       &       &       &       &       &       &       & T_5    & 0.01  &       &       &       &       & T_5    & 0.01 \\
\hline
\hline
\end{tabular}
\label{tab:addlabel}

\end{table*}

\begin{table*}
\centering
~~~~~~TABLE (IV) continued
\vspace{0.5cm}
\begin{tabular}{cLLccccLLccccLLc}
\hline 
\hline

$^{72}$Zn $\rightarrow$ $^{72}$Ga & 0^+_1 & T_1    & 1.00$\times 10^{0}$ &       & 100.00 & {6.81$\times 10^{-6}$} & 0^+_1 & T_1    & 0.14  &       & 100.00 & 6.59$\times 10^{-6}$ & 0^+_1 & T_1    & 0.14 \\
          &       & T_3    & 1.00$\times 10^{0}$ &       & 100.00 & 6.83$\times 10^{-6}$ &       & T_3    & 0.14  &       & 100.00 & 6.68$\times 10^{-6}$ &       & T_3    & 0.14 \\
          &       & T_5    & 1.00$\times 10^{0}$ &       & 100.00 & 6.84$\times 10^{-6}$ &       & T_5    & 0.14  &       & 100.00 & 6.71$\times 10^{-6}$ &       & T_5    & 0.14 \\
\cline{8-10} \cline{14-16}
          &       &       &       &       &       &       & 1,2   & T_1    & 37.85 &       &       &       & 1,2   & T_1    & 37.97 \\
          &       &       &       &       &       &       &       & T_3    & 37.84 &       &       &       &       & T_3    & 37.92 \\
          &       &       &       &       &       &       &       & T_5    & 37.83 &       &       &       &       & T_5    & 37.90 \\
\cline{8-10} \cline{14-16}
          &       &       &       &       &       &       & 1^+_2 & T_1    & 39.30 &       &       &       & 1^+_2 & T_1    & 39.31 \\
          &       &       &       &       &       &       &       & T_3    & 39.30 &       &       &       &       & T_3    & 39.31 \\
          &       &       &       &       &       &       &       & T_5    & 39.30 &       &       &       &       & T_5    & 39.31 \\
\cline{8-10} \cline{14-16}
          &       &       &       &       &       &       & 1^+_3 & T_1    & 22.71 &       &       &       & 1^+_3 & T_1    & 22.58 \\
          &       &       &       &       &       &       &       & T_3    & 22.72 &       &       &       &       & T_3    & 22.64 \\
          &       &       &       &       &       &       &       & T_5    & 22.73 &       &       &       &       & T_5    & 22.66 \\
\hline
$^{70}$Ga $\rightarrow$ $^{70}$Ge & 1^+_1 & T_1    & 1.00$\times 10^{0}$ &       & 100.00 & {5.49$\times 10^{-4}$} & 0^+_1 & T_1    & 96.85 &       & 100.00 & 5.46$\times 10^{-4}$ & 0^+_1 & T_1    & 96.88 \\
          &       & T_3    & 1.00$\times 10^{0}$ &       & 100.00 & 5.50$\times 10^{-4}$ &       & T_3    & 96.85 &       & 100.00 & 5.48$\times 10^{-4}$ &       & T_3    & 96.87 \\
          &       & T_5    & 1.00$\times 10^{0}$ &       & 100.00 & 5.50$\times 10^{-4}$ &       & T_5    & 96.84 &       & 100.00 & 5.48$\times 10^{-4}$ &       & T_5    & 96.87 \\
\cline{8-10} \cline{14-16}
          &       &       &       &       &       &       & 2^+_1 & T_1    & 2.75  &       &       &       & 2^+_1 & T_1    & 2.73 \\
          &       &       &       &       &       &       &       & T_3    & 2.76  &       &       &       &       & T_3    & 2.74 \\
          &       &       &       &       &       &       &       & T_5    & 2.76  &       &       &       &       & T_5    & 2.74 \\
\cline{8-10} \cline{14-16}
          &       &       &       &       &       &       & 0^+_2 & T_1    & 0.40  &       &       &       & 0^+_2 & T_1    & 0.39 \\
          &       &       &       &       &       &       &       & T_3    & 0.40  &       &       &       &       & T_3    & 0.39 \\
          &       &       &       &       &       &       &       & T_5    & 0.40  &       &       &       &       & T_5    & 0.39 \\
\hline
$^{75}$Ge $\rightarrow$ $^{75}$As & 1/2^-_1 & T_1    & 1.00$\times 10^{0}$ &       & 100.00 & 1.49$\times 10^{-4}$ & 3/2^-_1 & T_1    & 84.23 &       & 100.00 & 1.48$\times 10^{-4}$ & 3/2^-_1 & T_1    & 84.29 \\
          &       & T_3    & 9.78$\times 10^{-1}$ &       & 99.97 & 1.46$\times 10^{-4}$ &       & T_3    & 84.20 &       & 99.97 & 1.45$\times 10^{-4}$ &       & T_3    & 84.24 \\
          &       & T_5    & 8.03$\times 10^{-1}$ &       & 99.20 & 1.14$\times 10^{-4}$ &       & T_5    & 83.57 &       & 99.22 & 1.19$\times 10^{-4}$ &       & T_5    & 83.60 \\
\cline{8-10} \cline{14-16}
          &       &       &       &       &       &       & 1/2^-_1 & T_1    & 0.44  &       &       &       & 1/2^-_1 & T_1    & 0.44 \\
          &       &       &       &       &       &       &       & T_3    & 0.44  &       &       &       &       & T_3    & 0.44 \\
          &       &       &       &       &       &       &       & T_5    & 0.44  &       &       &       &       & T_5    & 0.44 \\
\cline{8-10} \cline{14-16}
          &       &       &       &       &       &       & 3/2^-_2 & T_1    & 11.14 &       &       &       & 3/2^-_2 & T_1    & 11.12 \\
          &       &       &       &       &       &       &       & T_3    & 11.13 &       &       &       &       & T_3    & 11.12 \\
          &       &       &       &       &       &       &       & T_5    & 11.05 &       &       &       &       & T_5    & 11.04 \\
\cline{8-10} \cline{14-16}
          &       &       &       &       &       &       & 1/2^-_2 & T_1    & 0.16  &       &       &       & 1/2^-_2 & T_1    & 0.16 \\
          &       &       &       &       &       &       &       & T_3    & 0.16  &       &       &       &       & T_3    & 0.16 \\
          &       &       &       &       &       &       &       & T_5    & 16.00 &       &       &       &       & T_5    & 0.16 \\
\cline{8-10} \cline{14-16}
          &       &       &       &       &       &       & 1/2^-_3 & T_1    & 3.53  &       &       &       & 1/2^-_3 & T_1    & {3.50} \\
          &       &       &       &       &       &       &       & T_3    & 3.53  &       &       &       &       & T_3    & {3.51} \\
          &       &       &       &       &       &       &       & T_5    & 3.50  &       &       &       &       & T_5    & {3.48} \\
\cline{8-10} \cline{14-16}
          &       &       &       &       &       &       & {3/2^-_3} & T_1    & 0.50  &       &       &       & {3/2^-_3} & T_1    & 0.49 \\
          &       &       &       &       &       &       &       & T_3    & 0.50  &       &       &       &       & T_3    & 0.49 \\
          &       &       &       &       &       &       &       & T_5    & 0.49  &       &       &       &       & T_5    & 0.49 \\
\cline{8-10} \cline{14-16}
          &       &       &       &       &       &       & 3/2^-_4 & T_1    & 0.01  &       &       &       & 3/2^-_4 & T_1    & 0.01 \\
          &       &       &       &       &       &       &       & T_3    & 0.01  &       &       &       &       & T_3    & 0.01 \\
          &       &       &       &       &       &       &       & T_5    & 0.01  &       &       &       &       & T_5    & 0.01 \\
\cline{2-16}       
   & 7/2^+_1 & T_1    & 0.00$\times 10^{0}$ &       & 0.00  & 0.00$\times 10^{0}$ & 5/2^+_1 & T_1    & 0.00  &       & 0.00  & 0.00$\times 10^{0}$     & 5/2^+_1 & T_1    & 0.00 \\
          &       & T_3    & 1.76$\times 10^{-2}$ &       & 0.02  & 1.97$\times 10^{-8}$ &       & T_3    & 0.01  &       & 0.01  & 1.96$\times 10^{-8}$ &       & T_3    & 0.01 \\
          &       & T_5    & 1.25$\times 10^{-1}$ &       & 0.12  & 1.41$\times 10^{-7}$ &       & T_5    & 0.04  &       & 0.12  & 1.40$\times 10^{-7}$ &       & T_5    & 0.04 \\
\cline{8-10} \cline{14-16}
          &       &       &       &       &       &       & (5/2^+_2) & T_1    & 0.00  &       &       &       & (5/2^+_2) & T_1    & 0.00 \\
          &       &       &       &       &       &       &       & T_3    & 0.01  &       &       &       &       & T_3    & 0.01 \\
          &       &       &       &       &       &       &       & T_5    & 0.06  &       &       &       &       & T_5    & 0.06 \\
\hline
\hline
\end{tabular}
\label{tab:addlabel}
\end{table*}


\begin{table*}
\centering
~~~~~~TABLE (IV) continued
\vspace{0.5cm}
\begin{tabular}{cLLccccLLccccLLc}
\hline 
\hline
            
  & 3/2^-_1 & T_1    & 0.00$\times 10^{0}$ &       & 0.00  & 0.00$\times 10^{0}$ & 3/2^-_1 & T_1    & 0.00  &       & 0.00  & 0.00$\times 10^{0}$     & 3/2^-_1 & T_1    & 0.00 \\
          &       & T_3    & 1.09$\times 10^{-4}$ &       & 0.01  & 1.37$\times 10^{-8}$ &       & T_3    & 0.01  &       & 0.01      &  1.36$\times 10^{-8}$     &       & T_3    & 0.01 \\
          &       & T_5    & 4.50$\times 10^{-3}$ &       & 0.47  & 5.67$\times 10^{-7}$ &       & T_5    & 0.28  &       & 0.47  & 5.64$\times 10^{-7}$ &       & T_5    & 0.28 \\
\cline{8-10} \cline{14-16}
          &       &       &       &       &       &       & 1/2^-_1 & T_1    & 0.00  &       &       &       & 1/2^-_1 & T_1    & 0.00 \\
          &       &       &       &       &       &       &       & T_3    & 0.00  &       &       &       &       & T_3    & 0.00  \\
          &       &       &       &       &       &       &       & T_5    & 0.13  &       &       &       &       & T_5    & 0.13 \\
\cline{8-10} \cline{14-16}
          &       &       &       &       &       &       & 3/2^-_2 & T_1    & 0.00  &       &       &       & 3/2^-_2 & T_1    & 0.00 \\
          &       &       &       &       &       &       &       & T_3    & 0.00  &       &       &       &       & T_3    & 0.00 \\
          &       &       &       &       &       &       &       & T_5    & 0.04  &       &       &       &       & T_5    & 0.04 \\
\hline
$^{78}$Ge $\rightarrow$ $^{78}$As & 0^+_1 & T_1    & 1.00$\times 10^{0}$ &       & 100.00 & 1.43$\times 10^{-4}$ & 1^+_1 & T_1    & 39.54 &       & 100.00 & 1.41$\times 10^{-4}$ & 1^+_1 & T_1    & 39.71 \\
          &       & T_3    & 1.00$\times 10^{0}$ &       & 100.00 & 1.44$\times 10^{-4}$ &       & T_3    & 39.52 &       & 100.00 & 1.42$\times 10^{-4}$ &       & T_3    & 39.64 \\
          &       & T_5    & 1.00$\times 10^{0}$ &       & 100.00 & 1.44$\times 10^{-4}$ &       & T_5    & 39.51 &       & 100.00 & 1.42$\times 10^{-4}$ &       & T_5    & 39.62 \\
\cline{8-10} \cline{14-16}
          &       &       &       &       &       &       & 1^+_2 & T_1    & 14.43 &       &       &       & 1^+_2 & T_1    & 14.49 \\
          &       &       &       &       &       &       &       & T_3    & 14.43 &       &       &       &       & T_3    & 14.47 \\
          &       &       &       &       &       &       &       & T_5    & 14.42 &       &       &       &       & T_5    & 14.46 \\
\cline{8-10} \cline{14-16}
          &       &       &       &       &       &       & 1^+_3 & T_1    & 46.03 &       &       &       & 1^+_3 & T_1    & 45.81 \\
          &       &       &       &       &       &       &       & T_3    & 46.05 &       &       &       &       & T_3    & 45.89 \\
          &       &       &       &       &       &       &       & T_5    & 46.06 &       &       &       &       & T_5    & 45.92 \\
\hline
$^{81}$Se $\rightarrow$ $^{81}$Br & 1/2^-_1 & T_1    & 1.00$\times 10^{0}$ &       & 100.00 & 5.86$\times 10^{-4}$ & 3/2^-_1 & T_1    & 98.74 &       & 100.00 & 5.95$\times 10^{-4}$ & 3/2^-_1 & T_1    & 98.74 \\
          &       & T_3    & 9.31$\times 10^{-1}$ &       & 99.99 & 5.46$\times 10^{-4}$ &       & T_3    & 98.72 &       & 99.99 & 5.55$\times 10^{-4}$ &       & T_3    & 98.73 \\
          &       & T_5    & 7.29$\times 10^{-1}$ &       &  99.90     & 4.28$\times 10^{-4}$ &       & T_5    & 98.65 &       & 99.92 & 4.35$\times 10^{-4}$ &       & T_5    & 98.66 \\
\cline{8-10} \cline{14-16} 
          &       &       &       &       &       &       & 1/2^-_1 & T_1    & 0.02  &       &       &       & 1/2^-_1 & T_1    & 0.02 \\
          &       &       &       &       &       &       &       & T_3    & 0.02  &       &       &       &       & T_3    & 0.02 \\
          &       &       &       &       &       &       &       & T_5    & 0.02  &       &       &       &       & T_5    & 0.02 \\
\cline{8-10} \cline{14-16}
          &       &       &       &       &       &       & 3/2^-_2 & T_1    & 0.10  &       &       &       & 3/2^-_2 & T_1    & 0.10 \\
          &       &       &       &       &       &       &       & T_3    & 0.10  &       &       &       &       & T_3    & 0.10 \\
          &       &       &       &       &       &       &       & T_5    & 0.10  &       &       &       &       & T_5    & 0.10 \\
\cline{8-10} \cline{14-16}
          &       &       &       &       &       &       & (3/2^-)_3 & T_1    & 0.18  &       &       &       & (3/2^-)_3 & T_1    & 0.18 \\
          &       &       &       &       &       &       &       & T_3    & 0.18  &       &       &       &       & T_3    & 0.18 \\
          &       &       &       &       &       &       &       & T_5    & 0.18  &       &       &       &       & T_5    & 0.18 \\
\cline{8-10} \cline{14-16}
          &       &       &       &       &       &       & 3/2^-_4 & T_1    & 0.92  &       &       &       & 3/2^-_4 & T_1    & 0.91 \\
          &       &       &       &       &       &       &       & T_3    & 0.92  &       &       &       &       & T_3    & 0.92 \\
          &       &       &       &       &       &       &       & T_5    & 0.92  &       &       &       &       & T_5    & 0.92 \\
\cline{8-10} \cline{14-16}
          &       &       &       &       &       &       & 1/2^-_2 & T_1    & 0.03  &       &       &       & 1/2^-_2 & T_1    & 0.03 \\
          &       &       &       &       &       &       &       & T_3    & 0.03  &       &       &       &       & T_3    & 0.03 \\
          &       &       &       &       &       &       &       & T_5    & 0.03  &       &       &       &       & T_5    & 0.03 \\
\hline 
\hline 

\end{tabular}
\label{tab:addlabel}
\end{table*}


\begin{table*}
\centering
\renewcommand\thetable{V.A}
\caption{Variation of bound state \bta decay contribution (in \%) for individual transition and to the total decay, with temperature at a free electron density. Column 5 and Column 7: Bound state \bta decay contribution in individual transition. Column 6 and  Column 8: contribution to total decay rate from bound state decay.  See text for more details.}
\vspace{1cm}
\begin{tabular}{cLLccMMcMM}
\hline
\hline  

\multicolumn{4}{c}{Transition Details} &       & \multicolumn{5}{c}{$n_e = 1{\times}10^{27}$ cm$^{-3}$} \\
\cline{6-10}
 \multirow{7}{*}{Decay} & \multirow{7}{*}{$J^\pi_p$} & \multirow{7}{*}{$J^\pi_d$} &  \multirow{7}{*} {$Q_n$(keV)}    &       & \multicolumn{2}{c}{$T = 1{\times}10^{8}$ K} &  & \multicolumn{2}{c}{$T = 5{\times}10^{8}$ K} \\
\cline{1-4}\cline{6-7}\cline{9-10}    

 &  &  &  &    & Bound State Decay Contribution (\%) in Individual Transition & Bound State Decay Contribution (\%) in Total \bta Decay &    & Bound State Decay Contribution (\%) in Individual Transition & Bound State Decay Contribution (\%) in Total \bta Decay \\
\hline    

$^{59}$Fe  $\rightarrow$ $^{59}$Co & 3/2^-_1 & 3/2^-_1 & 465.7 &       & 7.86  & 12.18 &       & 7.81  & 11.22 \\
          &       & 3/2^-_2 & 273.4 &       & 15.80 &       &       & 15.62 &  \\
          &       & 1/2^-_1 & 130.7 &       & 34.28 &       &       & 33.56 &  \\
          &       & 5/2^-_1 & 83.4  &       & 49.47 &       &       & 48.14 &  \\
\cline{1-4}\cline{6-6}\cline{9-9}         
 & 1/2^-_1 & 3/2^-_1 & 752.723 &       &       &       &       & 3.74  &  \\
          &       & 3/2^-_2 & 560.423 &       &       &       &       & 5.96  &  \\
          &       & 1/2^-_1 & 417.723 &       &       &       &       & 9.09  &  \\
\cline{1-4}\cline{6-6}\cline{9-9}         
 & 5/2^-_1 & 7/2^-_1 & 2037.87 &       &       &       &       & 0.56  &  \\
          &       & 3/2^-_1 & 938.57 &       &       &       &       & 2.56  &  \\
          &       & 3/2^-_2 & 746.27 &       &       &       &       & 3.79  &  \\
          &       & 5/2^-_1 & 556.27 &       &       &       &       & 6.03  &  \\
          &       & 7/2^-_2 & 293.18 &       &       &       &       & 14.37 &  \\
\hline   
$^{60}$Co  $\rightarrow$ $^{60}$Ni & 5^+_1 & 4^+_1 & 317.047 &       & 14.15 & 12.53 &       & 14.02 & 1.59 \\
\cline{2-4}\cline{6-6}\cline{9-9}         
 & 2^+_1 & 2^+_1 & 1549.19 &       & 1.08  &       &       & 1.08  &  \\
          &       & 2^+_2 & 723.09 &       & 4.36  &       &       & 4.34  &  \\
\cline{2-4}\cline{6-6}\cline{9-9}      
    & 4^+_1 & 4^+_1 & 594.247 &       &       &       &       & 5.90  &  \\
          &       & 3^+_1 & 473.94 &       &       &       &       & 8.23  &  \\
\cline{2-4}\cline{6-6}\cline{9-9}     
     & 3^+_1 & 2^+_1 & 1778.7 &       &       &       &       & 0.81  &  \\
          &       & 2^+_2 & 952.6 &       &       &       &       & 2.72  &  \\
          &       & 4^+_1 & 605.447 &       &       &       &       & 5.74  &  \\
          &       & 3^+_1 & 485.14 &       &       &       &       & 7.96  &  \\
\cline{2-3}\cline{6-6}\cline{9-9}     
     & 5^+_2 & 4^+_1 & 752.757 &       &       &       &       & 4.06  &  \\
          &       & 4^+_2 & 138.64 &       &       &       &       & 33.69 &  \\
\hline    
$^{61}$Co  $\rightarrow$ $^{61}$Ni & 7/2^-_1  & 5/2^-_1 & 1256.4 &       & 1.64  & 1.78  &      & 1.63  & 1.77 \\
          &       & 5/2^-_2 & 415.19 &       & 9.96  &       &       & 9.89  &  \\
          &       & 7/2^-_1 & 1320.8 &       & 10.26 &       &       & 10.18 &  \\
          &       & 7/2^-_2 & 308.56 &       & 14.63 &       &       & 14.49 &  \\
          &       & 5/2^-_3 & 191.46 &       & 25.12 &       &       & 24.74 &  \\
\hline   
$^{63}$Ni  $\rightarrow$ $^{63}$Cu & 1/2^-_1  & 3/2^-_1  & 66.945 &       & 62.03 & 61.91 &       & 60.38 & 28.43 \\
\cline{2-3}\cline{4-4}\cline{6-6}\cline{9-9}        
  & 5/2^-_1  & 3/2^-_1  & 154.095 &       & 32.94 &       &       & 32.34 &  \\
\cline{2-3}\cline{4-4}\cline{6-6}\cline{9-9}        
 & 3/2^-_1  & 3/2^-_1  & 222.495 &       & 22.78 &       &       & 22.48 &  \\
\hline  
$^{65}$Ni  $\rightarrow$ $^{65}$Cu & 5/2^-_1  & 3/2^-_1 & 2137.9 &       & 0.60  & 2.35  &       & 0.60  & 0.83 \\
          &       & 5/2^-_1 & 1022.3 &       & 2.61  &       &       & 2.60  &  \\
          &       & 7/2^-_1 & 656.1 &       & 5.50  &       &       & 5.47  &  \\
          &       & 5/2^-_2 & 514.5 &       & 7.93  &       &       & 7.89  &  \\
          &       & 3/2^-_2 & 2132.9 &      & 10.78 &       &       & 10.70 &  \\
          &       & 7/2^-_2 & 43.56 &       & 76.84 &       &       & 74.86 &  \\
          &       & 5/2^-_3 & 30.5  &       & 86.82 &       &       & 84.95 &  \\
\cline{2-3}\cline{4-4}\cline{6-6}\cline{9-9} 
        & 1/2^-_1  & 3/2^-_1 & 2201.27 &       & 0.56  &       &       & 0.56  &  \\
          &        & 1/2^-_1 & 1430.63 &       & 1.38  &       &       & 1.38  &  \\
          &        & 3/2^-_2 & 476.27  &       & 8.86  &       &       & 8.80  &  \\

\hline
\hline
\end{tabular}
\label{bound_contri_a}
\end{table*}


\begin{table*}
\centering
~~~~~~TABLE (V.A) continued
\vspace{0.5cm}
\begin{tabular}{cLLccMMcMM}
\hline
\hline  

\multicolumn{4}{c}{Transition Details} &       & \multicolumn{5}{c}{$n_e = 1{\times}10^{27}$ cm$^{-3}$} \\
\cline{6-10}

 \multirow{7}{*}{Decay} & \multirow{7}{*}{$J^\pi_p$} & \multirow{7}{*}{$J^\pi_d$} &  \multirow{7}{*} {$Q_n$(keV)}    &       & \multicolumn{2}{c}{$T = 1{\times}10^{8}$ K} &  & \multicolumn{2}{c}{$T = 5{\times}10^{8}$ K} \\
\cline{1-4}\cline{6-7}\cline{9-10}    

 &  &  &  &    & Bound State Decay Contribution (\%) in Individual Transition & Bound State Decay Contribution (\%) in Total \bta Decay &    & Bound State Decay Contribution (\%) in Individual Transition & Bound State Decay Contribution (\%) in Total \bta Decay \\
\hline    
$^{66}$Ni  $\rightarrow$ $^{66}$Cu & 0^+_1 & 1^+_1 & 251.9 &       & 19.86 & 19.86 &       & 19.63 & 19.63 \\
\hline
$^{64}$Cu  $\rightarrow$ $^{64}$Zn & 1^+_1 & 0^+_1 & 579.6 &       & 7.15  & 7.15  &       & 7.12  & 7.12 \\
\hline
$^{66}$Cu $\rightarrow$ $^{66}$Zn & {1^+_1} & {0^+_1} & 2640.9   &       & 0.41   & 0.59   &       & 0.41    & 0.59 \\
          &       & {2^+_1} & 1601.7  &                            & 1.20   &        &       & 1.19    &  \\
          &       & {2^+_2} & 768.1   &                            & 4.60   &        &       & 4.58    &  \\
          &       & {0^+_2} & 269.5   &                            & 19.65  &        &       & 19.43   &  \\
\hline
    $^{67}$Cu  $\rightarrow$ $^{67}$Zn & {3/2^-_1} & {5/2^-_1} & 561.7 &       & {7.49} & {10.57} &       & {7.45} & {10.51} \\
          &       & {1/2^-_1} & 468.4 &       & {9.72} &       &       & {9.66} &  \\
          &       & {3/2^-_1} & 377.1 &       & {13.01} &       &       & {12.90} &  \\
          &       & {3/2^-_2} & 168.2 &       & {32.00} &       &       & {31.47} &  \\
\hline
    $^{69}$Zn  $\rightarrow$ $^{69}$Ga & {1/2^-_1}  & {3/2^-_1} & 910.2 &       & {3.72}   & {3.72} &       & {3.71} & {3.71} \\
          &       & {1/2^-_1} & 591.8 &                                   & {7.43}   &        &       & {7.39} &  \\
          &       & {3/2^-_2} & 38.5  &                                   & {83.84}  &        &       & {82.03} &  \\
\hline
    $^{72}$Zn  $\rightarrow$ $^{72}$Ga & {0^+_1}  & {0^+_1} & 323.6 &       & {16.77} & {19.83} &       & {16.61} & {19.62} \\
          &       & {1^+_1} & 314.3 &       & {17.37} &       &       & {17.20} &  \\
          &       & {1^+_2} & 281.7 &       & {19.74} &       &       & {19.53} &  \\
          &       & {1^+_3} & 234.9 &       & {24.14} &       &       & {23.84} &  \\
\hline
    $^{70}$Ga  $\rightarrow$ $^{70}$Ge & {1^+_1} & {0^+_1} & 1651.7 &       & {1.31} & {1.52} &       & {1.31} & {1.52} \\
          &       & {2^+_1} & 612.2 &                                 & {7.55} &        &       & {7.51} &  \\
          &       & {0^+_2} & 436.1 &                                 & 12.21  &        &       & 12.12  &  \\
\hline
\hline
\end{tabular}
\end{table*}

\begin{table*}
\centering
~~~~~~TABLE (V.A) continued
\vspace{0.5cm}
\begin{tabular}{cLLccMMcMM}
\hline
\hline  

\multicolumn{4}{c}{Transition Details} &       & \multicolumn{5}{c}{$n_e = 1{\times}10^{27}$ cm$^{-3}$} \\
\cline{6-10}

 \multirow{7}{*}{Decay} & \multirow{7}{*}{$J^\pi_p$} & \multirow{7}{*}{$J^\pi_d$} &  \multirow{7}{*} {$Q_n$(keV)}    &       & \multicolumn{2}{c}{$T = 1{\times}10^{8}$ K} &  & \multicolumn{2}{c}{$T = 5{\times}10^{8}$ K} \\
\cline{1-4}\cline{6-7}\cline{9-10}    

 &  &  &  &    & Bound State Decay Contribution (\%) in Individual Transition & Bound State Decay Contribution (\%) in Total \bta Decay &    & Bound State Decay Contribution (\%) in Individual Transition & Bound State Decay Contribution (\%) in Total \bta Decay \\
\hline 

$^{75}$Ge  $\rightarrow$ $^{75}$As & 1/2^-_1 & 3/2^-_1 & 1177.2 &       & 2.71   & 3.13  &       & 2.70  & 3.12 \\
          &       & 1/2^-_1 & 978.594  &                          & 3.78   &       &       & 3.76  &  \\
          &       & 3/2^-_2 & 912.542  &                          & 4.26   &       &       & 4.25  &  \\
          &       & 1/2^-_2 & 708.46   &                          & 6.44   &       &       & 6.41  &  \\
          &       & 1/2^-_3 & 592.2    &                          & 8.46   &       &       & 8.42  &  \\
          &       & 3/2^-_3 & 559.52   &                          & 9.19   &       &       & 9.14  &  \\
          &       & 3/2^-_4 & 311.8    &                          & 19.66  &       &       & 19.47 &  \\
          &       & 3/2^-_5 & 113.9    &                          & 50.66  &       &       & 49.64 &  \\
          &       & 3/2^-_6 & 102.7    &                          & 54.47  &       &       & 53.33 &  \\
          &       & 1/2^-_4 & 50.2     &                          & 79.38  &       &       & 77.67 &  \\
          &       & 1/2^-_5 & 5.2      &                          & 100.00 &       &       & 100.00  &  \\
\cline{2-4}\cline{6-6}\cline{9-9}   
          & 7/2^+_1 & 9/2^+_1 & 1012.97 &       &       &       &       & 3.54  &  \\     
          &       & 5/2^+_1 & 916.232   &       &       &       &       & 4.22  &  \\
          &       & 5/2^+_2 & 236.09    &       &       &       &       & 26.36 &  \\
          &       & 5/2^+_3 & 216.69    &       &       &       &       & 28.74 &  \\
          &       & 9/2^+_2 & 55.88     &       &       &       &       & 74.37 &  \\
          &       & 5/2^+_4 & 14.59     &       &       &       &       & 98.63 &  \\
\cline{2-4}\cline{6-6}\cline{9-9}       
          & 3/2^-_1 & 3/2^-_1 & 1430.35 &       &       &       &       & 1.87  &  \\
          &         & 1/2^-_1 & 1231.74 &       &       &       &       & 2.49  &  \\
          &         & 3/2^-_2 & 1165.69 &       &       &       &       & 2.75  &  \\
          &         & 5/2^-_1 & 1150.81 &       &       &       &       & 2.82  &  \\
          &         & 1/2^-_2 & 961.61 &       &       &       &       & 3.88  &  \\
          &         & 5/2^-_2 & 857.94 &       &       &       &       & 4.71  &  \\
          &         & 1/2^-_3 & 845.35 &       &       &       &       & 4.83  &  \\
          &         & 3/2^-_3 & 812.67 &       &       &       &       & 5.15  &  \\
          &         & 3/2^-_4 & 564.95 &       &       &       &       & 9.01  &  \\
          &         & 3/2^-_5 & 367.05 &       &       &       &       & 16.03 &  \\
          &         & 3/2^-_6 & 355.85 &       &       &       &       & 16.65 &  \\
          &         & 1/2^-_4 & 303.35 &       &       &       &       & 20.09 &  \\
          &         & 1/2^-_5 & 258.35 &       &       &       &       & 23.99 &  \\
          &         & 3/2^-_7 & 226.85 &       &       &       &       & 27.45 &  \\
          &         & 3/2^-_8 & 80.95  &       &       &       &       & 61.83 &  \\
          &         & 3/2^-_9 & 59.55  &       &       &       &       & 73.32 &  \\
          &         & 5/2^-_3 & 10.15  &       &       &       &       & 99.75 &  \\  
\hline 
$^{78}$Ge $\rightarrow$ $^{78}$As  & 0^+_1 & 1^+_1 & 677.7 &       & 6.90  & 10.04 &       & 6.87  & 9.98 \\
          &       & 1^+_2 & 661.1 &                          & 7.17  &       &       & 7.14  &  \\
          &       & 1^+_3 & 419   &                          & 13.67 &       &       & 13.57 &  \\
\hline
$^{81}$Se $\rightarrow$ $^{81}$Br & 1/2^-_1 & 3/2^-_1 & 1585.3 &       & 1.75  & 1.81  &       & 1.75  & 1.82 \\
          &       & 1/2^-_1 & 1047.1                     &       & 3.82  &       &       & 3.81  &  \\
          &       & 3/2_2 & 1019.26                      &       & 4.01  &       &       & 3.99  &  \\
          &       & 3/2^-_3 & 935.4                      &       & 4.64  &       &       & 4.63  &  \\
          &       & 3/2^-_4 & 757.01                     &       & 6.57  &       &       & 6.55  &  \\
          &       & 1/2^-_2 & 480                        &       & 12.80 &       &       & 12.71 &  \\
          &       & 3/2^-_5 & 318.9                      &       & 21.27 &       &       & 21.07 &  \\
          &       & 3/2^-_6 & 49.4                       &       & 82.77 &       &       & 81.18 &  \\
          &       & 3/2^-_7 & 42.1                       &       & 86.86 &       &       & 85.34 &  \\
\hline
\hline
\end{tabular}
\end{table*}


\begin{table*}
\centering
\renewcommand\thetable{V.B}
\caption{Variation of bound state \bta decay contribution (in \%) for individual transition and to the total decay, with free electron density at a temperature. Column 5 and Column 7: Bound state \bta decay contribution (\%) in individual transition. Column 6 and  Column 8: contribution (\%) to total decay rate from bound state decay.  See text for more details.}
\vspace{1cm}
\begin{tabular}{cLLccMMcMM}
\hline
\hline  

\multicolumn{4}{c}{Transition Details} &       & \multicolumn{5}{c}{$T = 3{\times}10^{8}$ K} \\
\cline{6-10}

 \multirow{7}{*}{Decay} & \multirow{7}{*}{$J^\pi_p$} & \multirow{7}{*}{$J^\pi_d$} &  \multirow{7}{*} {$Q_n$(keV)}    &       & \multicolumn{2}{c}{$n_e = 1{\times}10^{26}$ cm$^{-3}$} &  & \multicolumn{2}{c}{$n_e = 1{\times}10^{27}$ cm$^{-3}$} \\
\cline{1-4}\cline{6-7}\cline{9-10}    

 &  &  &  &    & Bound State Decay Contribution (\%) in Individual Transition & Bound State Decay Contribution (\%) in Total \bta Decay &    & Bound State Decay Contribution (\%) in Individual Transition & Bound State Decay Contribution (\%) in Total \bta Decay \\
\hline    
 $^{59}$Fe  $\rightarrow$ $^{59}$Co & 3/2^-_1 & 3/2^-_1 & 465.7 &       & 8.02  & 12.36 &       & 7.83 & 12.11 \\
          &       & 3/2^-_2 & 273.4 &                             & 15.94 &       &       & 15.67 &  \\
          &       & 1/2^-_1 & 130.7 &                             & 33.90 &       &       & 33.75 &  \\
          &       & 5/2^-_1 & 83.4  &                             & 48.32 &       &       & 48.46 &  \\
\hline    
$^{60}$Co  $\rightarrow$ $^{60}$Ni & 5^+_1 & 4^+_1 & 317.047 &       & 14.32 & 2.18  &       & 14.05 & 2.11 \\
\cline{2-4}\cline{6-6}\cline{9-9}       
                             & 2^+_1 & 2^+_1 & 1549.19 &       & 1.12  &       &       & 1.08 &  \\
                             &       & 2^+_2 & 723.09  &       & 4.46  &       &       & 4.34 &  \\
\cline{2-4}\cline{6-6}\cline{9-9}        
                             & 4^+_1 & 4^+_1 & 594.247 &       & 6.07  &       &       & 5.91 &  \\
                             &       & 3^+_1 & 473.94  &       & 8.45  &       &       & 8.25 &  \\
\hline
$^{61}$Co  $\rightarrow$ $^{61}$Ni & 7/2^-_1  & 5/2^-_1 & 1256.4 &       & 1.69  & 1.83  &       & 1.63 & 1.77 \\
          &       & 5/2^-_2 & 415.19 &       & 10.27 &       &       & 9.91 &  \\
          &       & 7/2^-_1 & 1320.8 &       & 10.57 &       &       & 10.20 &  \\
          &       & 7/2^-_2 & 308.56 &       & 14.99 &       &       & 14.53 &  \\
          &       & 5/2^-_3 & 191.46 &       & 25.39 &       &       & 24.84 &  \\
\hline
$^{63}$Ni  $\rightarrow$ $^{63}$Cu & 1/2^-_1  & 3/2^-_1  & 66.945 &       & 60.36 & 35.45 &       & 60.76 & 35.20 \\
          & 5/2^-_1  & 3/2^-_1  & 154.095 &                         & 32.71 &       &       & 32.50 &  \\
          & 3/2^-_1  & 3/2^-_1  & 222.495 &                         & 22.85 &       &       & 22.56 &  \\
\hline
$^{65}$Ni  $\rightarrow$ $^{65}$Cu & 5/2^-_1  & 3/2^-_1 & 2137.9 &       & 0.62  & 1.19  &       & 0.60 & 1.14 \\
          &       & 5/2^-_1 & 1022.3 &                             & 2.68  &       &       & 2.60 &  \\
          &       & 7/2^-_1 & 656.1 &                              & 5.63  &       &       & 5.48 &  \\
          &       & 5/2^-_2 & 514.5 &                              & 8.09  &       &       & 7.90 &  \\
          &       & 3/2^-_2 & 2132.9 &                             & 10.96 &       &       & 10.73 &  \\
          &       & 7/2^-_2 & 43.56 &                              & 74.52 &       &       & 75.29 &  \\
          &       & 5/2^-_3 & 30.5  &                              & 84.42 &       &       & 85.35 &  \\
\cline{2-4}\cline{6-6}\cline{9-9}          
          & 1/2^-_1  & 3/2^-_1 & 2201.27 &       & 0.58   &       &       & 0.56 &  \\
          &          & 1/2^-_1 & 1430.63 &       & 1.42   &       &       & 1.38 &  \\
          &          & 3/2^-_2 & 476.27  &       & 9.03   &       &       & 8.82 &  \\
\hline
$^{66}$Ni  $\rightarrow$ $^{66}$Cu & 0^+_1 & 1^+_1 & 251.9 &       & 19.99 & 19.99 &       & 19.69 & 19.69 \\
\hline
$^{64}$Cu  $\rightarrow $ $^{64}$Zn & 1^+_1 & 0^+_1 & 579.6 &       & 7.31  & 7.31  &       & 7.12 & 7.12 \\
\hline
$^{66}$Cu $\rightarrow$ $^{66}$Zn & 1^+_1 & 0^+_1  & 2640.9 &       & 0.42  & 0.61  &       & 0.41  & 0.59 \\
          &                 & 2^+_1 & 1601.7 &                & 1.23  &       &       & 1.19  &  \\
          &                 & 2^+_2 & 768.7  &                & 4.72  &       &       & 4.59  &  \\
          &                 & 0^+_2 & 269.2  &                & 19.79 &       &       & 19.49 &  \\
\hline
$^{67}$Cu  $\rightarrow$ $^{67}$Zn & 3/2^-_1  & 5/2^-_1 & 561.7 &       & 7.65  & 10.77 &       & 7.46 & 10.52 \\
          &       & 1/2^-_1 & 468.4 &       & 9.91 &       &       & 9.68 &  \\
          &       & 3/2^-_1 & 377.1 &       & 13.20 &       &       & 12.93 &  \\
          &       & 3/2^-_2 & 168.2 &       & 31.85 &       &       & 31.61 &  \\
\hline
$^{69}$Zn  $\rightarrow$ $^{69}$Ga & 1/2^-_1 & 3/2^-_1 & 910.2 &       & 3.82  & 3.82  &       & 3.71 & 3.72 \\
          &       & 1/2^-_1 & 591.8 &                                & 7.72  &       &       & 7.40 &  \\
          &       & 3/2^-_2 & 38.5  &                                & 81.86 &       &       & 82.43 &  \\
\hline
    $^{72}$Zn  $\rightarrow$ $^{72}$Ga & 0^+_1 & 0^+_1 & 323.6 &       & 16.96 & 19.99 &       & 16.66 & 19.68 \\
          &                      & 1^+_1   & 314.3 &             & 17.55 &       &       & 17.25 &  \\
          &                      & 1^+_2   & 281.7 &             & 19.90 &       &       & 19.59 &  \\
          &                      & 1^+_3   & 234.9 &             & 24.23 &       &       & 23.92 &  \\
\hline
\hline
\end{tabular}
\label{bound_contri_b}
\end{table*}


\begin{table*}
\centering
~~~~~~TABLE (V.B) continued
\vspace{1cm}
\begin{tabular}{cLLccMMcMM}
\hline
\hline  

\multicolumn{4}{c}{Transition Details} &       & \multicolumn{5}{c}{$T = 3{\times}10^{8}$ K} \\
\cline{6-10}

 \multirow{7}{*}{Decay} & \multirow{7}{*}{$J^\pi_p$} & \multirow{7}{*}{$J^\pi_d$} &  \multirow{7}{*} {$Q_n$(keV)}    &       & \multicolumn{2}{c}{$n_e = 1{\times}10^{26}$ cm$^{-3}$} &  & \multicolumn{2}{c}{$n_e = 1{\times}10^{27}$ cm$^{-3}$} \\
\cline{1-4}\cline{6-7}\cline{9-10}    

 &  &  &  &    & Bound State Decay Contribution (\%) in Individual Transition & Bound State Decay Contribution (\%) in Total \bta Decay &    & Bound State Decay Contribution (\%) in Individual Transition & Bound State Decay Contribution (\%) in Total \bta Decay \\
\hline 

    $^{70}$Ga  $\rightarrow$ $^{70}$Ge & 1^+_1 & 0^+_1 & 1651.7 &       & 1.35  & 1.57  &       & 1.31  & 1.52 \\
          &                              & 2^+_1 & 612.2  &       & 7.72  &       &       & 7.52  &  \\
          &                              & 0^+_2 & 436.1  &       & 12.41 &       &       & 12.15 &  \\
\hline
    $^{75}$Ge  $\rightarrow$ $^{75}$As & 1/2^-_1 & 3/2^-_1   & 1177.2  &       & 2.79  & 3.22  &       & 2.71 & 3.12 \\
          &                                & 1/2^-_1   & 978.594 &       & 3.88  &       &       & 3.77 &  \\
          &                                & 3/2^-_2   & 912.542 &       & 4.37  &       &       & 4.25 &  \\
          &                                & 1/2^-_2   & 708.46  &       & 6.59  &       &       & 6.42 &  \\
          &                                & 1/2^-_3   & 592.2   &       & 8.64  &       &       & 8.43 &  \\
          &                                & 3/2^-_3   & 559.52  &       & 9.37  &       &       & 9.15 &  \\
          &                                & 3/2^-_4   & 311.8   &       & 19.83 &       &       & 19.52 &  \\
          &                                & 3/2^-_5   & 113.9   &       & 49.81 &       &       & 49.90 &  \\
          &                                & 3/2^-_6   & 102.7   &       & 53.43 &       &       & 53.61 &  \\
          &                                & 1/2^-_4   & 50.2    &       & 77.25 &       &       & 78.06 &  \\
          &                                & 1/2^-_5   & 5.2     &       & 100.00 &       &       & 100.00 &  \\
\cline{2-4}\cline{6-7}\cline{9-10} 
          & 7/2^+_1 & 9/2^+_1 & 1012.97  &       & 3.65  &       &       & 3.55 &  \\       
          &         & 5/2^+_1 & 916.232  &       & 4.34  &       &       & 4.22 &  \\
          &       & 5/2^+_2 & 236.09     &       & 26.75 &       &       & 26.45 &  \\
          &       & 5/2^+_3 & 216.69     &       & 29.12 &       &       & 28.84 &  \\
          &       & 9/2^+_2 & 55.88      &       & 74.02 &       &       & 74.75 &  \\
          &       & 5/2^+_4 & 14.59      &       & 98.26 &       &       & 98.76 &  \\     
\cline{2-4}\cline{6-7}\cline{9-10}      
          & 3/2^-_1 & 3/2^-_1 & 1430.35 &       & 1.93  &       &       & 1.87 &  \\
          &       & 1/2^-_1 & 1231.74 &       & 2.56  &       &       & 2.49 &  \\
          &       & 3/2^-_2 & 1165.69 &       & 2.84  &       &       & 2.76 &  \\
          &       & 5/2^-_1 & 1150.81 &       & 2.91  &       &       & 2.82 &  \\
          &       & 1/2^-_2 & 961.61 &        & 4.00  &       &       & 3.88 &  \\
          &       & 5/2^-_2 & 857.94 &        & 4.85  &       &       & 4.71 &  \\
          &       & 1/2^-_3 & 845.35 &        & 4.97  &       &       & 4.83 &  \\
          &       & 3/2^-_3 & 812.67 &       & 5.30  &       &       & 5.16 &  \\
          &       & 3/2^-_4 & 564.95 &       & 9.24  &       &       & 9.03 &  \\
          &       & 3/2^-_5 & 367.05 &        & 16.36 &       &       & 16.07 &  \\
          &       & 3/2^-_6 & 355.85 &        & 16.99 &       &       & 16.69  &  \\
          &       & 1/2^-_4 & 303.35 &        & 20.46 &       &       & 20.15 &  \\
          &       & 1/2^-_5 & 258.35 &       & 24.37 &       &       & 24.06 &  \\
          &       & 3/2^-_7 & 226.85 &       & 27.84 &       &       & 27.55 &  \\
          &       & 3/2^-_8 & 80.95 &          & 61.76 &       &       & 62.17 &  \\
          &       & 3/2^-_9 & 59.55 &       & 72.02 &       &       & 72.71 &  \\
          &       & 5/2^-_3 & 10.15 &        & 99.57 &       &       & 99.79 &  \\
\hline
$^{78}$Ge $\rightarrow$ $^{78}$As  & 0^+_1 & 1^+_1 & 677.7 &       & 7.06  & 10.24 &       & 6.88 & 10.00 \\
          &       & 1^+_2 & 661.1 &       & 7.33  &       &       & 7.15 &  \\
          &       & 1^+_3 & 419   &       & 13.88 &       &       & 13.59 &  \\
\hline
$^{81}$Se $\rightarrow$ $^{81}$Br & 1/2^-_1 & 3/2^-_1 & 1585.3 &       & 1.81  & 1.87  &       & 1.75 & 1.81 \\
          &       & 1/2^-_1 & 1047.1                     &       & 3.92  &       &       & 3.81 &  \\
          &       & 3/2^-_2 & 1019.26                    &       & 4.11  &       &       & 4.00 &  \\
          &       & 3/2^-_3 & 935.4                      &       & 4.77  &       &       & 4.63 &  \\
          &       & 3/2^-_4 & 757.01                     &       & 6.73  &       &       & 6.55 &  \\
          &       & 1/2^-_2 & 480                        &       & 13.02 &       &       & 12.74 &  \\
          &       & 3/2^-_5 & 318.9                      &       & 21.46 &       &       & 21.12 &  \\
          &       & 3/2^-_6 & 49.4                       &       & 80.81 &       &       & 81.55 &  \\
          &       & 3/2^-_7 & 42.1                       &       & 84.89 &       &       & 85.68 &  \\
\hline
\hline
\end{tabular}
\end{table*}


\begin{table*}
\centering
\renewcommand\thetable{VI}
\caption{Total \bta decay rate ($\lambda_{bare(s)}$ in s$^{-1}$) and half-life ($T_{1/2(bare(s))}$) of bare atoms for different density - temperature combinations and comparison with previous results (CT \cite{Cosner_ASS_1981} and TY \cite{Takahashi_ADNDT_1987}). Half-life of a nucleus is given in unit of min/ hr/ d /yr, as the abbreviations of minute/ hour/ day/ year, respectively. Here, $T_1$ stands for temperature $T = 1{\times}10^{8}$ K, $T_3$ for $T = 3{\times}10^{8} $ K and $T_5$ for $T = 5{\times}10^{8} $ K. R is the ratio of calculated neutral atom terrestrial half-life to bare atom stellar half-life, i.e., $R = T_{1/2(t)}/ T_{1/2(bare(s))}$. See text for details.}
\vspace{1.0cm}
\begin{tabular}{cLccccccccMM}
\hline
\hline
          &    \multicolumn{8}{c}{This Work} &   & \multicolumn{2}{c}{Previous Results} \\
          
\cline{1-9}\cline{11-12}          

&     & \multicolumn{3}{c}{$n_e = 1{\times}10^{26}$ cm$^{-3}$} &    & \multicolumn{3}{c}{$n_e = 1{\times}10^{27}$ cm$^{-3}$} &    & {$n_e =1{\times}10^{27}$ cm$^{-3}$} & {$n_e = 1{\times}10^{26}$ cm$^{-3}$} \\

\cline{3-5}\cline{7-9}\cline{11-12}    

{Decay} &       & $\lambda_{bare(s)}$ & $T_{1/2(bare(s))}$ & R &   & $\lambda_{bare(s)}$ & $T_{1/2(bare(s))}$ & R &   & $\lambda_{bare(s)}(CT)$ & $\lambda_{bare(s)}(TY)$ \\
\hline

$^{59}$Fe $\rightarrow$ $^{59}$Co & T_1    & 2.13$\times 10^{-7}$ & 37.65 d & 1.095 &       & 2.08$\times 10^{-7}$ & 38.59 d & 1.068 &       &       & {1.80$\times 10^{-7}$} \\
          & T_3    & 2.14$\times 10^{-7}$ & 37.55d & 1.097 &       & 2.10$\times 10^{-7}$ & 38.21 d & 1.079 &       &       &  \\
          & T_5    & 2.31$\times 10^{-7}$ & 34.71 d & 1.187 &       & 2.28$\times 10^{-7}$ & 35.19 d & 1.171 &       &       & {1.88$\times 10^{-7}$} \\
\hline
$^{60}$Co $\rightarrow$ $^{60}$Ni & T_1    & 8.50$\times 10^{-9}$ & 943.21 d & 1.271 &       & 8.25$\times 10^{-9}$ & 971.70d & 1.234 &       & {1.30$\times 10^{-8}$} & {5.56$\times 10^{-9}$} \\
          & T_3    & 1.03$\times 10^{-7}$ & 78.22 d & 15.324 &       & 1.02$\times 10^{-7}$ & 78.59 d & 15.253 &       &       &  \\
          & T_5    & 2.37$\times 10^{-7}$ & 33.82 d & 35.448 &       & 2.36$\times 10^{-7}$ & 33.93 d & 35.327 &       & {1.60$\times 10^{-6}$} & {2.92$\times 10^{-7}$} \\
\hline
$^{61}$Co $\rightarrow$ $^{61}$Ni & T_1    & 1.03$\times 10^{-4}$ & 1.86 hr & 1.005 &       & 1.03$\times 10^{-4}$ & 1.87 hr & 0.999 &       &       & {1.17$\times 10^{-4}$} \\
          & T_3    & 1.03$\times 10^{-4}$ & 1.86 hr & 1.006 &       & 1.03$\times 10^{-4}$ & 1.87 hr & 1.001 &       &       &  \\
          & T_5    & 1.03$\times 10^{-4}$ & 1.86 hr & 1.006 &       & 1.03$\times 10^{-4}$ & 1.87 hr & 1.002 &       &       & {1.17$\times 10^{-4}$} \\
\hline
$^{63}$Ni $\rightarrow$ $^{63}$Cu & T_1    & 1.01$\times 10^{-9}$ & 21.75 yr & 2.055 &       & 9.25$\times 10^{-10}$ & 23.75 yr & 1.881 &       & {2.20$\times 10^{-8}$} & {3.69$\times 10^{-10}$} \\
          & T_3    & 5.17$\times 10^{-9}$ & 4.25 yr & 10.515 &       & 4.97$\times 10^{-9}$ & 4.42 yr & 10.118 &       &       &  \\
          & T_5    & 1.96$\times 10^{-8}$ & 1.12 yr & 39.812 &       & 1.90$\times 10^{-8}$ & 1.15 yr & 38.742 &       & {3.80$\times 10^{-7}$} & {8.19$\times 10^{-9}$} \\
\hline
$^{65}$Ni $\rightarrow$ $^{65}$Cu & T_1    & 7.41$\times 10^{-5}$ & 2.60 hr & 1.026 &       & 7.36$\times 10^{-5}$ & 2.65 hr & 1.019 &       &       & {7.64$\times 10^{-5}$} \\
          & T_3    & 2.23$\times 10^{-4}$ & 51.74 min & 3.093 &       & 2.23$\times 10^{-4}$ & 51.91 min & 3.083 &       &       &  \\
          & T_5    & 4.56$\times 10^{-4}$ & 25.31 min & 6.321 &       & 4.55$\times 10^{-4}$ & 25.37 min & 6.307 &       &       & {7.72$\times 10^{-5}$} \\
\hline
$^{66}$Ni $\rightarrow$ $^{66}$Cu & T_1    & 3.91$\times 10^{-6}$ & 49.25 hr & 1.177 &       & 3.78$\times 10^{-6}$ & 50.98 hr & 1.137 &       &       & {3.51$\times 10^{-6}$} \\
          & T_3    & 3.92$\times 10^{-6}$ & 49.07 hr & 1.181 &       & 3.83$\times 10^{-6}$ & 50.28 hr & 1.153 &       &       &  \\
          & T_5    & 3.93$\times 10^{-6}$ & 49.00 hr & 1.183 &       & 3.85$\times 10^{-6}$ & 50.03 hr & 1.159 &       &       & {3.51$\times 10^{-6}$} \\
\hline
$^{64}$Cu $\rightarrow$ $^{64}$Zn & T_1    & 5.42$\times 10^{-6}$ & 35.54 hr & 1.049 &       & 5.34$\times 10^{-6}$ & 36.08 hr & 1.033 &       & {1.20$\times 10^{-5}$} & {6.06$\times 10^{-6}$} \\
          & T_3    & 5.43$\times 10^{-6}$ & 35.48 hr & 1.050 &       & 5.37$\times 10^{-6}$ & 35.87 hr & 1.039 &       &       &  \\
          & T_5    & 5.43$\times 10^{-6}$ & 35.46 hr & 1.051 &       & 5.38$\times 10^{-6}$ & 35.79 hr & 1.041 &       & {1.30$\times 10^{-5}$} & {5.81$\times 10^{-6}$} \\
\hline
$^{66}$Cu $\rightarrow$ $^{66}$Zn & T_1    & 1.55$\times 10^{-3}$ & 7.45 min & 0.996 &       & 1.55$\times 10^{-3}$ & 7.47 min & 0.993 &       &       & 2.27$\times 10^{-3}$ \\
          & T_3    & 1.55$\times 10^{-3}$ & 7.45 min & 0.995 &       & 1.55$\times 10^{-3}$ & 7.47 min & 0.993 &       &       &  \\
          & T_5    & 1.52$\times 10^{-3}$ & 7.61 min & 0.975 &       & 1.52$\times 10^{-3}$ & 7.62 min & 0.973 &       &       & 2.21$\times 10^{-3}$ \\
\hline
$^{67}$Cu $\rightarrow$ $^{67}$Zn & T_1    & 3.03$\times 10^{-6}$ & 63.52 hr & 1.080 &       & 2.97$\times 10^{-6}$ & 64.81 hr & 1.059 &       &       & 3.11$\times 10^{-6}$ \\
          & T_3    & 3.04$\times 10^{-6}$ & 63.37 hr & 1.083 &       & 2.99$\times 10^{-6}$ & 64.31 hr & 1.067 &       &       &  \\
          & T_5    & 3.04$\times 10^{-6}$ & 63.32 hr & 1.084 &       & 3.00$\times 10^{-6}$ & 64.11 hr & 1.070 &       &       & 3.11$\times 10^{-6}$ \\
\hline
$^{69}$Zn $\rightarrow$ $^{69}$Ga & T_1    & 2.19$\times 10^{-4}$ & 52.74 min & 1.020 &       & 2.17$\times 10^{-4}$ & 53.24 min & 1.010 &       &       & 2.06$\times 10^{-4}$ \\
          & T_3    & 2.19$\times 10^{-4}$ & 52.68 min & 1.021 &       & 2.18$\times 10^{-4}$ & 53.04 min & 1.014 &       &       &  \\
          & T_5    & 2.19$\times 10^{-4}$ & 52.65 min & 1.021 &       & 2.18$\times 10^{-4}$ & 52.96 min & 1.015 &       &       & 2.06$\times 10^{-4}$ \\
\hline
$^{72}$Zn $\rightarrow$ $^{72}$Ga & T_1    & 6.81$\times 10^{-6}$ & 28.26 hr & 1.931 &       & 6.59$\times 10^{-6}$ & 29.19 hr & 1.869 &       &       & 4.14$\times 10^{-6}$ \\
          & T_3    & 6.83$\times 10^{-6}$ & 28.17 hr & 1.937 &       & 6.68$\times 10^{-6}$ & 28.83 hr & 1.892 &       &       &  \\
          & T_5    & 6.84$\times 10^{-6}$ & 28.13 hr & 1.940 &       & 6.71$\times 10^{-6}$ & 28.69 hr & 1.902 &       &       & 4.14$\times 10^{-6}$ \\
\hline
$^{70}$Ga $\rightarrow$ $^{70}$Ge & T_1    & 5.49$\times 10^{-4}$ & 21.03 min & 1.004 &       & 5.46$\times 10^{-4}$ & 21.14 min & 0.998 &       &       & 5.46$\times 10^{-4}$ \\
          & T_3    & 5.50$\times 10^{-4}$ & 21.01 min & 1.004 &       & 5.48$\times 10^{-4}$ & 21.10 min & 1.000 &       &       &  \\
          & T_5    & 5.50$\times 10^{-4}$ & 21.01 min & 1.004 &       & 5.48$\times 10^{-4}$ & 21.08 min & 1.001 &       &       & 5.46$\times 10^{-4}$ \\
\hline
$^{75}$Ge $\rightarrow$ $^{75}$As & T_1    & 1.49$\times 10^{-4}$ & 1.29 hr & 1.014 &       & 1.48$\times 10^{-4}$ & 1.30 hr & 1.006 &       &       & 1.40$\times 10^{-4}$ \\
          & T_3    & 1.46$\times 10^{-4}$ & 1.32 hr & 0.994 &       & 1.45$\times 10^{-4}$ & 1.33 hr & 0.988 &       &       &  \\
          & T_5    & 1.20$\times 10^{-4}$ & 1.59 hr & 0.816 &       & 1.20$\times 10^{-4}$ & 1.60 hr & 0.818 &       &       & 1.17$\times 10^{-4}$ \\
\hline
$^{78}$Ge $\rightarrow$ $^{78}$As & T_1    & 1.43$\times 10^{-4}$ & 1.34 hr & 1.076 &       & 1.41$\times 10^{-4}$ & 1.37 hr & 1.056 &       &       & 1.33$\times 10^{-4}$ \\
          & T_3    & 1.44$\times 10^{-4}$ & 1.34 hr & 1.078 &       & 1.42$\times 10^{-4}$ & 1.36 hr & 1.063 &       &       &  \\
          & T_5    & 1.44$\times 10^{-4}$ & 1.34 hr & 1.079 &       & 1.42$\times 10^{-4}$ & 1.35 hr & 1.066 &       &       & 1.33$\times 10^{-4}$ \\
\hline
$^{81}$Se $\rightarrow $ $^{81}$Br & T_1    & 5.98$\times 10^{-4}$ & 19.31 min & 1.004 &       & 5.83$\times 10^{-4}$ & 19.42 min & 0.998 &       &       & 6.24$\times 10^{-4}$ \\
          & T_3    & 5.57$\times 10^{-4}$ & 20.73 min & 0.936 &       & 5.44$\times 10^{-4}$ & 20.82 min & 0.932 &       &       &  \\
          & T_5    & 4.37$\times 10^{-4}$ & 26.43 min & 0.734 &       & 4.35$\times 10^{-4}$ & 26.53 min & 0.731 &       &       & 4.57$\times 10^{-4}$ \\
\hline
\hline
\end{tabular}
\label{total_decay}

\vspace{1cm}
\footnotetext{} {In case of $^{60}$Co, $^{75}$Ge, and $^{81}$Se the value of R is determined taking the contribution of ground state terrestrial half-life only.}
\end{table*}
\begin{table*}
\renewcommand\thetable{AI}
\centering
\caption{Details of Shell-Model calculations: particle partitions used. fp single particle state (sps) ordering is ($1f_{7/2}2p_{3/2}1f_{5/2}2p_{1/2}$), for fpg, sps ordering is ($1f_{5/2}2p_{3/2}2p_{1/2}1g_{9/2}$). In the particle partitions (mentioned in the last two columns) the numbers in the parentheses are minimum and maxinum number of particles, respectively, in sps. p: proton, n: neutron.}
\vspace{1cm}
\label{partition}
\resizebox{14.5cm}{!}{
\begin{tabular}{cccccc} 
\hline
\hline                                                                                                                                                                              
Parent $\rightarrow$ Daughter & Interaction & Code & Formalism & Parent   & Daughter  \\ 
&&&& Configuration  & Configuration \\ 
\hline
\hline
$^{59}$Fe $\rightarrow$ $^{59}$Co & fpd6        & OXBASH    & JT        & (14, 16), (0, 5),  & (14, 16), (0, 5), \\ 

&&&& (0, 5), (0, 4);  &  (0, 5), (0, 4);   \\ 
\hline
$^{60}$Co $\rightarrow$  $^{60}$Ni  & fpd6pn     & NuShellX  & PN        & p: (6, 7), (0, 1), & p: (6, 8), (0, 2),  \\

&&&&  (0, 1), (0, 0); &  (0, 0), (0, 2);  \\

                               &             &           &           & n: (7, 8), (3, 4), &  n: (6, 8), (2, 4),  \\
                               
&&&&  (0, 2), (0, 1); &   (0, 2), (0, 2);  \\  
\hline
$^{61}$Co $\rightarrow$ $^{61}$Ni  & fpd6n       & OXBASH    & JT        & (15, 16), (0, 6),   & (15, 16), (0, 6),  \\ 

&&&& (0, 6), (0, 4);   & (0, 6), (0, 4);  \\ 
\hline
$^{63}$Ni $\rightarrow$  $^{63}$Cu  & fpd6npn     & NuShellX  & PN        & p: (7, 8), (0, 1), & p: (8, 8), (0, 0),  \\

&&&&  (0, 1), (0, 1); &  (1, 1), (0, 0);  \\

                               &             &           &           & n: (8, 8), (0, 4), &  n: (8, 8), (0, 4),  \\
                               
&&&&  (0, 6), (0, 2); &   (0, 6), (0, 2);  \\                               
\hline
$^{65}$Ni $\rightarrow$  $^{65}$Cu & jun45  & NuShellX  & PN        & Untruncated     & Untruncated  \\ 
\hline
$^{66}$Ni $\rightarrow$ $^{66}$Cu     & fpd6n       & OXBASH    & JT   & (15, 16), (0, 8),  & (15, 16), (0, 8),  \\ 

&&&& (0, 11), (0, 4);    & (0, 11), (0, 4)  \\

\hline
$^{64}$Cu $\rightarrow$ $^{64}$Zn    & jun45      & NuShellX  & PN        & Untruncated   & Untruncated   \\ 
\hline
$^{66}$Cu $\rightarrow$ $^{66}$Zn    & jun45      & NuShellX  & PN        & Untruncated      & Untruncated  \\
\hline
$^{67}$Cu $\rightarrow$ $^{67}$Zn & fpd6n  & OXBASH    & JT  & (15, 16), (0, 8),   & (15, 16), (0, 8),  \\ 

&&&& (0, 12), (0, 4);   & (0, 12), (0, 4);  \\
\hline
$^{69}$Zn $\rightarrow$ $^{69}$Ga & fpd6n   & OXBASH    & JT        & (15, 16), (0, 8),  & (15, 16), (0, 8), \\ 
   
&&&& (1, 12), (0, 4);   & (1, 12), (0, 4); \\   
\hline
$^{72}$Zn $\rightarrow$ $^{72}$Ga     & jun45      & NuShellX  & PN  & Untruncated  & p: (0, 3), (0, 3),  \\

&&&&& (0, 2), (0, 1);  \\

&&&&& n: (0, 6), (0, 4),   \\ 

&&&&& (0, 2), (0, 10);   \\ 
\hline
 $^{70}$Ga $\rightarrow$ $^{70}$Ge   & gx1         & OXBASH    & JT        & (14, 16), (6, 8),  & (14, 16), (6,8),  \\ 
 
&&&& (4, 6), (0, 2);  & (4, 6), (0, 2);  \\ 
\hline
$^{75}$Ge $\rightarrow$ $^{75}$As & jun45  & NuShellX  & PN   & p: (0, 3), (0, 3),  & p: (0, 6), (0, 4),  \\

&&&& (0, 1), (0, 1);  & (0, 0), (0, 0);  \\

&&&& n: (0, 6), (0, 4) & n: (0, 6), (0, 4),  \\ 
 
&&&& (0, 2), (0, 6) & (0, 2), (0, 10)  \\ 
\hline
$^{78}$Ge $\rightarrow$ $^{78}$As & jun45      & NuShellX  & PN        & p: (0, 6), (0, 4),   & p: (0, 6), (0, 4),\\

&&&& (0, 1), (0, 1);   & (0, 1), (0, 1); \\

&&&& n: (0, 6), (0, 4),  & n: (0, 6), (0, 4), \\ 
                               
&&&& (0, 2), (0, 10);  & (0, 2), (0, 10);      \\ 
\hline
$^{81}$Se $\rightarrow$ $^{81}$Br        & jun45      & NuShellX  & PN        & p: (0, 6), (0, 4)   & p: (0, 6), (0, 4),  \\

&&&&   (0, 2), (0, 5);   &    (0, 2), (0, 0);  \\

&&&& n: (0, 6), (0, 4), & n: (0, 6), (0, 4) \\
                               
&&&& (0, 2), (0, 10);   &  (0, 2), (0, 10)                                                    \\
\hline
\hline
\end{tabular}}
\end{table*}

\end{document}